\pdfoutput=1

\documentclass{aa}

\usepackage{graphicx}
\usepackage{natbib}
\usepackage{ulem}

\usepackage[varg]{txfonts}

\usepackage{multirow}

\bibpunct{(}{)}{;}{a}{}{,} 

\usepackage{epsfig}

\usepackage{amssymb}

\usepackage{amsmath}
 
\usepackage{wasysym} 

\usepackage{xspace} 

\everymath{\displaystyle}

\usepackage{longtable}

\newcommand{\ind}[1]{_{\mathrm{#1}}}

\defcitealias{Alicavus_Soydugan_2014}{AS14} 

\defcitealias{Alicavus_2017}{Ali17} 

\defcitealias{Armstrong_2014}{Arm14} 

\defcitealias{Balaji_2015}{Blj15} 

\defcitealias{Balona_2011}{Bal11} 

\defcitealias{Balona_2013}{Bal13} 

\defcitealias{Balona_2014}{Bal14} 

\defcitealias{Balona_2015}{Bal15} 

\defcitealias{Balona_2015b}{Bal15b} 

\defcitealias{Beck_2014}{Bck14} 

\defcitealias{Bedding_2011}{Bed11} 

\defcitealias{Bell_2019}{Bel19} 

\defcitealias{Bognar_2015}{Bog15} 

\defcitealias{Bowman_2016}{Bow16} 

\defcitealias{Biersteke_2017}{Bie17} 

\defcitealias{Borucki_2011}{Bor11} 

\defcitealias{Borkovits_2014}{Brk14}

\defcitealias{Borkovits_2016}{Brk16}

\defcitealias{Bradley_2015}{Bra15} 

\defcitealias{Brogaard_2018}{Bro18} 

\defcitealias{cakirli_2013}{cak13}

\defcitealias{Ceillier_2017}{Cei17} 

\defcitealias{Conroy_2014}{Con14}

\defcitealias{Coughlin_2011}{Cou11}

\defcitealias{Debosscher_2011}{Deb11} 

\defcitealias{Debosscher_2013}{Deb13} 

\defcitealias{Demircan_2015}{Dem15} 

\defcitealias{Dimitrov_2017}{Dim17} 

\defcitealias{Dodson_Robinson_2012}{DR12} 

\defcitealias{Dong_2013}{Don13} 

\defcitealias{Frandsen_2013}{Fra13}

\defcitealias{Gaulme_2013}{Gau13}

\defcitealias{Gaulme_2014}{Gau14} 

\defcitealias{Gaulme_2016a}{Gau16}

\defcitealias{Gaulme_Guzik_2014}{GG14}

\defcitealias{Gies_2015}{Gie15} 

\defcitealias{Guo_2016}{Guo16} 

\defcitealias{Guo_2017}{Guo17} 

\defcitealias{Guo_2017b}{Guo17b} 

\defcitealias{Faigler_2015}{Fai15} 

\defcitealias{George_2018}{Geo18} 

\defcitealias{Getley_2017}{Get17} 

\defcitealias{Hekker_2010}{Hek10} 

\defcitealias{Hambleton_2013}{Ham13} 

\defcitealias{Hambleton_2016}{Ham16} 

\defcitealias{Hambleton_2018}{Ham18} 

\defcitealias{Helminiak_2016}{Hel16} 

\defcitealias{Helminiak_2017}{Hel17} 

\defcitealias{Helminiak_2017b}{Hel17b}

\defcitealias{Hummerich_2018}{Hum18}

\defcitealias{Karoff_2016}{Kar16}  

\defcitealias{Katsova_Nizamov_2018}{KN18} 

\defcitealias{Kjurkchieva_2016}{Kju16} 

\defcitealias{Kjurkchieva_2016b}{Kju16b} 

\defcitealias{Kochoska_2017}{Koc17}

\defcitealias{Kouzuma_2018}{Kou18}

\defcitealias{Kuszlewicz_2019}{Kus19} 

\defcitealias{Kurtz_2015b}{Kur15} 

\defcitealias{Lee_2014}{Lee14} 

\defcitealias{Lee_2015}{Lee15} 

\defcitealias{Lee_2016}{Lee16a} 

\defcitealias{Lee_2016b}{Lee16b} 

\defcitealias{Lee_2017}{Lee17} 

\defcitealias{Lehmann_2013}{Leh13} 

\defcitealias{LilloBox_2014}{LiB14} 

\defcitealias{Lurie_2017}{Lur17} 

\defcitealias{Liakos_2017}{Lia17} 

\defcitealias{Liakos_2018}{Lia18} 

\defcitealias{Liakos_Niarchos_2017}{LN17} 

\defcitealias{Maceroni_2014}{Mac14} 

\defcitealias{Manuel_Hambleton_2018}{MH18} 

\defcitealias{Masuda_2017}{Mas17} 

\defcitealias{Matson_2017}{Mat17} 

\defcitealias{McNamara_2012}{McN12} 

\defcitealias{Moya_2017}{Moy17} 

\defcitealias{Morton_2016}{Mor16} 

\defcitealias{Murphy_2015}{Mur15} 

\defcitealias{Murphy_2018}{Mur18} 

\defcitealias{Niemczura_2015}{Nie15} 

\defcitealias{Niemczura_2017}{Nie17} 

\defcitealias{Ozdarcan_2017}{Ozd17} 

\defcitealias{Orosz_2015}{Oro15} 

\defcitealias{Pablo_2012}{Pab12} 

\defcitealias{Ramsay_2014}{Ram14} 

\defcitealias{Rappaport_2013}{Rap13} 

\defcitealias{Rappaport_2015}{Rap15} 

\defcitealias{Rawls_2016}{Raw16} 

\defcitealias{Rowe_2010}{Row10} 

\defcitealias{Rowe_2015}{Row15} 

\defcitealias{Sowicka_2017}{Sow17} 

\defcitealias{Saio_2018}{Sai18} 

\defcitealias{Sandquist_2016}{Snd16} 

\defcitealias{Santerne_2016}{San16} 

\defcitealias{Schmitt_2014}{Sch14} 

\defcitealias{Shporer_2016}{Shp16} 

\defcitealias{Smullen_2015}{Smu15} 

\defcitealias{Southworth_2011}{Sou11} 

\defcitealias{Southworth_2015}{Sou15} 

\defcitealias{Themessl_2018}{The18} 

\defcitealias{Thompson_2012}{Tho12} 

\defcitealias{Turner_Maynard_2016}{TM16} 

\defcitealias{Turner_2015}{Tur15} 

\defcitealias{Turner_2019}{Tur19} 

\defcitealias{Uytterhoeven_2011}{Uyt11} 

\defcitealias{Welsh_2011}{Wel11} 

\defcitealias{Yakut_2015}{Yak15}

\defcitealias{Zhang_2017}{Zha17}

\defcitealias{Zhang_2017b}{Zha17b}

\defcitealias{Zhang_XB_2017}{ZXB17}

\defcitealias{Zhou_2010}{Zho10}

\defcitealias{Zimmerman_2017}{Zim17} 

\begin{document}

\title{Systematic search for stellar pulsators in the eclipsing binaries observed by Kepler}

\author{Patrick Gaulme\inst{\ref{inst1},\ref{inst2}} 
\and Joyce A. Guzik\inst{\ref{inst3}} 
}

\institute{Max-Planck-Institut f\"{u}r Sonnensystemforschung, Justus-von-Liebig-Weg 3, 37077, G\"{o}ttingen, Germany \email{gaulme@mps.mpg.de}\label{inst1}
\and 
Department of Astronomy, New Mexico State University, P.O. Box 30001, MSC 4500, Las Cruces, NM 88003-8001, USA \label{inst2}
\and
Los Alamos National Laboratory, XTD-NTA, MS T086, Los Alamos, NM 87545-2345, USA \label{inst3}
}

\titlerunning{Kepler Stellar Pulsators in Eclipsing Binaries}
\authorrunning {Gaulme \& Guzik}
\abstract{
Eclipsing binaries (EBs) are unique targets for measuring precise stellar properties and constrain stellar evolution models. In particular, it is possible to measure at the percent level masses and radii of both components of a double-lined spectroscopic EB. Since the advent of high-precision photometric space missions (MOST, CoRoT, Kepler, BRITE, TESS), the use of stellar pulsation properties to infer stellar interiors and dynamics constitutes a revolution for low-mass star studies. The Kepler mission has led to the discovery of thousands of classical pulsators such as $\delta$ Scuti and solar-like oscillators (main sequence and evolved), but also almost 3000 EBs with orbital periods shorter than 1100 days. We report the first systematic search for stellar pulsators in the entire Kepler eclipsing binary catalog. The focus is mainly aimed at discovering $\delta$ Scuti, $\gamma$ Doradus, red giant, and tidally excited pulsators. We developed a data inspection tool (DIT) that automatically produces a series of plots from the Kepler light-curves that allows us to visually identify whether stellar oscillations are present in a given time series. We applied the DIT to the whole Kepler eclipsing binary database and identified 303 systems whose light curves display oscillations, including 163 new discoveries. A total of 149 stars are flagged as $\delta$ Scuti (100 from this paper), 115 stars as $\gamma$ Doradus (69 new), 85 stars as red giants (27 new), 59 as tidally excited oscillators (29 new). There is some overlap among these groups, as some display several types of oscillations. Despite many of these systems are likely to be false positives, i. e., when an EB light curve is blended with a pulsator, this catalog gathers a vast sample of systems that are valuable for a better understanding of stellar evolution.}

\maketitle

\section{Introduction}
It is widely agreed that asteroseismology has become the most reliable way to infer global and internal properties of solar-like stars, from the main sequence to the red giant phase, since the remarkable success of the space-borne photometers CoRoT, Kepler, and TESS \citep{Baglin_2009,Borucki_2010,Ricker_2015}. The simplest and most popular application of asteroseismology consists of comparing the oscillation global properties  of a given star to the Sun's, and retrieving its mass and radius from the asteroseismic scaling relations \citep{Kjeldsen_Bedding_1995}. Since masses and radii lead to ages and distances, solar-like oscillators represent key tracers of the properties of stellar populations. This approach has been used to characterize large samples of stars observed by CoRoT and Kepler, which detected thousands of solar-like oscillation spectra for main-sequence (MS) to red-giant (RG) stars \citep{Chaplin_Miglio_2013, Chiappini_2015}. It also will play a central role in the ESA PLATO mission, for which asteroseismic inference is expected to constrain stellar and planetary properties of tens of thousands of systems.
  
Besides, classical pulsators such as  $\gamma$ Doradus ($\gamma$ Dor), $\delta$ Scuti ($\delta$ Sct), $\beta$ Cephei, slowly pulsating B-type stars (SPB), pulsating B stars exhibiting emission lines (Be), or rapidly oscillating Ap stars are becoming of prime importance too. These stars are more massive and hotter than the Sun and display pressure modes (e.g., $\delta$ Sct), gravity modes (e.g., $\gamma$ Dor), or both (hybrid). So far, asteroseismology for $\gamma$ Dor and $\delta$ Sct stars has been hindered because of difficulty of mode identification, rotational splitting, combination frequencies, mode selection, and mismatch between observed mode spectrum and theoretical predictions. However, very significant progress has been made in the last few years in deciphering the pulsation spectra, providing hope that asteroseismology may be feasible with these stars.  \citet{Bedding_2015} use \'echelle diagrams to study $\gamma$ Dor $g$-mode period spacings, and identify sequences of rotationally split multiplets of degrees 1 and 2.  \citet{Van_Reeth_2015a,Van_Reeth_2015b} show how to use nonuniform period spacings in Kepler $\gamma$ Dor stars as a seismic diagnostic. \citet{Garcia_Hernandez_2016} report progress in developing a method to use low-order $p$ modes to determine the mean density of $\delta$ Sct stars, and they apply it successfully to a CoRoT star, and \citet{Mirouh_2019} report theoretical studies about period spacing in fast-rotating $\delta$-Sct stars.  \citet{Kurtz_2015} show that complex frequency spectra of $\gamma$ Dor, SPB, and Be stars can be explained by just a few $g$-mode frequencies plus their combination frequencies.  It appears that the community is on the verge of a breakthrough in interpreting the frequency spectra of main-sequence pulsators and developing asteroseismic techniques to derive interior structure properties.

Given the importance of asteroseismology for both solar-like stars and classical pulsators, it is fundamental to identify a set of benchmarks to refine the stellar models on. Such benchmarks are stars whose main physical properties are known with high precision, especially mass, radius, metallicity and temperature. 
In the past decades, eclipsing binaries (EBs) have become very popular as benchmarks for stellar physics, as they provide accurate ways to measure masses, radii, and distances. It is possible to determine the mass and radius of each component of a double-line spectroscopic binary (SB2) from measurements of eclipse photometry and radial velocities. It is also possible to measure the mass of stars belonging to highly eccentric binary systems nicknamed ``heartbeat'' (HB) stars because their lightcurves recall electrocardiograms \citep[e.g.,][]{Welsh_2011}, hierarchical triple systems \citep[HTs, e.g.,][]{Borkovits_2016}, and visual binaries \citep[e.g.,][]{Marcadon_2018}.

Recent space missions based on high-precision photometry drastically changed the number of known eclipsing binaries, as well as the sensitivity of measurements  \citep[e.g.,][]{Prsa_2011, Debosscher_2011,Coughlin_2011}. The original Kepler mission has discovered 2,925 EBs\footnote{according to Villanova's Kepler EB catalog, updated on May 6, 2018, which includes 2909 systems (\url{http://keplerebs.villanova.edu/}), and \citet{Coughlin_2011} which lists 82 systems among which 16 are not in the Villanova catalog.}, including over 800 with periods from 10 to 1100 days, over 50 HB systems \citep[e.g.,][]{Thompson_2012,Beck_2014,Gaulme_2014,Shporer_2016}, and 222 triples displaying eclipse timing variations (ETVs) \citep[][]{Borkovits_2016}. Beyond Kepler, the CoRoT mission has discovered a few thousands of EB systems (Cilia Damiani, priv. comm.), and Kepler's extended mission, K2, has found almost 700 EBs in the first six fields of view (according to Villanova's database). 

As regards classical pulsators, much research about binary systems including a pulsator has been led for the past thirty years \citep{Szatmary_1990}. The majority concerns $\delta$-Sct pulsators. We refer the reader to the comprehensive reviews available in \citet{Liakos_Niarchos_2017} and \citet{Kahraman_2017}. In brief, there are 199 confirmed cases of binary systems containing at least a $\delta$-Sct pulsator \citep{Liakos_Niarchos_2017}, 87 of which being detached or semidetached eclipsing binaries, the other being visual, ellipsoidal variables and spectroscopic binaries. Among the catalog of 199 $\delta$-Sct in binaries, 29 are targets of the NASA original Kepler mission, among which 16 are eclipsing binaries. \citet{Debosscher_2011}, who were the first to look for oscillators in eclipsing binaries observed by Kepler, detected 14 classical pulsators in EBs, of which five are either $\gamma$ Dor or SPB. In addition, \citet{Gaulme_Guzik_2014} reported the identification of eight bona-fide classical pulsators among the Kepler EBs, one including a $\gamma$ Dor and seven $\delta$ Sct. Note that in addition to the eclipsing binaries, the measurement of fine frequency modulation of classical pulsators observed by Kepler led to the discovery of 341 new binary systems \citep{Shibahashi_Kurtz_2012,Murphy_2018}. These systems are wide binaries -- otherwise no fluctuation of mode frequencies would be measurable -- and are likely not to be eclipsing. 

As regards \textit{solar-like oscillators} belonging to EBs, all are red giants (RGs) detected by the Kepler mission \citep[][Benbakoura et al., in prep.]{Hekker_2010, Gaulme_2013, Gaulme_2014, Beck_2014,Beck_2015, Kuszlewicz_2019}. So far, eleven wide SB2 EBs including an oscillating RG have been fully characterized with the help of radial-velocity ground-based support \citep{Frandsen_2013,Rawls_2016, Gaulme_2016a, Brogaard_2018,Themessl_2018} and three more are part of the upcoming paper of Benbakoura et al. (in prep.). Note that an equivalent number of RG displaying oscillations have been detected in HB systems \citep[][]{Gaulme_2013, Gaulme_2014, Beck_2014, Beck_2015, Kuszlewicz_2019}, but most do not show eclipses and are single-line spectroscopic binaries (SB1s).

In the context of the preparation of the new ESA space mission PLATO where asteroseismology plays a key role, we estimate that it is a good time for an inventory of the stellar pulsators in eclipsing binaries, which is a unique class of stellar benchmarks. This motivated us to lead the first systematic search for any stellar pulsators in eclipsing binaries in the Kepler data. We focus on the original Kepler mission because we can legitimately consider the list of EBs to be complete, whereas the catalog of the extended K2 mission is not complete. We consider the sample of EBs that is publicly available on the Villanova webpage, in its most recent update (May 2018), and the systems from the \citet{Coughlin_2011} paper. The sum of the two catalogs contains 2925 systems, with orbital periods ranging from 0.05 to 1087 days. This global catalog is not in a strict sense a catalog of eclipsing binaries as about 600 ellipsoidal binaries are included. Ellipsoidal binaries are non-eclipsing tight binaries where the stellar oblateness (ellipsoidal shape) introduces a periodic modulation of the light curve. We focus on $\gamma$ Dor, $\delta$ Sct, and RG oscillators, but we also list all other types of pulsators that we are able to identify, in particular tidal pulsators, and a handful of possible SPB or white dwarf (WD) pulsators.

Our goal is to indicate whether oscillations are detected in the Kepler EBs. We classify the oscillators according to their types, but we do not model each eclipse light curve or attempt to identify the oscillation modes that we detect. The objective is to motivate future in-depth studies of the most interesting cases, which would involve complementary radial velocity (RV) measurements. In terms of organization we review the main properties of the Villanova EBs and we describe the methods employed to disentangle the stellar pulsations from the eclipse signal in Section 2.  We then present the detection of 303 systems where pulsations are detected, including 163 that have been discovered in the present study (Sect 3) before discussing the results (Section 4). 

\section{Method}
\subsection{Source of information}
\label{sect_source_info}
Most of the paper is based on the Villanova eclipsing binary database \citep[e.g.,][]{Prsa_2011, Slawson_2011, Matijevic_2012, Conroy_2014, Kirk_2016}, which contains information about 2909 systems that were extracted with the ``eclipsing binaries via artificial intelligence'' (EBAI) pipeline \citep{Prsa_2008}. EBAI  is a pipeline based on an artificial neural network that automatically extracts the sum of relative radii  $(R_1+R_2)/a$, temperature ratio $T_2/T_1$, orbital argument of the periastron $\omega$, eccentricity $e$ and inclination angle $i$ of a binary system. From the database, it is possible to download estimates of the orbital period, the date of primary eclipse, primary and secondary (if any) eclipse depths, durations (widths), relative separation, eclipse morphology, Kepler magnitude, and effective temperature from up to three different sources (Kepler Input Catalog, \citealt{Pinsonneault_2012}, and \citealt{Casagrande_2011} catalogs). Regarding data availability and history, it is also indicated whether short cadence (1 minute) or long cadence (30 minutes) data are available, and publications mentioning a given system exist. Besides, the detection of ETVs -- an indicator of interacting triple systems -- is flagged too. We optimized our data processing tools for long-cadence data and do not consider possible short-cadence data in this paper.

\begin{figure}[t]
\begin{center}
\includegraphics[width=9cm]{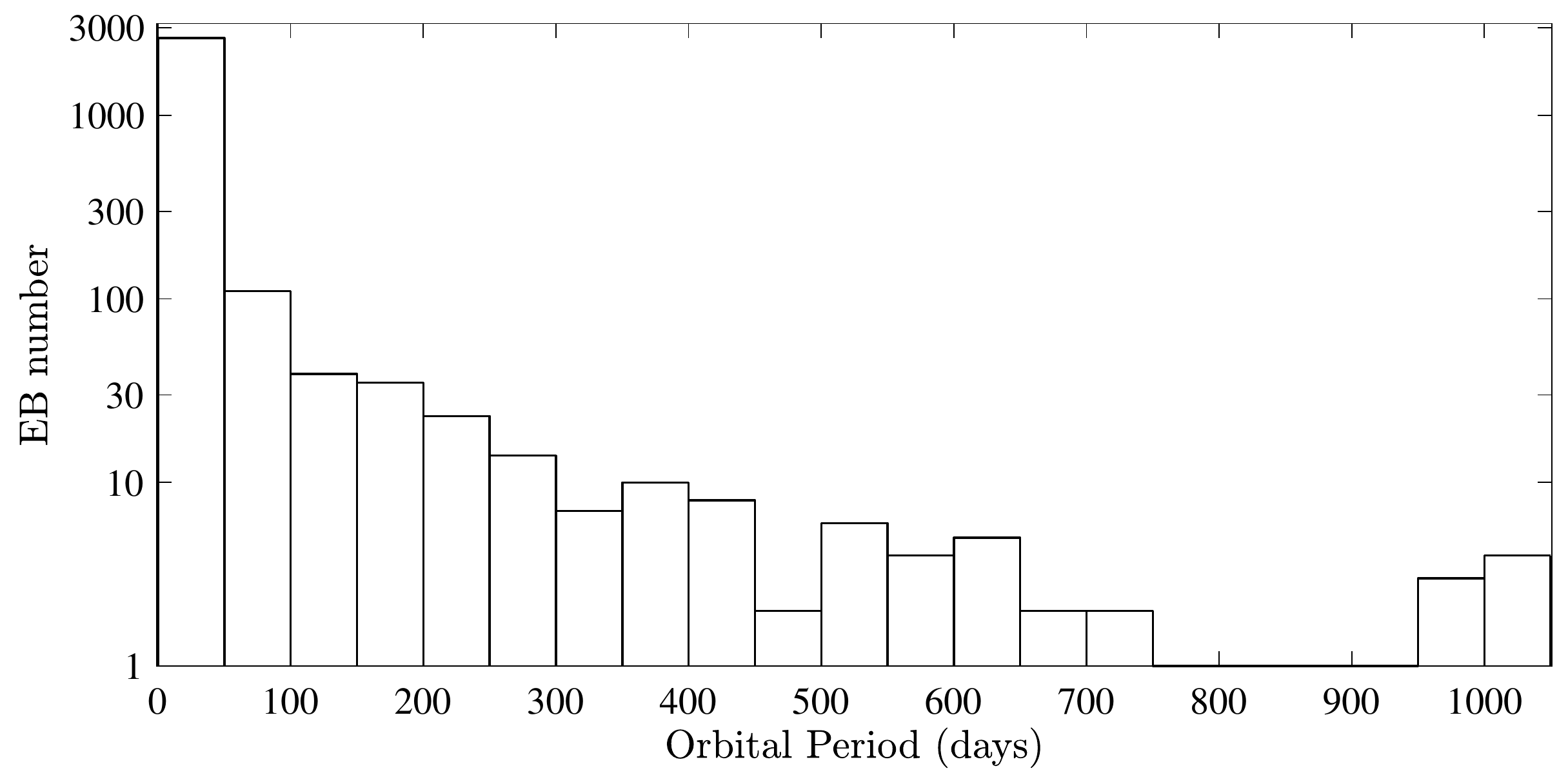}
\end{center}
\caption{Orbital period histogram of the Villanova eclipsing binary systems. The $y$-axis sampling is the logarithm with base ten of the number of systems per step. Out of 2909 systems total, 2631 (i.e., 90.4\,\%) display an orbital period less than 50 days. }\label{fig_histo_Porb}
\end{figure}

\begin{figure}[t]
\begin{center}
\includegraphics[width=9cm]{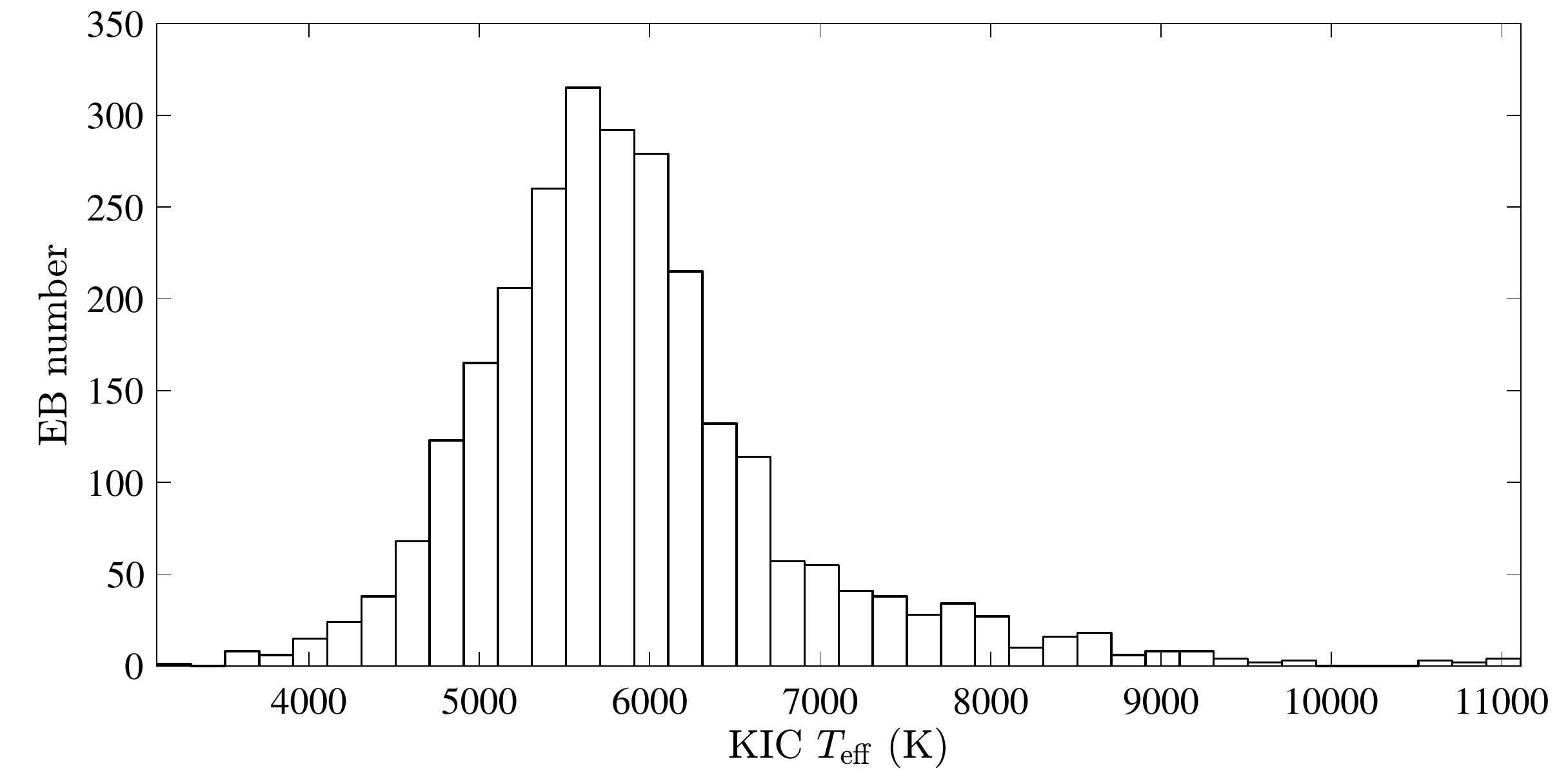}
\end{center}
\caption{Effective temperatures of the Villanova eclipsing binaries. Temperatures come from the Kepler Input Catalog (KIC) and are available for 2625 out of the 2909 systems (90.2\,\%). }\label{fig_histo_Teff} 
\end{figure}

\begin{figure}[t]
\begin{center}
\includegraphics[width=9cm]{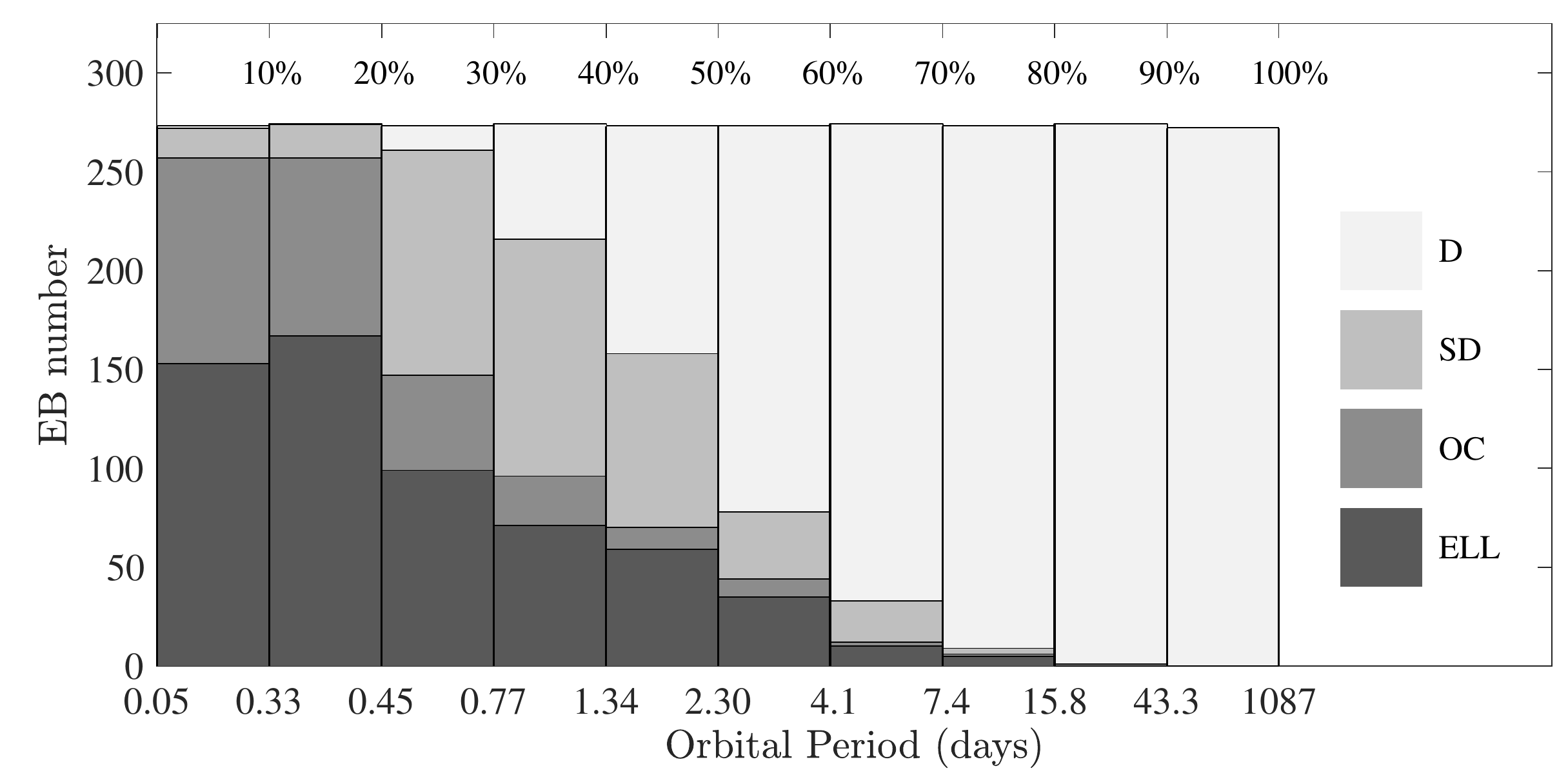}
\end{center}
\caption{Eclipse light curve classification according to \citet{Matijevic_2012} as a function of orbital period. The size of each bloc is adjusted to include 10\,\% of the total number of systems that were classified.  The total number of classified systems is 2734 and each chunk consists of $273\pm1$ systems. The median orbital period is 2.30 days, while the 10 and 90 percentiles are 0.33 and 43.3 days respectively. The labels ``D'', ``SD'', ``OC'', and ``ELL'' refer to detached, semi-detached, overcontact and ellipsoidal systems. A total of 1431 systems are D, 413 SD, 290 OC, and EBs 
599 ELL, i.e., a total of 2135 systems are classified as eclipsing binaries.  }\label{fig_type_syst}
\end{figure}

In Figs. \ref{fig_histo_Porb} and \ref{fig_histo_Teff}, we present the distribution of the EB sample as a function of orbital period and effective temperature. The shortest orbital period is 0.05 day, the maximum is 1087 days. Ten percent of the stars display a period shorter than 0.33 day (tenth percentile), and 10\,\% longer than 43.3 days. The median orbital period is 2.30 days (Fig. \ref{fig_type_syst}). Out of the 2625 systems with an estimated effective temperature, the histogram reveals that 545 are in the $[4500, 5300]$ K range, in which we expect red giant oscillators -- together with K dwarfs --, and 785 lie in the [6000-7500] K range where both $\delta$ Sct and $\gamma$ Dor stars are expected.

The eclipse light curves were classified by \citet{Matijevic_2012} who introduce the ``morphology'' parameter $c$, which is a measure of ``detachedness'' of the system. According to them, all systems with $c < 0.5$ are predominantly detached.  This parameter is built upon a grid of simulated light curves that range from well detached binaries to overcontact stars. The range of $c$ for semidetached systems broadly lies in the $0.5 < c < 0.7$ range. Overcontact systems dominate the  $0.7 < c < 0.8$ region, after which a mixture of ellipsoidal variables and systems with uncertain classification sets in, including many HB systems. Figure \ref{fig_type_syst} shows the histogram of the typology of the systems as a function of orbital period, based on this rough classification. Note that among the 2909 systems, 175 are not classified. Out of the 2734 systems that are classified, 1431 (52.3\,\%) are detached systems, 413 are semi-detached (15.1\,\%), 290 (10.6\,\%) contact binaries and 599 (21.9\,\%) are classified as ellipsoidal or miscellaneous. It is very likely that most ``ellipsoidal'' binaries with period shorter than one day are either contact or semi-detached systems. It is worth noticing that detached systems dominate the sample starting from the median period ($P\geq2.30$ day).

Even though most of the paper is based on the Villanova database, a handful of systems that were published by \citet{Coughlin_2011} have never been included in the Villanova database. We count 16 of these systems and include them in our analysis. We also add the RG in an eclipsing HB system recently studied by \citet{Kuszlewicz_2019}, which is not listed in the Villanova catalog. We performed our study with Kepler public light curves available on the Mikulski Archive for Space Telescopes (MAST)\footnote{\url{http://archive.stsci.edu/kepler/}}. We work with both the Simple Aperture Photometry (SAP, i.e., raw data) and the Pre-search Data Conditioning Simple Aperture Photometry (PDCSAP) light curves. The latter consist of time series that corrected from discontinuities, systematic errors and excess flux due to aperture crowding \citep{Twicken_2010}. Note that we do not seek for complementary information from the GAIA data release 2 catalog because binary stars are not included, and if some are, their parameters may be biased.

\subsection{Disentangling eclipses, surface activity, and oscillations}

Searching for pulsations in eclipsing binaries entails removing the photometric variability caused by binarity: eclipses and phase effects mainly. The methods we use are described in \citet{Gaulme_2013,Gaulme_2014,Gaulme_2016a} and here we provide a summary.

 The major challenge in concatenating light curves is to ensure photometric continuity before and after each interruption. The main cause of photometric jumps from quarter to quarter is the fact that the \textit{Kepler} telescope rotated four times a year, which implied that a given star would fall on four different chips. However, the pointing was fine enough that a star repeatedly covered the same group of pixels every four quarters. Light curves are obtained by adding the pixels of the masks that were designed for every star of the field of view. For  a given star, a mask was designed for each of \textit{Kepler}'s positions. Because of the photo response non-uniformity (PRNU) of the pixels and the changing size of the masks, the recorded flux changes. Both PRNU and varying mask areas lead to flux discontinuity that should be adjusted in a multiplicative way. The first correction we apply is therefore a normalization that turns the photoelectric counts into relative flux, by dividing each quarter's light curve by its average. A median is actually more appropriate than a mean as outliers and large photometric jumps can bias the mean. If photometric variations would only be generated by PRNU and masks, this process should be enough. As a matter of fact, this is true for systems where no stellar activity is measurable, if we exclude the effect of the differential velocity aberration.

Issues arise with the systems that display strong pseudo-periodic luminosity fluctuations. For those, the average (or median) over a quarter is biased by the fact that the number of pseudo periods is too small to be averaged out.  Therefore, the median is not a perfect estimator of the mean photometry. This is an intrinsic limitation of the light curve photometric accuracy. In such cases, jumps remain with amplitudes within a few percent. Given that the remaining jumps are caused by a biased normalization, the second layer of adjustment to be applied should still be done in a multiplicative way. However, this is not possible in practice because none of the quarters can be considered as an absolute reference. The only corrections we may apply are additive, to ensure a smooth aspect of the light curve and to minimize their effects in the Fourier domain. 

We employ two ways to smooth the remaining discontinuities once quarters are divided by their median. When a gap is short with respect to the photometric variability timescale, each side of the gap is adjusted accordingly. When a gap is longer than the variability period, we simply adjust the photometry with the difference of the means of each chunk surrounding the gap. Once the complete time series is leveled and concatenated, a linear fit is subtracted from it to take into account the decreasing instrumental sensitivity. Finally, when working with SAP light curves, we compensate for the differential velocity aberration -- the motion of the target across a fixed aperture smaller than the point spread function -- caused by the pixel scale breathing along the satellite's orbit (372.5 days), whose peak-to-peak amplitude ranges from 0.5\,\% to $\sim10\,\%$. This is done by subtracting from each light curve a 372.5-day period sine fitting and a first harmonic, which is enough to reduce its amplitude to less than 0.5\,\%.

We search for stellar pulsations in the power density spectra of the light curves.  To minimize the effects of the incomplete duty cycle, we perform gap fillings and make use of the fast Fourier transform.  All short gaps (only several missing points) are interpolated with a second order polynomial estimated from the nearby data points. Long gaps are filled with zeros. To reduce the impact of abrupt discontinuities around long gaps, the edges of each section in between gaps are apodized with a cosine function. This is particularly important when significant variability is detected.  

In the case of well detached systems, in which the time spent during eclipses is less than 20\,\% of the time, we remove the data corresponding to the eclipses and bridge them with a second-order polynomial, constrained by the surrounding data. In the case where time spent during eclipses is more than that, we prefer folding the light curve on the orbital period and subtract the average signal to each orbital period. Another option could have been to model the eclipse light curve with a fitting  routine, such as PHOEBE \citep{Prsa_2018} or JKTEBOP \citep{Southworth_2013}. However the advantage of detrending the light curves with a mean folded light curve avoids having to model the signal, which represents a huge amount of work for a sample of 3000 targets. In addition, residuals of a light curve minus a model are often significant enough to alter the signal. Still there is a big drawback when opting to subtract the folded light curve: because of pointing jitter and PRNU, the amplitude of eclipses may vary from quarter to quarter, or even during a quarter. Then, there are significant residuals in the light curves, and a signature of the eclipse signal is still quite visible. In such cases, we then remove all harmonics of the orbital period from the Fourier spectra. The latter method is rather gross, but can still help in detecting classical pulsator oscillations, as illustrated in Fig. \ref{fig_classical_puls} (middle panel, gray vs. black curve). 

\subsection{Data inspection tool}
\begin{figure*}[t]
\begin{center}
\includegraphics[width=16.5cm]{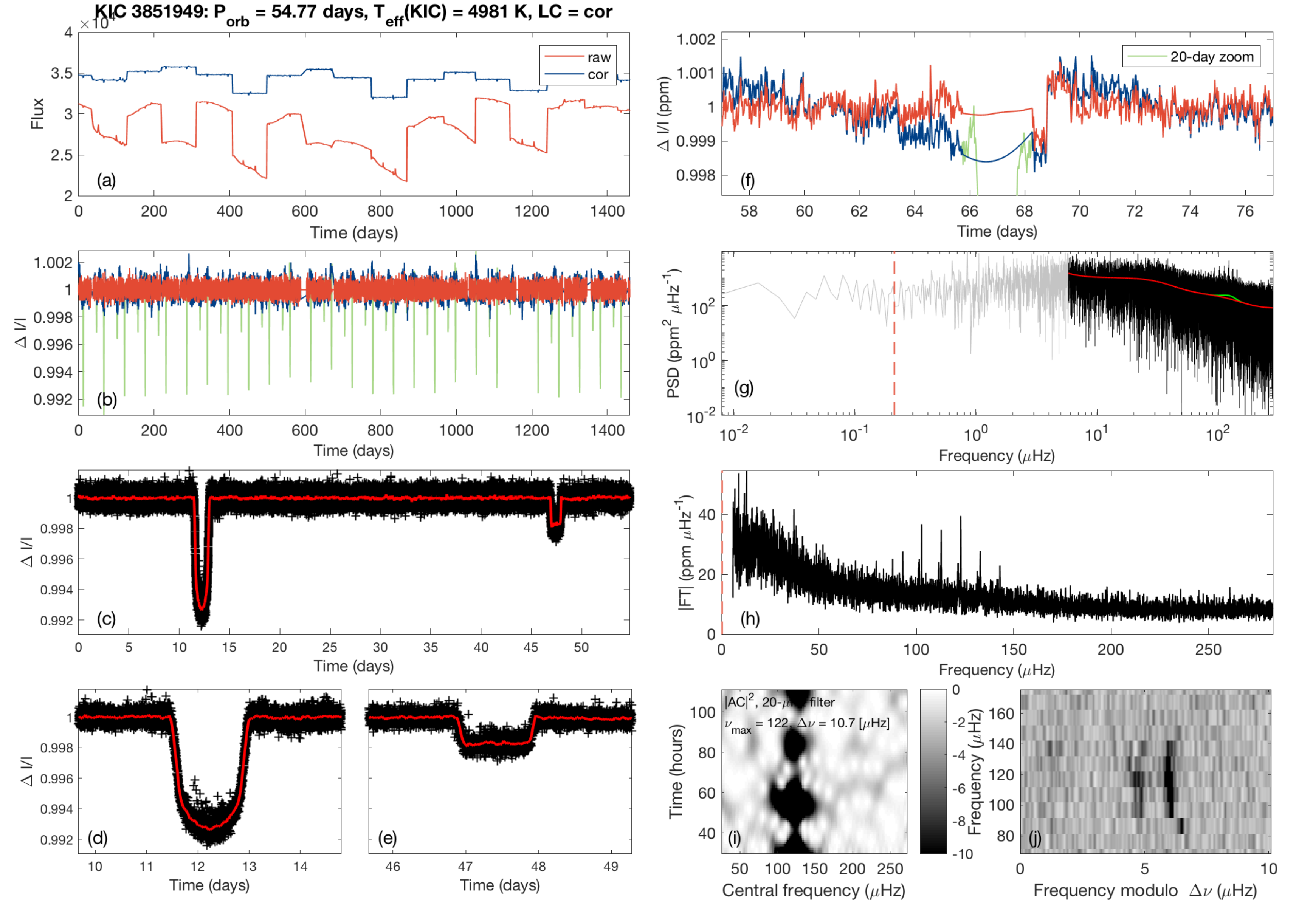}
\end{center}
\caption{Data inspection tool applied to the eclipsing binary system KIC 3851949, which includes a solar-like oscillator with $\nu\ind{max}\approx 120\ \mu$Hz and depleted $\ell=1$ modes. Left column from top to bottom. Panel a: Kepler light curves as a function of time (days), where raw stands for SAP and cor for PDCSAP. Panel b: light curves expressed in relative fluxes, where the blue line contains the stellar activity and oscillations (eclipses clipped out), the green line is the eclipse signal (activity filtered out), and the red curve is optimized for oscillation search (activity and eclipses filtered out).  Panel c: eclipse light curve folded on the orbital period. Panels d and e: zooms of the folded light curves around the eclipses. Right column from top to bottom. Panel f: zoom of the light curve over 20 days. Panel g: log-log scale display of the power spectral density of the time series expressed in ppm$^2$ $\mu$Hz$^{-1}$ as a function of frequency ($\mu$Hz). Panel h: amplitude Fourier spectrum of the time series as a function of frequency. Panel i: envelope of the autocorrelation function (EACF) as a function of frequency and time. Panel j: \'echelle diagram associated with the large frequency spacing automatically determined from the EACF plot. The $x$-axis is the frequency modulo the large frequency spacing (i.e., from 0 to $\Delta\nu$), and the $y$-axis is frequency.  }\label{fig_visualization_tool}
\end{figure*}

We developed a quick look tool that helps the user classify a system at a glance (Fig. \ref{fig_visualization_tool})\footnote{The ultimate goal of the DIT is to be public, but it is still under development and needs more time.}. For each star, it is composed of page divided into a series of ten plots. First, the original Kepler light curves (SAP and PDCSAP) are displayed to look for any major issue regarding the data (panel a). It is useful in the case where a quarter or two are outliers with respect to the others and deserve to be manually removed before reprocessing the light curve. In a second panel (b), the three types of light curves described in the previous subsection (variability, eclipse, oscillation types) are overplotted to make sure that the disentangling is correct. Especially, the user checks whether discontinuities are still present in the oscillation-optimized light curve. Then, a fold of the complete light curve is a good indicator of any misestimate of the orbital period, or of the presence of a third body eclipsing or causing ETV (panel c). A zoom on each eclipse is also displayed for refining this analysis for long orbits, where the eclipses represent less than 10\,\% of the period (panels d, e).
We also plot a zoom on a relatively short range (20 days) at a random location, to help the user visually identify slow pulsations in the time domain, such as $\gamma$-Dor or tidally induced (panel f). 

The last four plots are the power spectral density (PSD) in log-log scale to identify surface activity (typical of solar-like stars) and outstanding peaks indicating oscillations (panel g). It is also useful to check the quality of the background fitting that is performed to whiten the PSD. The background fitting is done following the prescription of \citet{Kallinger_2014} for solar-like oscillators. 
The next plot is the square root of the previous one, i.e., the Fourier transform module, which is more appropriate to identify pulsators with a large variety of oscillation amplitude, such as $\delta$-Scuti (panel h). The last two plots are more focused on the search for solar-like oscillators, as they display the envelope of the autocorrelation of the time series \citep[EACF, see][]{Mosser_Appourchaux_2009}, which is usually considered the best performing tool to detect solar-like oscillations and measure the large frequency spacing $\Delta\nu$, even in low signal-to-noise (SNR) conditions (panel i). The last panel is an \'echelle diagram of the PSD, based on the $\Delta\nu$ deduced from the EACF computation (panel j).

Besides the plots, the title includes the orbital period, the data sampling (long or short cadence), the light curve type (SAP, PDCSAP), and the effective temperature. The latter is particularly useful when oscillations are detected. For example, an effective temperature of 4900 K is compatible with the detection of red giant oscillations, and a temperature of 7000 K with a $\delta$ Scuti. We are aware that published temperatures may be off beyond the fact that they are often not available, but it is valuable support information.

\section{Results}

We ran the data inspection tool (DIT) on all of the targets from the 2018 update of the Villanova catalog, except for the RGs in EB or HB systems published by \citet{Hekker_2010,Gaulme_2013,Gaulme_2016a,Beck_2014,Beck_2015} and the forthcoming paper Benbakoura et al. (in prep). Indeed, these systems were already well characterized, and first author Gaulme already had processed them. As regards the classical pulsators, we were not sure about the exact number of systems already known and we did not exclude them a priori from the analysis. The total amount of stars for which we applied the DIT pipeline and that we visually inspected is 2875 out of 2925 total.

We first inspected every light curve to check whether orbital parameters were off. In 100 cases, we had to modify the ephemeris published on the Villanova database. Many were small inaccuracies regarding eclipse timing and duration. Also, for 36 systems, only primary eclipses were detected. Thanks to the visual inspection of the light curves folded on the orbital period, we identified secondary eclipses for 36 new systems. 

Even though it can sometimes be difficult to distinguish the detection of oscillations from imperfect light curve cleaning, we iteratively converged to the detection of pulsations in 303 out of the 2925 light curves, i.e., in about 10.4\,\% of the cases. To do so, first author Gaulme produced a first screening by inspecting all of the DIT files, and selected about 350 of them. He attempted a preliminary classification according to pulsator type. Then co-author Guzik reviewed all of them and commented on each case with no instruction from Gaulme. Then Gaulme led a systematic search of every target on a web search engine by entering, one by one, ``KIC'' and the KIC number of each star, with quotes and without quotes. Except for pictures of soccer players \textit{KICking} balls, it ended up being more efficient than only through publication search tools such as NASA ADS or SIMBAD. The result of this search is certainly not fully exhaustive but is pretty much complete. In total, 187 systems were already studied in one way or another in a peer-review paper or a conference proceeding. The level of study was highly variable, from simple notes in large tables containing many Kepler targets, to dedicated papers about single binary systems. Having 187 systems previously cited in the literature does not mean that all were characterized both as binary and pulsators. Indeed, many were studied as double or triple systems but there was no mention of any stellar pulsation. Actually, only 140 systems were already published as both multiple-star system and stellar pulsator. The main result of the present paper is thus to have identified 163 new pulsating stars in multiple star systems. It represents 54\,\% of the 303 stellar pulsators identified in the Kepler EB catalogs, and an increase of 116\,\% of the known stellar pulsators in EBs among the Kepler data.

Now, not all of the pulsators we list are actual pulsators in multiple-star systems. Many are false positives, i.e., pulsators that are either aligned (within the point spread function or pixel) with an eclipsing binary or bright stellar pulsators whose light leaks into the considered EB mask. Disentangling those is a huge work that goes well beyond the scope of the present work. To do so, one should first check the Kepler imagettes and check whether the maximum of intensity matches the maximum of amplitude of the oscillations and the maximum depth of the eclipses. Such an approach was used by \citet{Gaulme_2013} for 70 RG candidates in EBs. The second step would consist of getting existing or acquiring new RV measurements. 
However, we can get a very rough idea of the likelihood that a pulsator belongs to a given binary system. For example, if we detect very clear oscillation peaks in the Fourier domain, while the photometric variability associated with the eclipses is significantly less than 1\,\%, we can assume that the pulsator is not a member of the eclipsing binary. Still, it can be a triple system. Another example, a 10-$R_\odot$ RG associated with a binary system that orbits in less than 5 days is very unlikely. Indeed, in such systems, stars are synchronized and the rotation rate of the star could make it disintegrate, depending on the orbital parameters. For RG specifically, \citet{Gaulme_2013} studied the question and used this type of consideration to classify many RG apparently in EBs as false positives. 

Our results are displayed in Table \ref{tab:maintable}, which includes the 303 systems where we are confident to have detected oscillations, plus the system identified by \citet{Kuszlewicz_2019}. The table is sorted by increasing KIC number and is optimized to not take too much room and cannot include all the existing information. We therefore refer to the Villanova catalog and the \citet{Coughlin_2011} paper to get all details about orbital parameters. We only list the main properties an observer would need to decide which target to study and observe: effective temperature, Kepler magnitude, orbital period, deepest eclipse depth, phase separation in between primary and secondary eclipses, ratio of eclipse durations, sum of eclipse durations relative to orbital period, and pulsator type. We indicate the number of eclipses per orbital period, which tells whether it is a true eclipsing binary or an ellipsoidal-variation binary, or even an HB star. We also list the references we have identified and we indicate with ``Y'' is this is the first time a system is identified as both a pulsator and a binary. Then we add some notes in the last column to point out random relevant information.

\section{Discussion}
In this section, we highlight the main findings of the current study and list actions that could be done based on this sample for future studies.

\begin{figure*}[t]
\begin{center}
\includegraphics[width=6cm]{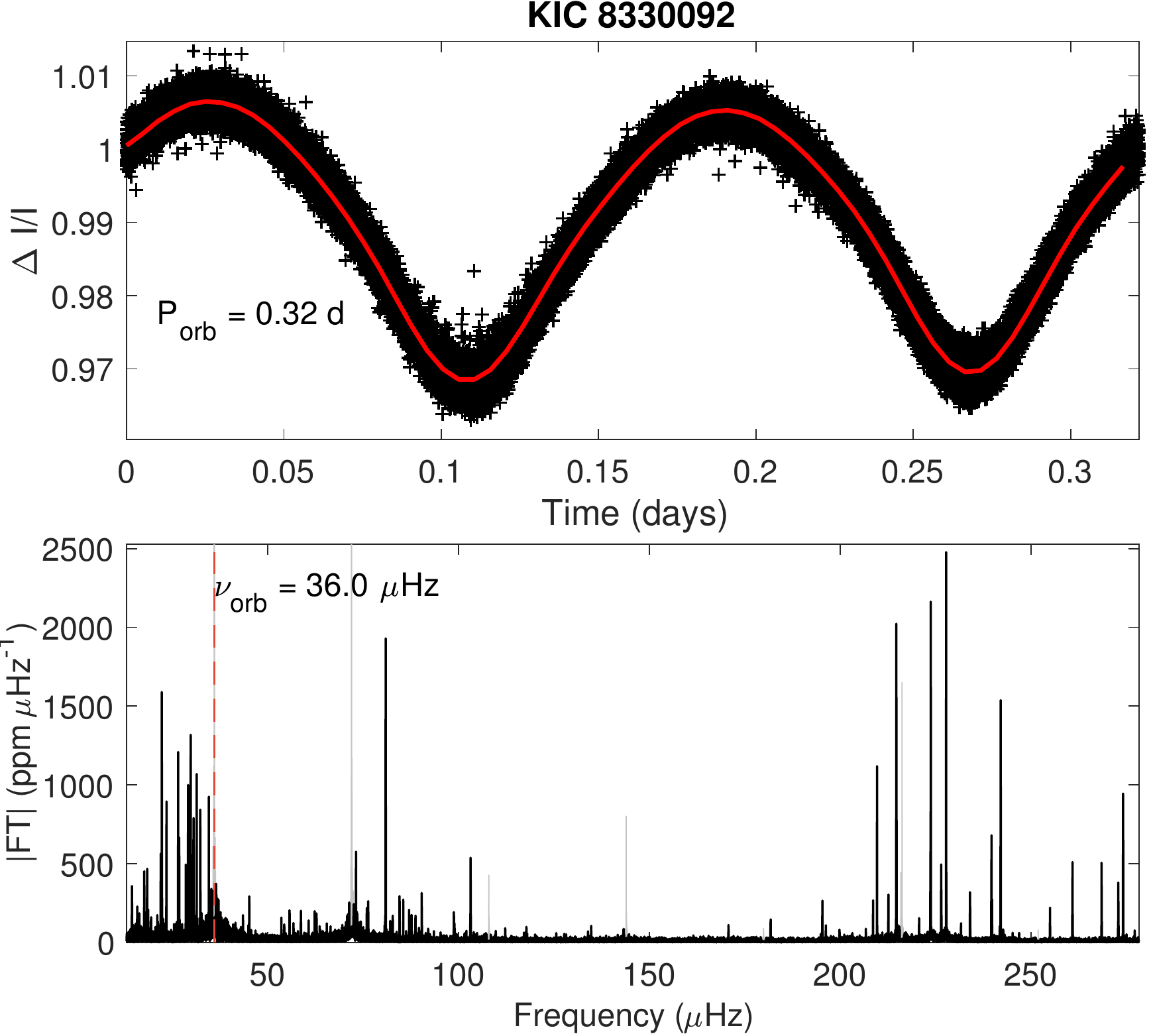}
\includegraphics[width=6cm]{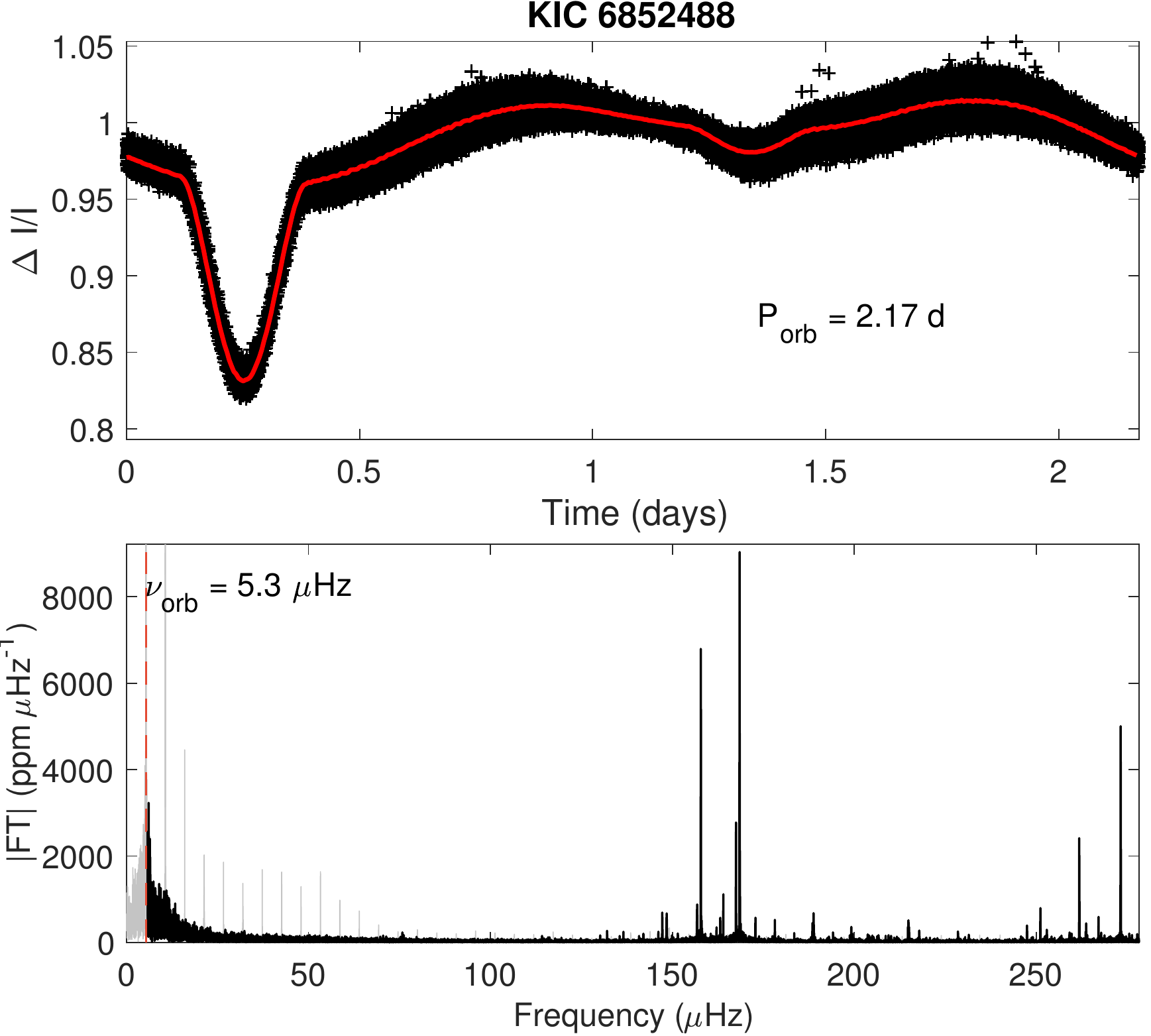}
\includegraphics[width=6cm]{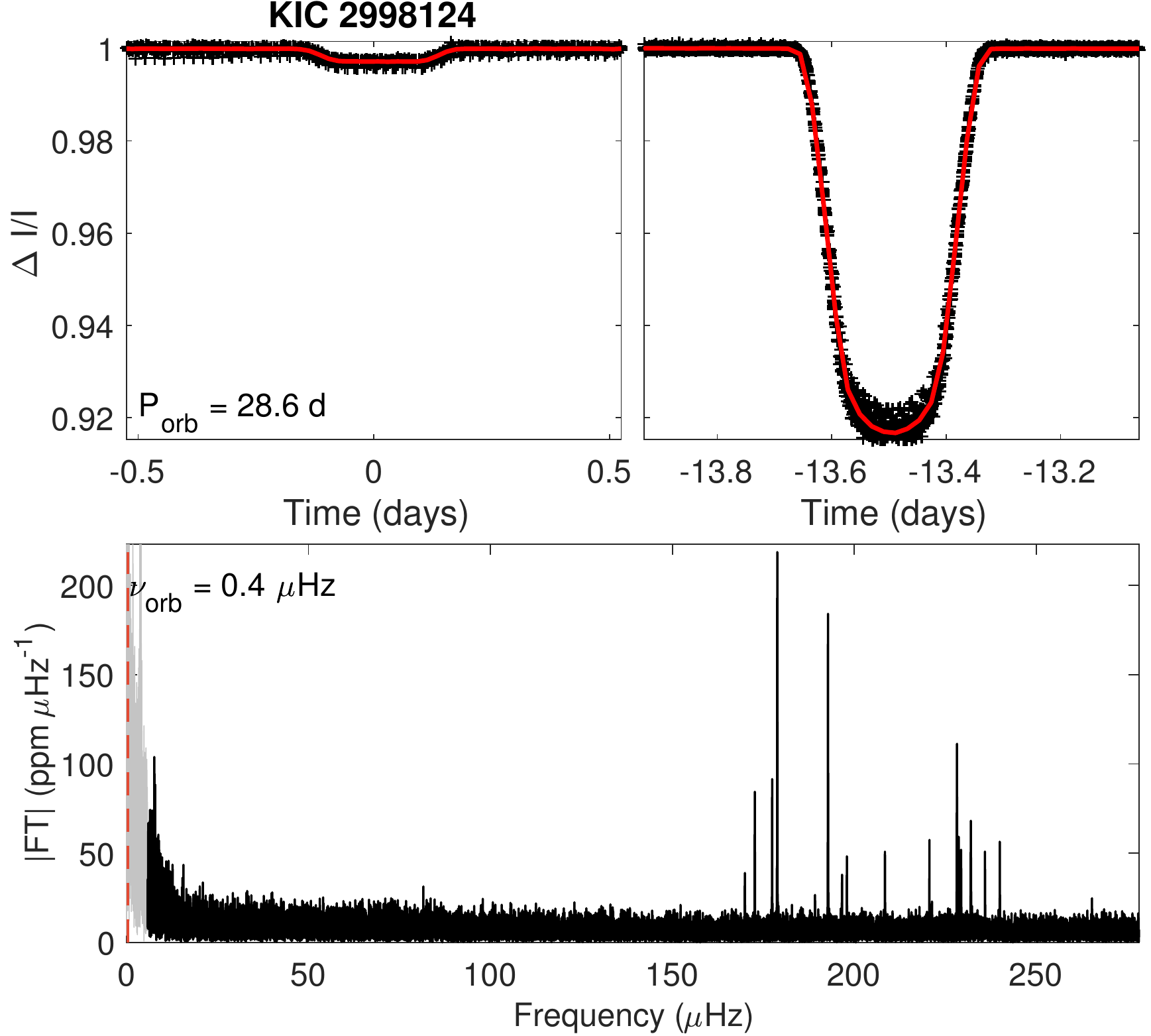}
\end{center}
\caption{Examples of classical pulsators observed in the light curves of binary systems. Left is a 0.32-day orbit OC binary (KIC 8330092) displaying $\gamma$ Dor and maybe Rossby modes at $\nu\lesssim50\ \mu$Hz and clear $\delta$ Sct oscillations above. Middle is the short period detached system KIC 6852488 in which $\delta$ Sct oscillations are clearly visible at $\nu\gtrsim120\ \mu$Hz. The asymmetrical dispersion of points around the rebinned folded light curve is caused by the presence of spots that modulate the photometric variations as a function of time. Right is the ``long''-period detached EB KIC 2998124 displaying  $\delta$ Sct oscillations at $\nu\gtrsim170\ \mu$Hz. Vertical dashed red lines indicate the orbital frequency $\nu\ind{orb}\equiv P\ind{orb}^{-1}$. For left and middle plots, the gray curve is the modulus of the Fourier transform prior to removing harmonics of the orbital period.}\label{fig_classical_puls}
\end{figure*}

\subsection{Classical pulsators}
Classical pulsators represent the majority of the pulsators in eclipsing binaries, as we expect from the effective temperature histogram (see Sect. \ref{sect_source_info}). A total of 190 $\gamma$ Dor, $\delta$ Sct or hybrid are identified, including 122 that were previously unknown. There is more uncertainty regarding the detection of $\gamma$ Dor oscillations with respect to $\delta$ Sct because of the frequency range that often overlaps harmonics of the orbital frequency. Because of the large amount of data, we did not look for regular period spacings or whatever type of pattern that would for sure indicate that a pulsator is a $\gamma$ Dor. We just flagged as $\gamma$ Dor pulsators stars that display frequencies less than $\sim50\,\mu$Hz ($\sim5$ c/d) and possibly typical $\gamma$ Dor features in the time domain \citep[e.g.,][]{Kurtz_2015}. In addition, any information about effective temperature helped us assessing $\gamma$ Dor pulsators. We flagged $\delta$ Sct stars whose oscillation spectra looked like typical $\delta$ Sct \citep[e.g.,][]{Baglin_1973,Balona_2015b}, with frequencies larger than 50 $\mu$Hz. However, in some cases, oscillation spectra are sparse (few peaks), and there could be some doubts with tidally excited oscillations when peaks are regularly spaced. In some other cases, oscillation spectra look much like a $\delta$ Sct but the stellar effective temperatures are a little hot. In these cases we write a note in the table, but we keep in mind that effective temperatures from the KIC may be off, and especially for close binary systems. We display three examples of $\gamma$ Dor and $\delta$ Sct pulsators in Fig. \ref{fig_classical_puls}.

A rather surprising finding of this study is the relatively large number of $\gamma$ Dor and/or $\delta$ Sct pulsators in very short period systems. A total of 13 systems with orbital periods less than 0.5 days display clear oscillations. We naturally can assume them to all be either false positives or triple systems. Actually, three of them are indeed triple systems as ETVs were measured and published in previous articles \citep{Gaulme_2013,Borkovits_2016}. However, it would be worth checking more into details whether the ten remaining systems are false positives, triples or genuine eclipsing binaries. 

The first panel of Fig. \ref{fig_classical_puls} is an example of our short-period classical pulsators with the 0.32-day OC binary KIC 8330092. The oscillation spectrum is very clear, with a low frequency part which we assume to be $\gamma$ Dor and higher frequencies that are typical of $\delta$ Sct. Besides, the crowded aspect of the low frequency part could be partially composed of Rossby modes as proposed by \citet{Saio_2018}. In the specific case of KIC 8330092, since the system is an OC, its rotation period is equal to its orbital period. From the light curve, the system seems to be composed of two similar stars (similar eclipse depth and shape), which we assume to be A0 to F5-type stars, i.e., with radii and masses of about $1.4\ R_\odot$ and  $1.4\ M_\odot$. Within such an assumption ($M_1=M_2$, $R_1=R_2$), the rotational velocity at equator is $V\ind{eq} \approx 220$ km s$^{-1}$, whereas the escape velocity from the gravitational field is $V\ind{esc} \approx 730$ km s$^{-1}$, which means that such a star is far from disintegrating because of fast rotation. Moreover, still within same assumptions, Kepler's third law indicates that the semi-major axis of the system would be $a\approx2.8 R_\odot$. Of course, these numbers are very rough estimates, but they indicate that a star such as a $\delta$ Sct can exist in such a tight contact binary.

Discovering classical pulsators in short period binary systems is not something new per se: several papers report detections of $\delta$ Sct or $\gamma$ Dor pulsators in tight binary systems. We here mention a few examples. \citet{Aerts_2002} report the detection of $\delta$ Sct pulsations in the 1.15-day orbit ellipsoidal binary XX Pyx. \citet{Dal_Sipahi_2013} report the detection of $\delta$ Sct oscillations in the 0.69-day orbit ellipsoidal binary V1464 Aql, and \citet{Sipahi_Dal_2014}  the detection of $\gamma$ Dor oscillations in the 0.93-day orbit close binary systems. \citet{Zhang_XB_2015} reported the detection of $\delta$ Sct oscillations in a component of the 0.65-day orbit near-contact binary V392 Ori. From the present study, the ten systems displaying  $\gamma$ Dor and/or $\delta$ Sct with $P\ind{orb} < 0.5$ day are odd as no oscillations have been detected so far, as far as we know, in such short period systems. In other words, what is new here is the relative amount of short period binaries among the oscillators (7\,\% with $P\ind{orb}<0.5$ d) but it is also the first time that we identify pulsators in systems with periods less than 0.65 days. Although some may be false positives, it is still very likely that we have identified the $\delta$ Sct in a binary system with the shortest orbit ever.

\begin{figure}[h!]
\begin{center}
\includegraphics[width=9cm]{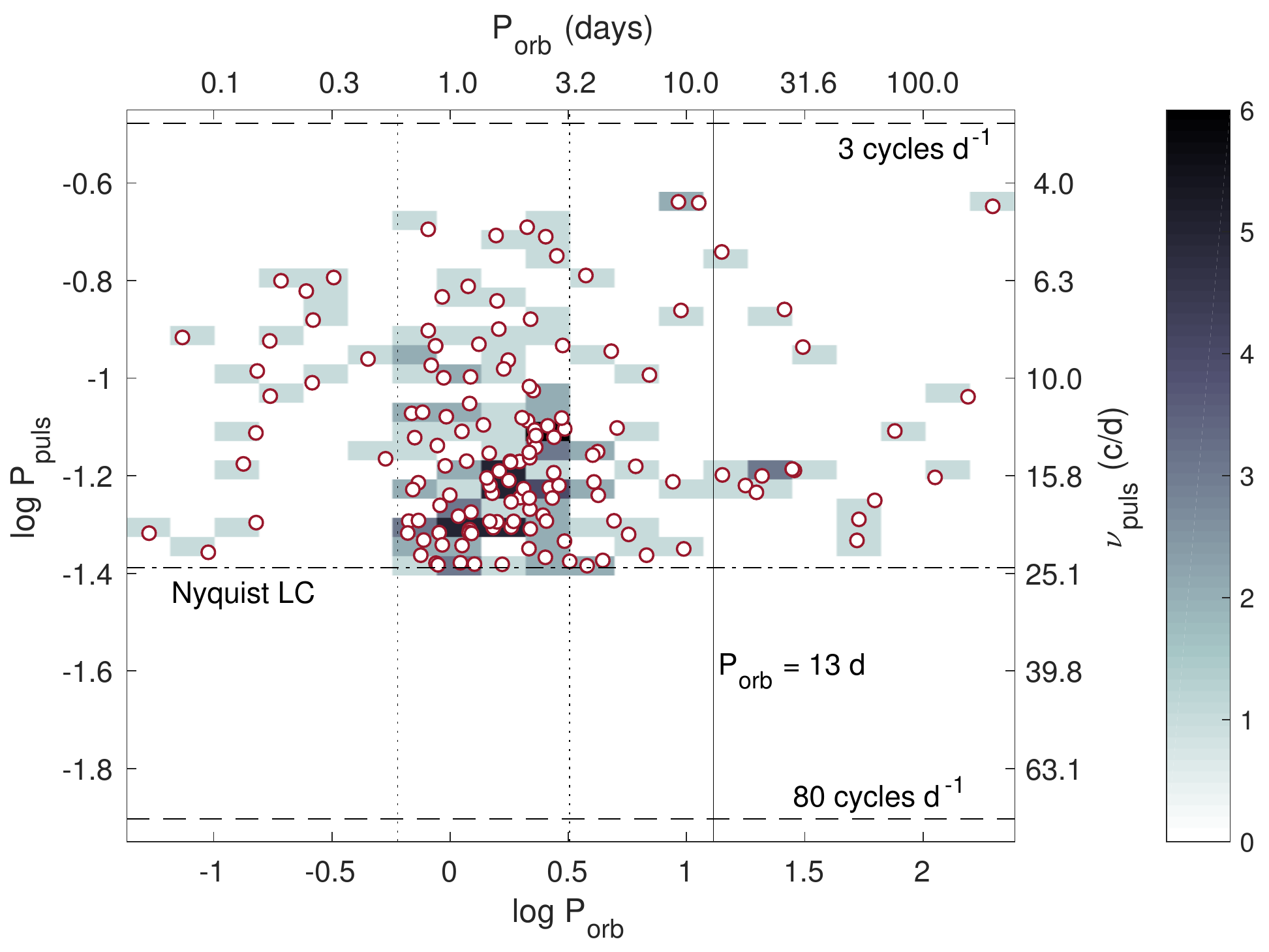}\\
\includegraphics[width=9cm]{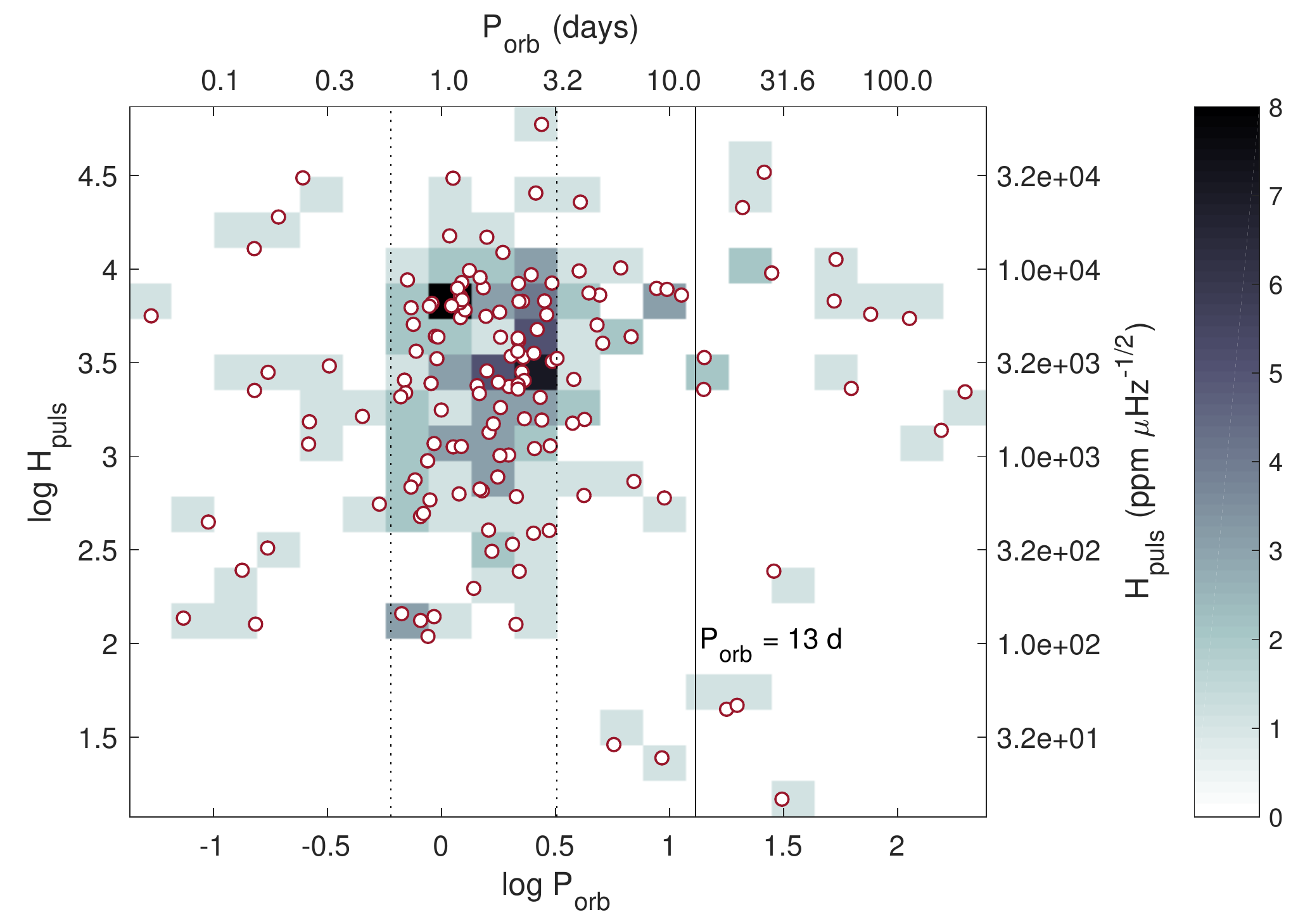}
\end{center}
\caption{Pulsation properties of the 148 $\delta$ Sct EB candidates identified in this work. Top: pulsation periods of oscillation spectra's highest peak $P\ind{puls}$ as a function of orbital periods $P\ind{orb}$. Bottom and left $x$- and $y$-axes are in the same units as in \citet{Liakos_Niarchos_2017}. As in their paper, the vertical line indicate an 13-day orbital period, while the horizontal lines indicate pulsation frequencies of 3 and 80 cycles per day (c/d). Top and right  $x$- and $y$-axes are the same in days for $P\ind{orb}$ and in frequency for pulsations. The bottom panel represents the height of the highest peak per $\delta$ Sct spectrum as a function of $P\ind{orb}$. In both panels the color scale indicates the number of systems per square, to help the reader look for trends, such as possible clustering. The two vertical dotted lines indicate orbital periods from 0.6 to 3.2 days.  }\label{fig_delta_Sct}
\end{figure}

Beyond short period systems and thanks to the large amount of $\delta$ Sct that we have, we can test observations that have been done in previous works. As presented first in \citet{Soydugan_2006} then confirmed on a larger sample by \citet{Liakos_Niarchos_2017} there exists a correlation between the pulsation frequency and the orbital period. For systems with $P\ind{orb}$ less than a threshold located in between 13 and 40 days, the dominant $\delta$ Sct pulsation tends to increase with decreasing orbital period (see Fig. 4 in \citealt{Liakos_Niarchos_2017}). In Fig. \ref{fig_delta_Sct}, we test whether we find the same trend from the 149 possible $\delta$ Sct that we identify. For all systems flagged as possible $\delta$ Sct pulsator, we measured the frequency and height of the largest peak belonging to the $\delta$ Sct domain, i.e., by excluding the $\gamma$ Dor region for the hybrids. It arises that there is no obvious trend visible from our sample. In the most populated part of the diagram, we note a small trend though: for $0.6<P\ind{orb}<1.4$ d the most likely pulsation frequency is 20 c/d,  for $1.4<P\ind{orb}<2.1$ d is is 16 c/d, and for $2.1<P\ind{orb}<3.2$ d it is 13.5 c/d.

 Several reasons may be responsible for not confirming that observation. Firstly, we have no way to know which systems are false-positive $\delta$ Sct binaries, while \citealt{Liakos_Niarchos_2017} used systems that were well characterized. Secondly, we have not looked at the possible short-cadence data for part of our systems. Therefore, our maximum frequency available is the Kepler long-cadence Nyquist frequency ($\simeq 283\ \mu$Hz). This means that a fraction of the frequencies that we present are likely to be aliases of higher frequencies. To emphasize this bias and help  the comparison, we kept the $y$-axis boundaries to be the same as that in \citealt{Liakos_Niarchos_2017} and we plotted the logarithm in base 10 of the pulsation and orbital periods. Thirdly, the Kepler sample lacks of long-period systems, which prevents us to have a view on long-orbit systems (8 $\delta$ Sct with $P\ind{orb}>30$ days). Once the sample we list here is fully characterized, in particular with the help of complementary RV measurements, it will be possible to lead such a kind of study. 
 
In addition to the pulsation dominant frequency, we tried to see whether binarity suppresses pulsation of A/F stars, as was observed by \citet{Gaulme_2014} for RGs in close binary systems. We looked for a dependence of the pulsation amplitude as a function of orbital frequency. The bottom panel of Fig. \ref{fig_delta_Sct} does not seem to show any correlation. 
More generally it would be interesting to see whether the fraction of binaries that do not pulsate is different than the fraction of stars in the instability strip that do not pulsate. Answering this question is tricky as the boundaries of the A/F pulsators is not very clear because the stellar parameters are not precise enough. 

\begin{figure*}[t]
\begin{center}
\includegraphics[width=6cm]{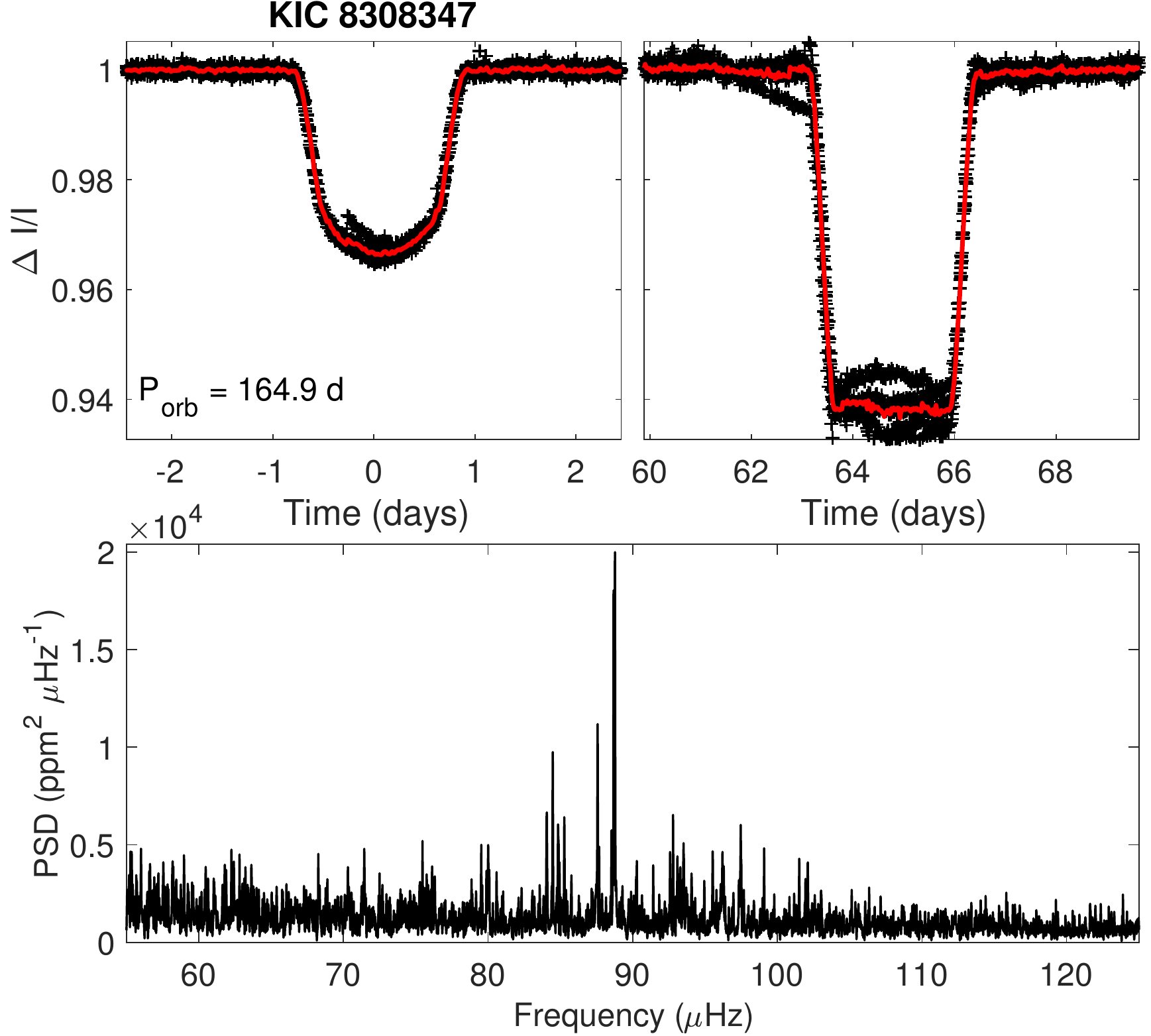}
\includegraphics[width=6cm]{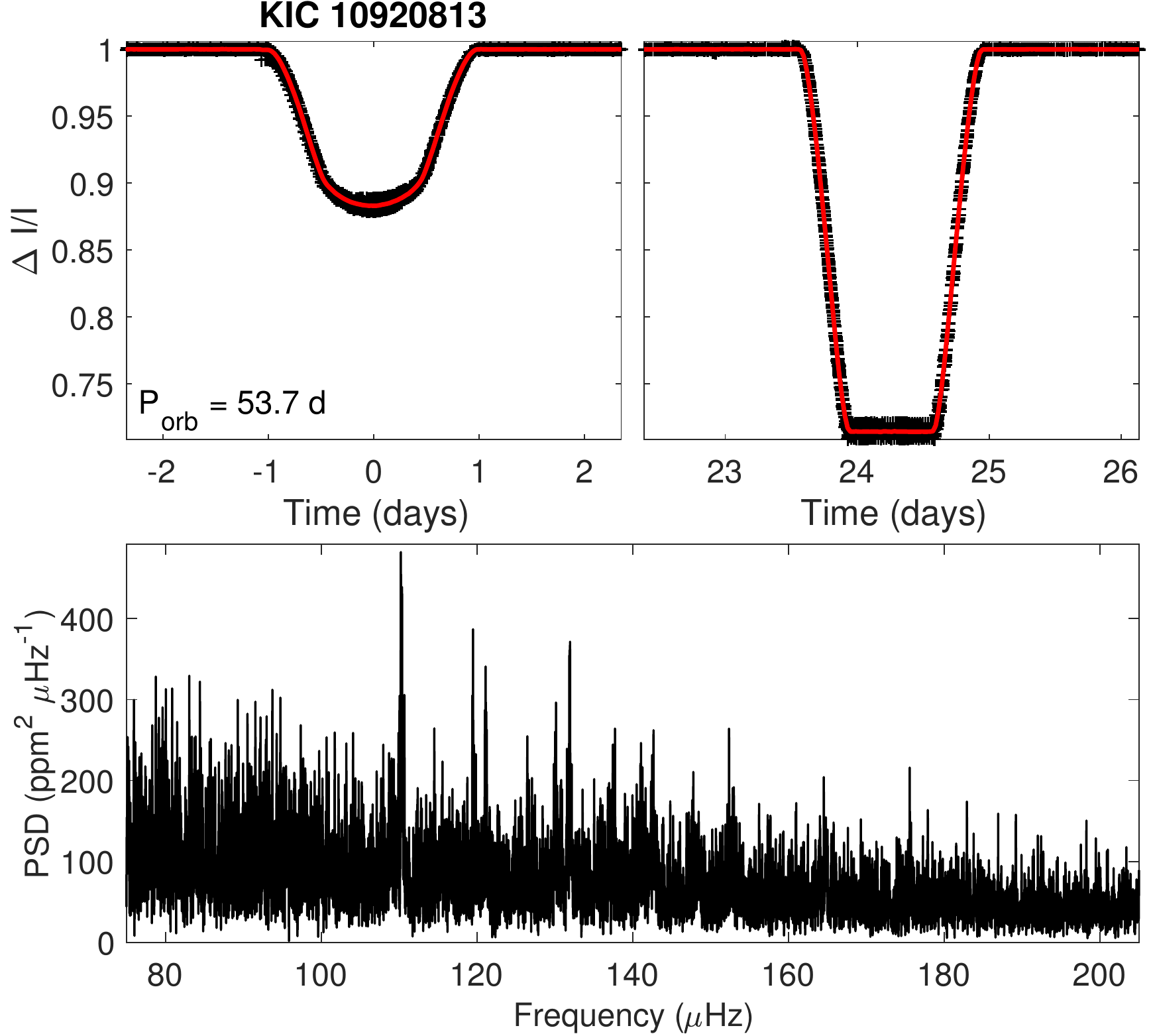}
\includegraphics[width=6cm]{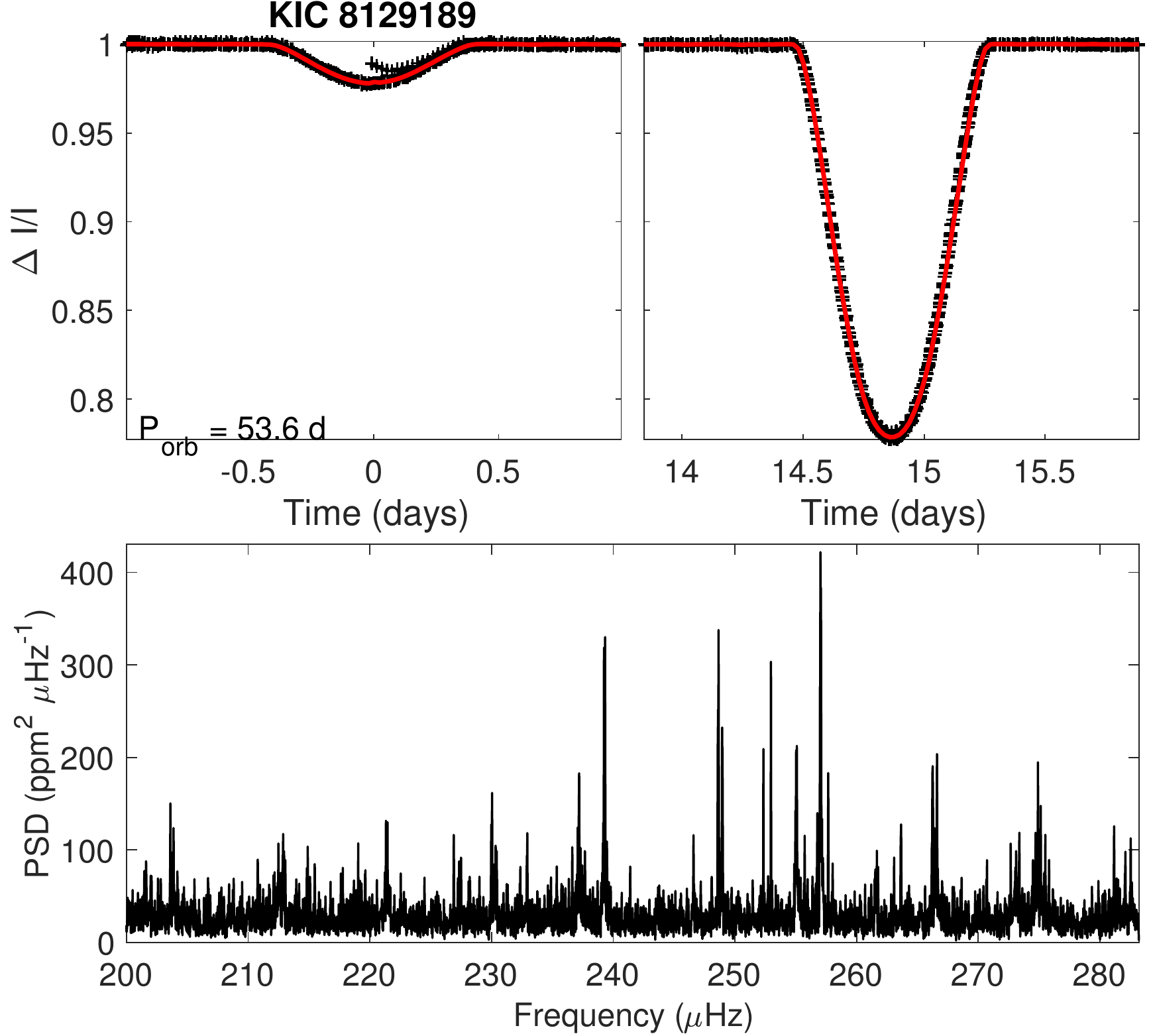}
\end{center}
\caption{Bona fide RG/EB that are possibly SB2 discovered in the present work. Top panels: Kepler light curves folded over the orbital period (black markers), with a rebinned version of if over 30 min bins. Bottom panels: power spectral density (PSD) as a function of frequency of the Kepler light curves -- after eclipse removal -- centered around the RG oscillations modes. PSD was smoothed with a triangular weighted moving average (3-bin wide at half maximum) to highlight the oscillations mode visibility.}\label{fig_RGEB}
\end{figure*}
\subsection{Solar-like oscillators}
As indicated in the introduction, all solar-like oscillators are RGs. We detect 85 RG oscillation spectra among all the binaries listed in the Villanova catalog and the \citet{Coughlin_2011} table. Among those, 27 are new to be both identified as oscillating RG and a binary (eclipsing, ellipsoidal or HB). As mentioned earlier, a large fraction of them must be either false positives or triple systems. Even though only RV measurements and close up imaging could allow us to make sure of their nature, we can already get some ideas on the nature of these systems (false positive, binary, triple). 

First of all, by following the conclusions of \citet{Gaulme_2013}, we are suspicious with RGs associated with a system of orbital period less than 10 days, especially if is oscillating \citep{Gaulme_2014}. As an example, so far, the shortest oscillating RG in a binary system is the 19.38-day orbit EB KIC 8702921 \citep{Gaulme_2016a}. Secondly, the shape of eclipses is a strong insight on the nature of the system. Since the RG branch (RGB) represents a short period compared to the overall star lifetime, it is rather unlikely to have both components on the same evolutionary state. Therefore, as observed in practice, the large majority of RGs that belong to binary systems are on the RGB and are composed of a main-sequence and an evolved star \citep[e.g.,][]{Gaulme_2013, Beck_2014}. Therefore one of the two stars is much smaller than the other, which means that one eclipse displays a flat bottom, while the other has a round shape caused by the limb darkening law of the RG star. Similar light curves can be produced with systems composed of a white dwarf and a main sequence star, with an M-dwarf and a subgiant, or occasionally by a hot Jupiter transiting a faint star provided the stellar light reflected by the planet is enough to cause a secondary transit. The system effective temperature, even if possibly biased, is a complementary information that can help confirm that a system hosts an RG.

Beyond eclipsing systems, many ``false positive'' RG/EBs are actually HT systems, i.e., where a close eclipsing binary orbits an RG. In rare cases, the close EB also eclipses the RG \citep{Derekas_2011}, but in general it is not the case \citep[e.g.,][]{Gaulme_2013}. In such kind of systems, the RG tends to cause ETVs. The detection of ETVs is a strong indication that the RG belongs to the triple system. Overall, when no ETVs are measured and the orbital period is too short to be an EB including an RG component, we classify a target as ``likely FP or triple''. Note that even if we do not detect ETVs, we cannot totally exclude a system to be triple, as ETVs may exist in wide hierarchical triple systems at a level that is not detectable (low ETVs amplitude, slow variations).

Based on these criteria, we estimate that 15 out of the 27 systems flagged as displaying RG oscillations are bona fide pulsators in multiple-star systems. Specifically, 10 are EBs, one is an HB with no eclipse, and 4 are HTs. Another system is a possible RG/EB (KIC 10858117), but the oscillation SNR is so poor that it is hard to be fully convinced that there are indeed RG oscillations in the Fourier spectrum of the time series. 

As reminded earlier, solar-like oscillators in EBs are unique targets for testing and calibrating asteroseismology. For that goal, individual masses and radii are needed, which implies that the EB systems must be double-lined spectroscopic binaries (SB2) to be considered as asteroseismic test benches. Given that most systems are composed of an MS star with an RG, the flux contrast in between both components is large: the companion usually account for a maximum of 10\,\% of the total flux. In practice, for a system to be SB2, the companion star must be hotter than the RG, i.e. a G or F type dwarf \citep[e.g.,][]{Gaulme_2016a, Helminiak_2016,Helminiak_2017b}. With the present work, we can estimate how many RG/EB/SB2 are present in total in the Kepler sample. So far, 11 RB/EB/SB2s have been identified and studied with complementary RV data \citep{Frandsen_2013, Rawls_2016,Gaulme_2016a,Helminiak_2015,Helminiak_2016,Brogaard_2018,Themessl_2018}. In the forthcoming paper by Benbakoura et al., three more systems are identified. Actually among those three, one shows no secondary eclipses due to a large eccentricity and a grazing primary, and only masses can be estimated. We are thus at 14 RG/EB/SB2. In addition, one system from the \citet{Gaulme_2016a} paper (KIC 8054233) which has a 1058 day orbit was classified as SB1, but it could turn into an SB2 with higher SNR observations in the future. Hitherto, the total amount is thus 15 at best.  

Among the new systems identified in this paper, only KICs 8308347 and 10920813, with orbital periods of 165 and 54 days respectively, show evidence that the companion's temperature is larger than the giant's (``$T_2>T_1$'' in Table  \ref{tab:maintable}). Another case, KIC 8129189 (54-day period), displays deep partial eclipses and a clear oscillation signal with $\nu\ind{max}\approx 250\ \mu$Hz, which indicates it could be composed of a small RG (for which oscillations are detected) and a smaller RG or a subgiant star. This latter could be an SB2 too. The case of KIC 10491544 displays the same kind of eclipses -- partial --, with $T_2\sim T_1$,  an RG of about $6.6\pm0.3 R_\odot$, and an orbit of 23 days could be a bona fide RG. However, their depths vary from quarter to quarter in between 2 and 5\,\% and the oscillations SNR is very high, whereas it has been observed that modes are much depleted in such short-orbit systems \citep{Gaulme_2014}. This latter could be a false positive. Overall, it appears that the number of RG/EB/SB2s from the Kepler data will be at best 19, which is a small statistical ensemble to test a tool as widely used as asteroseismology.
 
 To complement these systems, triple systems displaying ETVs may be a good option. \citet{Borkovits_2016} showed that it is possible to estimate the mass of the RG and the total mass of the tight binary component of a HT system, just based on ETV measurements. However, the precision of these estimate is rather poor (e.g., $1.5\pm0.5 M_\odot$ for the very clear ETV KIC 7955301). To improve the precision, complementary RV measurements even in the case of SB1 systems should drastically reduce the error bars on masses, to the percent level that is required to calibrate asteroseismology. Note that no radii measurements arise from HT systems that are not triply eclipsing. The total amount of HT systems for which ETVs have sure be detected is 13, including 4 from the present study. By considering both EB/SB2s and HTs, the total number of systems that could help calibrating asteroseismic measurements of red giant stars is about 30 (maximum 32) from the original Kepler mission.

Overall, to extend the sample of asteroseismic calibrators, we could use non eclipsing binaries that are both astrometric and visual binaries. \citet{Marcadon_2018}  studied such a system hosting a main-sequence solar-like oscillator. In this specific case though, the SNR of the RV measurement was not high enough to estimate the oscillating star's mass to better than 5\,\%, which was not sufficient. Similarly to visual binaries, bright EB could be resolved with interferometric measurements and be able to measure the proper motion of each companion, leading to the same result as for visual ones. So far typical magnitude limitation is about eight in the visible. A few bright non-eclipsing binaries may be present in the Kepler sample, and should be revealed by the ESA Gaia mission in the coming years (data release 3 or 4 planned in the early 2020s).

Besides, more data from other space missions can be considered too, but none are comparable to Kepler. K2 is limited to 90 days, most TESS fields of view are limited to 27 days, and CoRoT fields were lasting 180 days at maximum. Given that \citet{Gaulme_2014} showed that RG/EBs with orbits shorter than 120 days tend to show oscillation suppression, these three other options (K2, TESS, CoRoT) are less promising than Kepler, even though it will be necessary to consider them too. The ESA PLATO mission, whose launch is planned in late 2026, will provide for sure new interesting targets, given its large field of view and long exposures (2 years).


\begin{figure*}[t]
\begin{center}
\includegraphics[width=6cm]{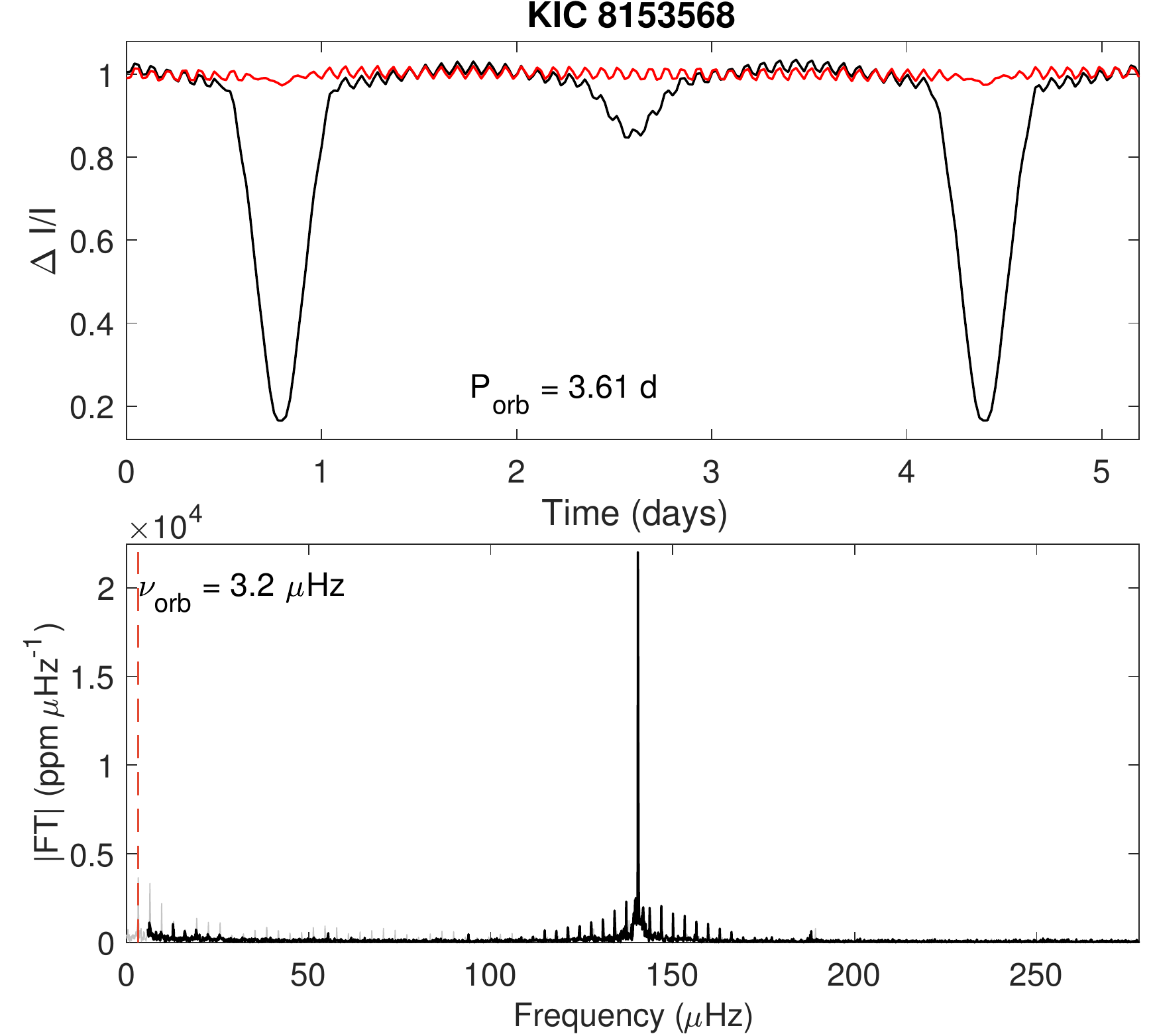}
\includegraphics[width=6cm]{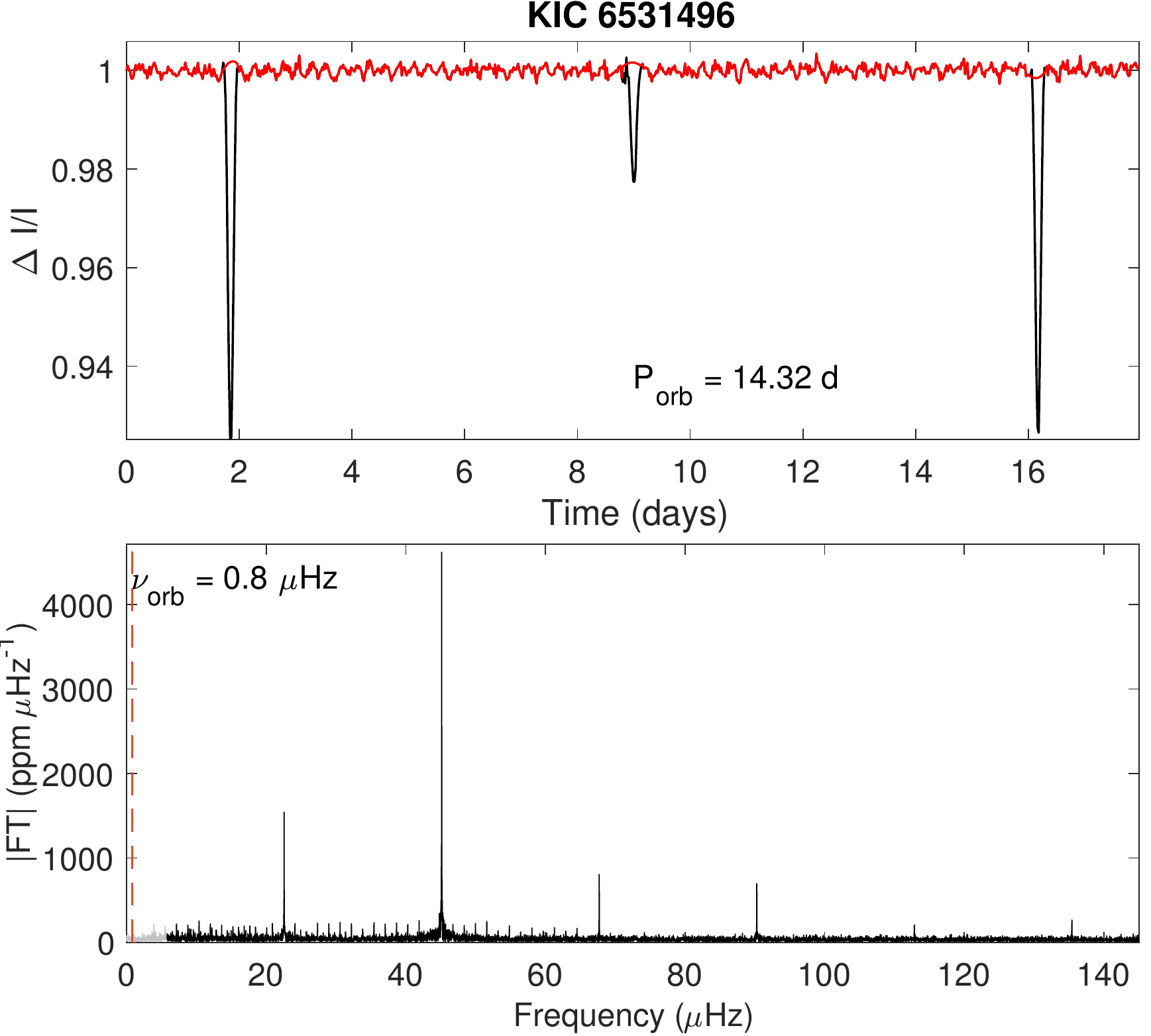}
\includegraphics[width=6cm]{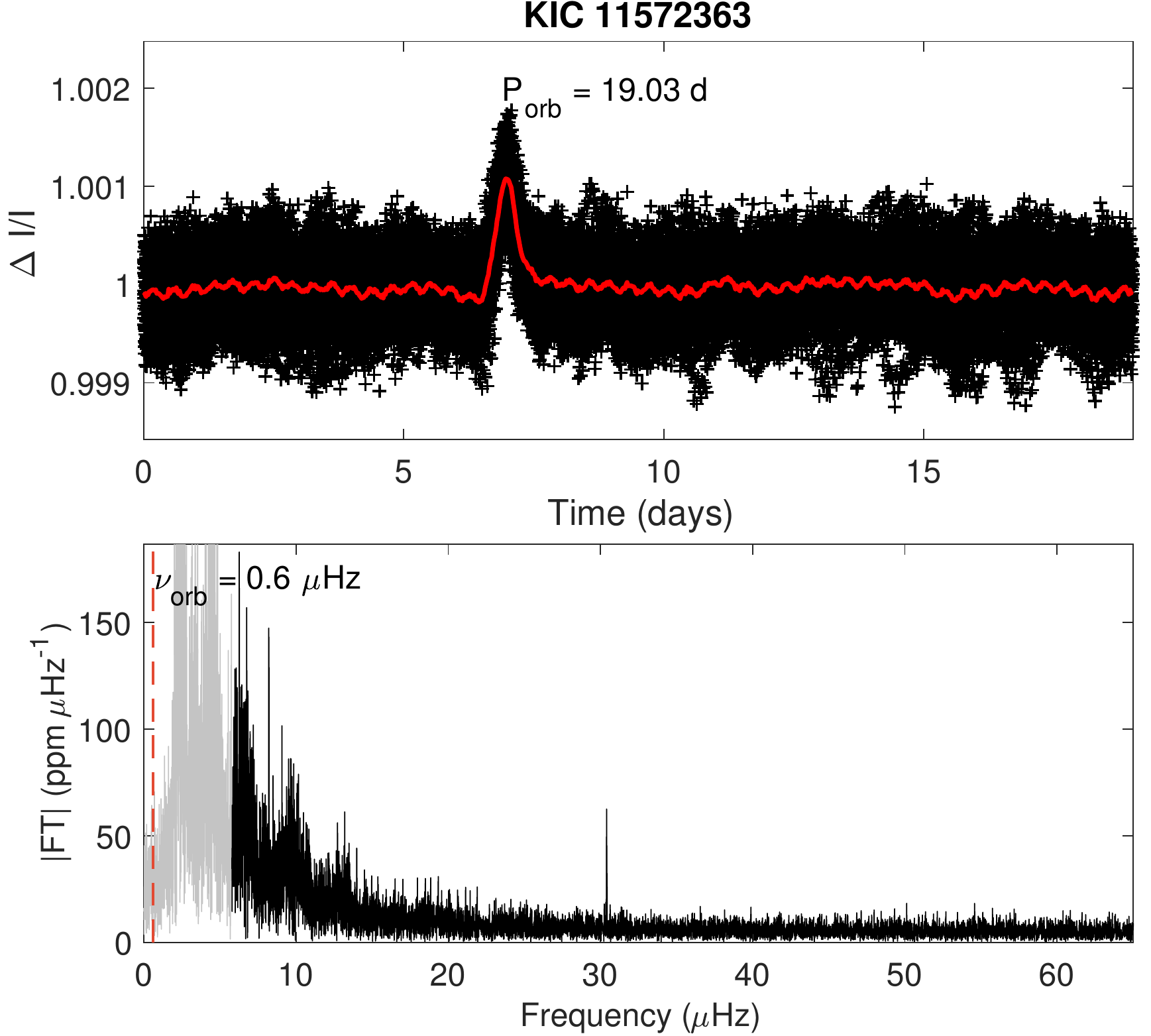}
\end{center}
\caption{\small Example of pulsators that either mimic tidal pulsators (KICs 8153568 and 6531496) or are actual tidal pulsators (KIC 11572363). In the cases of KIC 8153568 and 6531496 we do not plot a folded light curve but an extract of it over a little more than an orbital period (black) as the oscillations are not exactly in phase with the orbital period. The red curve is the original light curve minus the folded signal. In the case of KIC 8153568, the oscillations occur on the the star hidden during the deeper eclipses, as the modulation is suppressed during them. The right panel represents the light curve and Fourier spectrum of KIC 11572363, a 19-day HB systems with tidally excited oscillations. }\label{fig_tidal_puls}
\end{figure*}
\begin{figure}[t]
\begin{center}
\includegraphics[width=8cm]{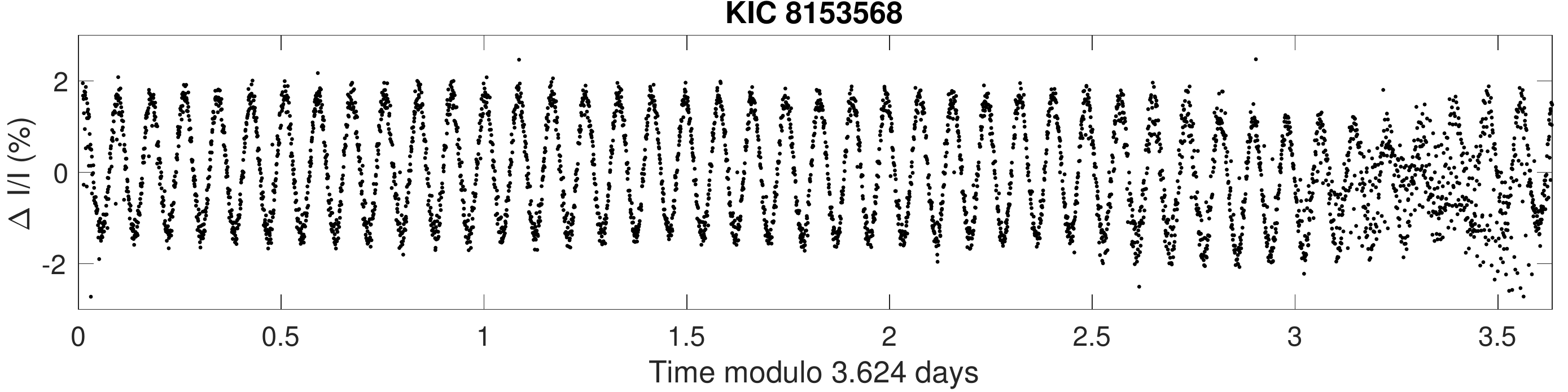}\\
\includegraphics[width=8cm]{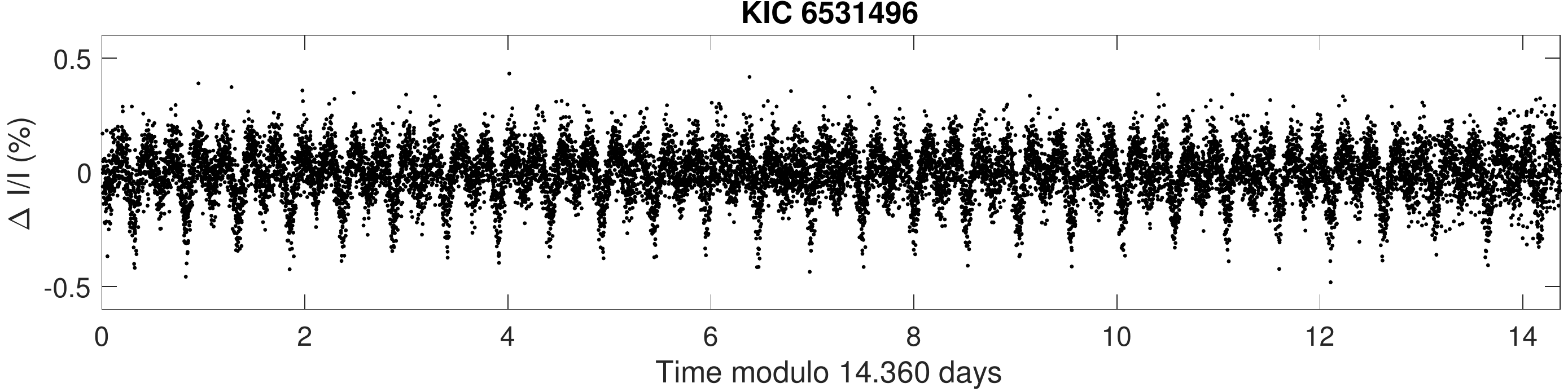}\\
\includegraphics[width=8cm]{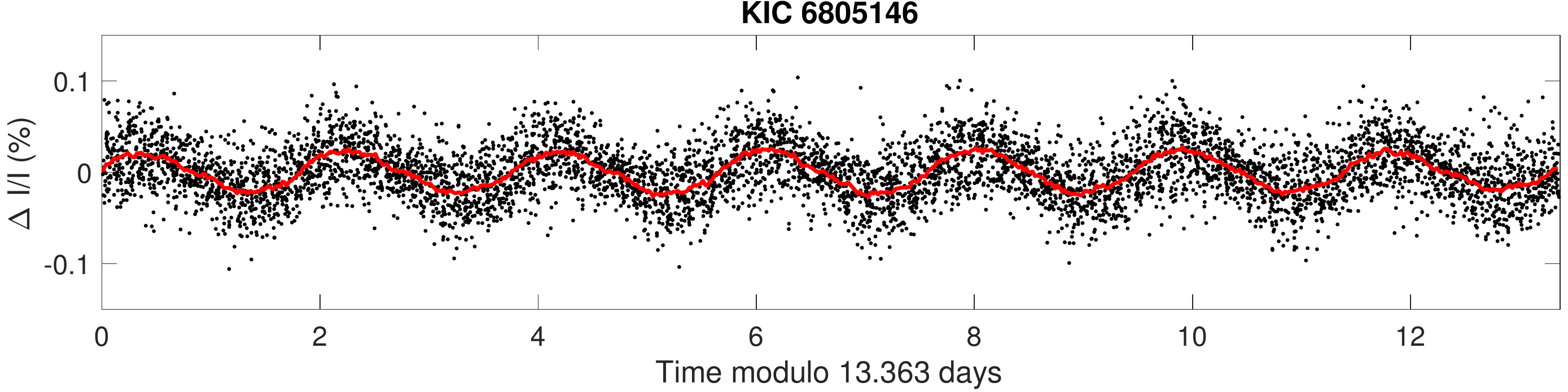}
\end{center}
\caption{\small Top: light curve of KIC 8153568 from Fig. \ref{fig_tidal_puls} folded over 44 times the period corresponding to the main peak in the Fourier spectrum (3.624 days).  The average eclipse signal was removed from the time series prior to folding. The region in between $t = [3, 3.6]$ days corresponds with the primary eclipses. Middle: same with the light curve of KIC 6531496 from Fig. \ref{fig_tidal_puls} folded over 28 times the period corresponding to the main peak in the Fourier spectrum (14.360 days). Bottom: light curve of KIC 6805146 folded over 7 times the period corresponding to the main peak in the Fourier spectrum (13.363 days). \label{fig_8153568}}
\end{figure}
\subsection{Other types of pulsators}
\label{sect_other_types}
Beyond $\gamma$ Dor, $\delta$ Sct, and RG we have identified 62 systems where other types of oscillations are possibly detected, most of which (59) being tidally excited. 
The Kepler mission has led to discover a large number of tidal pulsators and attracted much interest regarding it \citep[e.g.,][]{Welsh_2011, Dong_2013, Maceroni_2014, Smullen_2015,  Hambleton_2013, Hambleton_2016, Hambleton_2018,Kjurkchieva_2016, Lee_2016,Lee_2016b, Shporer_2016, Lee_2017,Dimitrov_2017}. Not surprisingly, among the 59 possibly tidal pulsators that we list in Table \ref{tab:maintable} about half (30) were already identified as such. Note that many (35) of them are not only tidal pulsators, but also $\gamma$ Dor or $\delta$ Sct, or both. For example, KIC 4544587 was studied in detail by \citet{Hambleton_2013}: they noticed that there were self-excited pressure and gravity modes ($\delta$ Sct types and $\gamma$ Dor respectively), but also tidally excited modes and tidally influenced p-modes, which were identified because their frequencies were harmonics of the orbital period. Their work perfectly illustrates how complex the identification of tidally excited oscillations might be: several types of oscillation modes can be present and it is hard to identify and make sure of the tidally-excited nature of some of them. Besides, harmonics of the orbital frequency can also be caused by insufficient removal of the eclipse signal. This difficulty of identifying tidal oscillations was already pointed out by \citet{Aerts_2010}, prior to the first Kepler results. 

As an example, Fig. \ref{fig_tidal_puls} shows three examples of pulsators whose oscillation spectra look like tidal ones, but where two are not. The left panel shows an extract of the light curve and Fourier spectrum of KIC 8153568, a detached EB orbiting in 3.607 day, which looks a typical tidal pulsator where oscillations of constant amplitude are observed all along the orbit. We note that the oscillation amplitude is suppressed during the deeper eclipses, which means that only one of the two components of the binary system oscillates. However, when folding the light curve over the orbital period, the periodic modulation averages out. The only peak that dominates the Fourier spectrum has a frequency of $\nu\ind{peak} = 140.511\ \mu$Hz, which is not an integer multiple of the orbital frequency (factor 43.79). Tidal oscillations are expected at integer multiples of the orbital frequency because excited modes do not oscillate at their natural frequencies, but rather at tidal forcing frequencies, i.e., orbital. When a pulsation is observed at not an exact integer multiple of the orbital frequency, it is almost certainly not a tidally excited pulsation. Also, tidally excited pulsations are almost always found in eccentric systems, which is not the case here.
The effective temperature of KIC 8153568 (6800 K) makes it compatible with a $\delta$ Sct, even though the oscillation spectrum does not look like a $\delta$ Sct one, with only one peak surrounded by a few aliases caused by the amplitude modulation of the pulsations during the primary eclipses. Figure \ref{fig_8153568} displays the time series (where the eclipse signal is removed) folded over $44 \times 1/\nu\ind{peak}$ (i.e., 3.624 days) and it shows how constant and coherent is the pulsation of that star.

A handful of cases look similar to KIC 8153568: they also are ``false'' tidal oscillators, but for different reasons. For example, KIC 6531496  -- a 14.32-day orbit detached system (Fig. \ref{fig_8153568}, middle) -- whose light curve looks like that of KIC 8153568 is of a different nature after further investigation. The time series displays a photometric modulation along the orbit, which appears as a series of peaks in the power spectrum. The dominant peak corresponds to a period of 0.5129 days, which is close to being a 28th of the orbital period. When we fold the light curve over $28\times 0.5129$ days (14.36 days), we observe a very coherent periodic signal as with KIC 8153568 (Fig. \ref{fig_8153568}, middle panel). However, the 0.5129-day modulation is not a sine curve but a typical OC binary light curve, which means that KIC 6531496 has no tidal oscillations but is instead either a quadruple system or two binary systems blending into each other. One could then argue that KIC 8153568 is not a binary system including a stellar pulsator, but another quadruple or another pair of blended binaries. However, KIC 8153568's pulsation period is 0.08 days which seems a little short to be a binary, even though the Villanova catalog presents some systems with periods down to 0.05 days. Besides, the amplitude of the photometric modulation of KIC 8153568 disappears during one of the two eclipses, which proves that it is intrinsic to one of the two stars.

Another system where the origin of a photometric modulation is unclear is the 13.78-day orbit detached binary KIC 6805146, which shows a sine modulation at about but not exactly $P\ind{orb}/7$ (Fig. \ref{fig_8153568}, bottom panel). Again, it cannot be a tidally excited mode because it is not an integer multiple of the orbital period, so the most likely option is that it is a contaminating ellipsoidal binary with period of 1.909 days. A last example is KIC 4677321 (described in Table \ref{tab:maintable} but not displayed in any figure), where we observe two prominent peaks at frequencies $(18 \pm 1/4) \nu\ind{orb}$, which suggest a rotational splitting where rotation period would be $\approx 4 m P\ind{orb}$ where $m$ is the azimuthal order of the excited mode. Assessing the nature of these tidally-looking but tidally-unlikely pulsations requires some more studies, involving high resolution spectroscopy. 

Despite the series of examples of misleading cases that resemble tidally excited pulsators, most of the systems we flagged as hosting tidally excited pulsations are HB stars. The right panel of Fig. \ref{fig_tidal_puls} displays an HB system (KIC 11572363) for which tidally excited modes are reported for the first time. 
There is a distinction in between the HB signal and tidally excited oscillations. The HB signal is the result of strong gravitational distortions and heating during periastron passage and it does not last the whole orbital period. Tidally excited modes are oscillations driven by the tidal force onto the internal structure of the star. It has been observed that some HB stars display tidal oscillations, which are often at exact multiples of the orbital frequency (e.g., Fig. \ref{fig_tidal_puls}, right panel), and some others do not. 
The latter case is often seen with HB systems including an RG star, as observed by \citet{Beck_2014}. Note that as observed by \citet{Thompson_2012}, it may be sometimes hard to determine what fraction of the peaks is due to the HB shape and what fraction is due to stellar pulsation.

\begin{figure*}[t]
\begin{center}
\includegraphics[width=6cm]{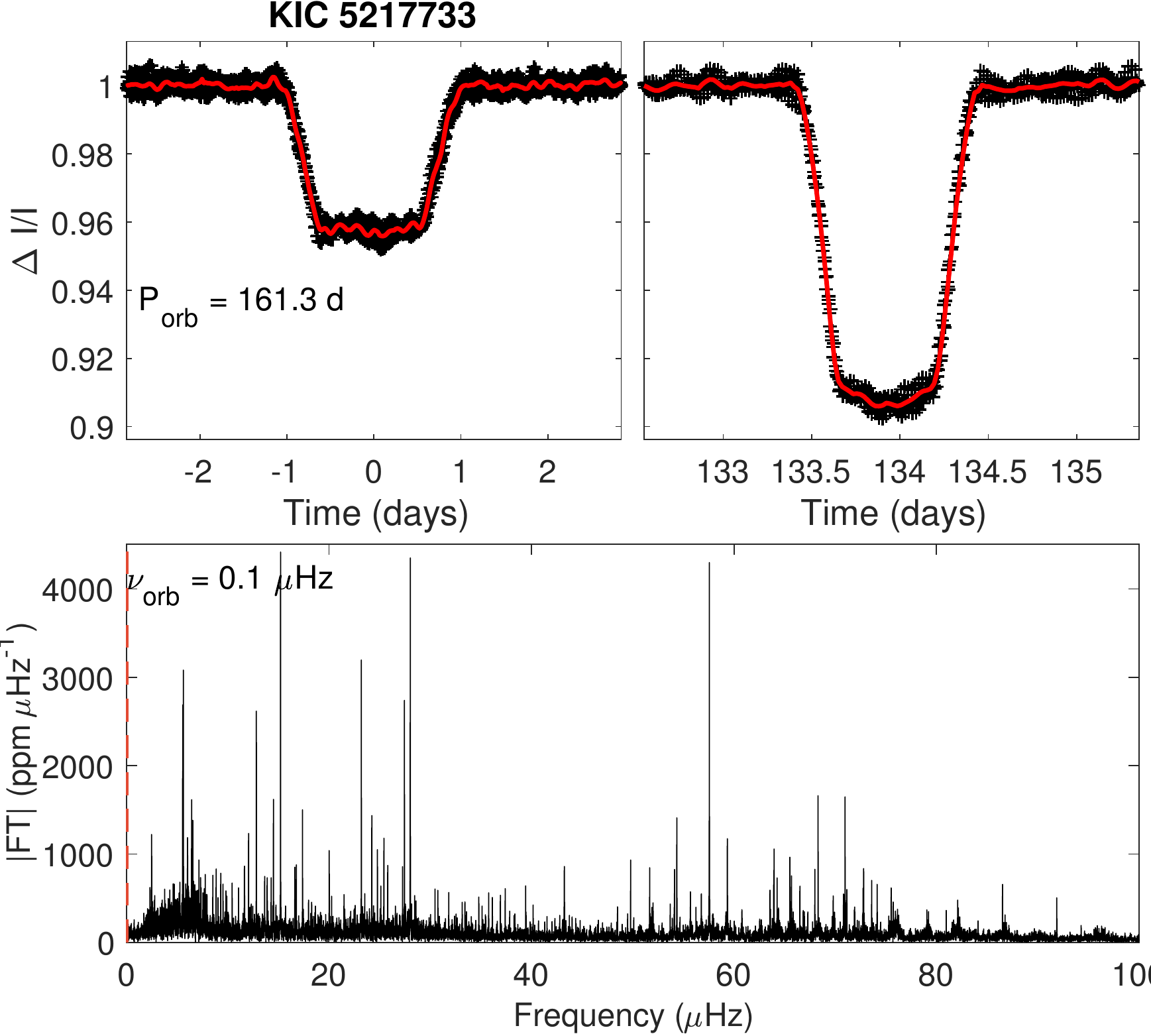}
\includegraphics[width=6cm]{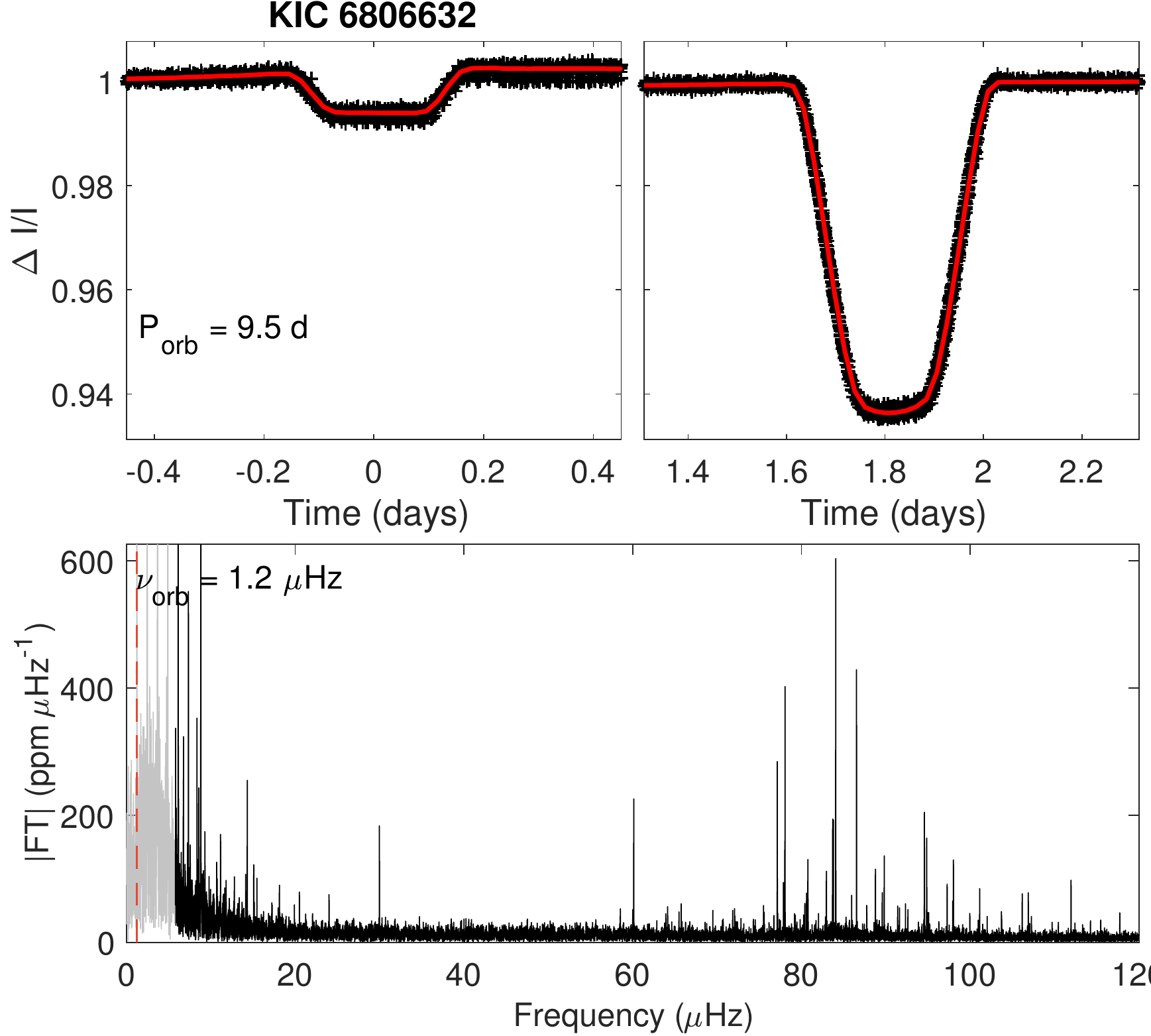}\\
\includegraphics[width=6cm]{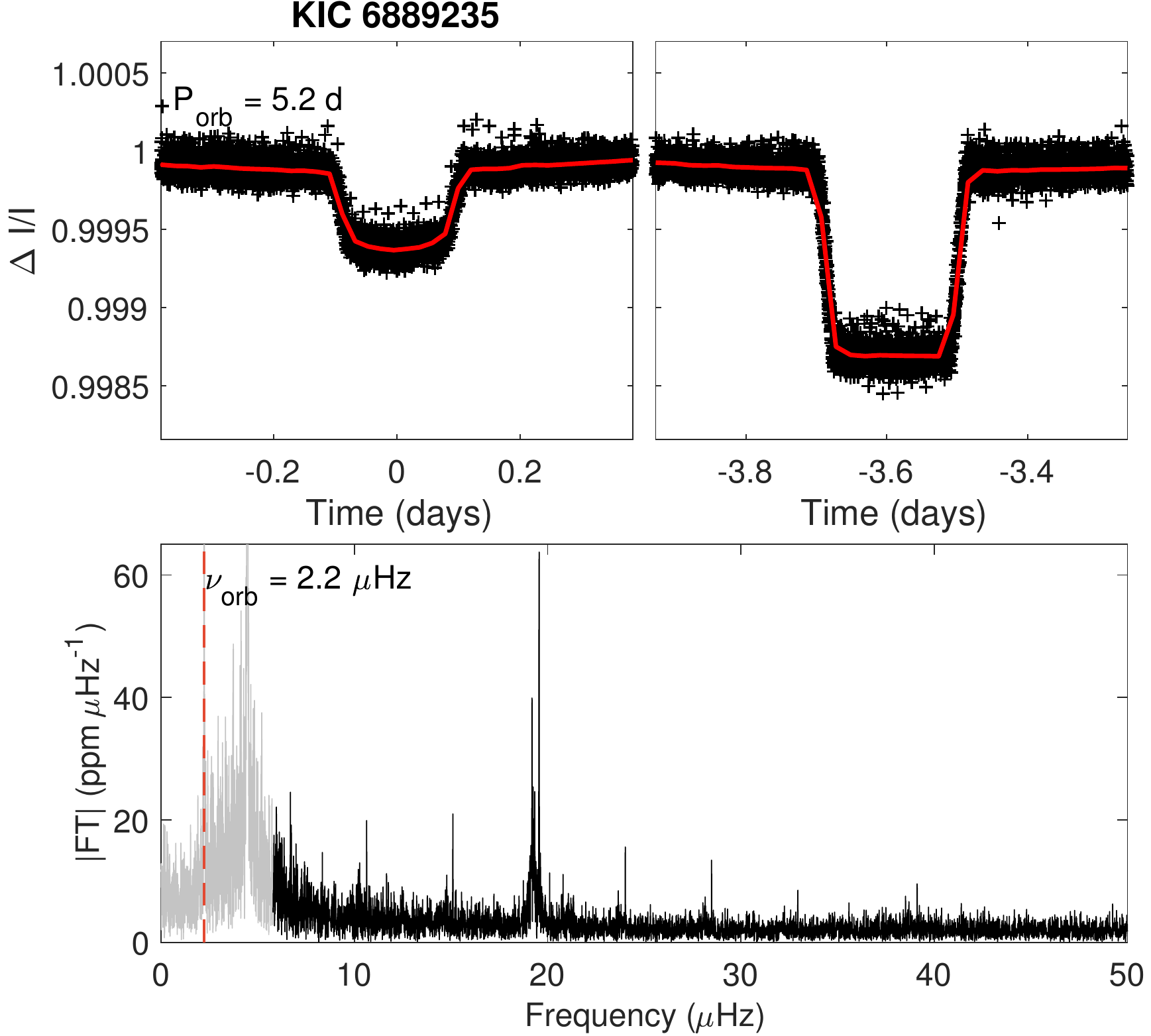}
\includegraphics[width=6cm]{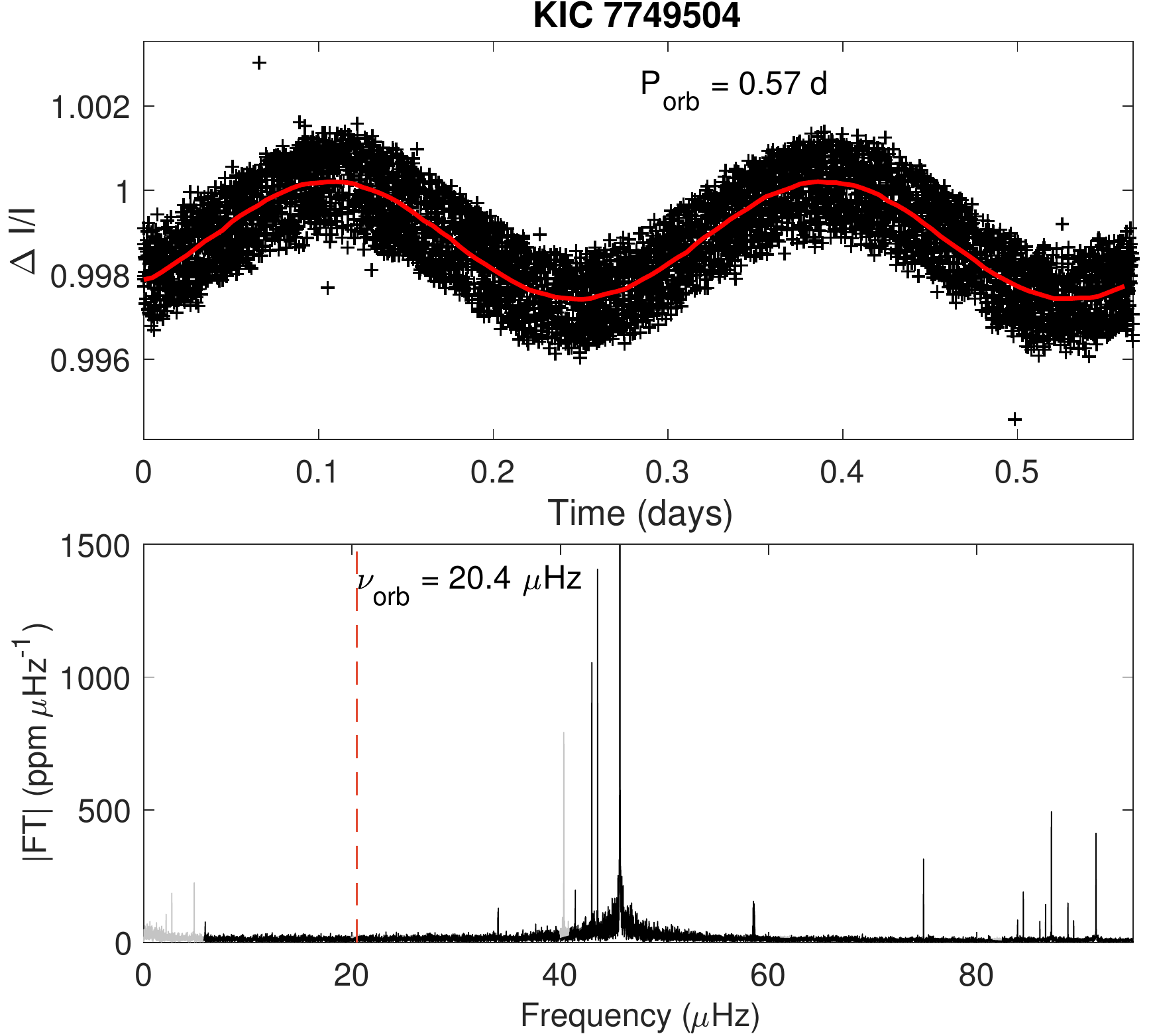}
\end{center}
\caption{\small The four systems in which we identify pulsations but where we have doubts about their nature (see Sect.\ref{sect_other_types}). KICs 5217733 and 6806632 could be either SPB or $\gamma$ Dor/$\delta$ Sct. KIC 6889235 also known as KOI 74 is likely white dwarfs orbiting an A star with tidally excited oscillations. KIC 7749504 is either a rotational or a Maia variable.}\label{fig_weird_puls}
\end{figure*}

Finally a handful of systems display different types of oscillations. 
KICs 5217733 and 6806632 display oscillations that look like those of $\gamma$ Dor/$\delta$ Sct and $\delta$ Sct respectively but their effective temperatures of $\approx 9200$ K according to the KIC are a little large for such kind of pulsators (Fig. \ref{fig_weird_puls}). It could be either a $\delta$ Sct with an overestimation of its $T\ind{eff}$ or an SPB with an underestimated $T\ind{eff}$, which we expect to be about 11,000 K.
KICs 6889235 and 8223868, also known as KOI 74 and 81 were identified as  likely white dwarfs orbiting an A and B star respectively \citep{Rowe_2010}. In both cases, oscillations are visible and are likely of tidal nature. KIC 7749504 is an ellipsoidal binary orbiting in 0.57 days: it displays clear peaks (last panel in Fig. \ref{fig_weird_puls}) and $T\ind{eff} \approx 11 000$ K. This system was part of \citet{Balona_2015}'s search for Maia variables that are postulated between $\delta$ Sct and SPB, and the author concluded it was a rotational variable. We confirm the presence of a series of multiple peaks at more than twice and four times the orbital period, without being able to completely discard the hypothesis of non-rotational peaks. Finally, KIC 11179657 is the only sdB pulsator observed among the EB catalog and was classified as such by \citet{Pablo_2012}.

\section{Conclusion}
To summarize, we led the first systematic search for stellar pulsators in the likely-to-be complete Kepler eclipsing binary catalog, which includes 2925 systems among which about 600 are actually ellipsoidal binaries. For this, we developed a dedicated data inspection tool that automatically processes the Kepler light curves, provided elementary information about EB properties are fed in (orbital period, epochs and duration of eclipses). We focused on three main classes of stellar oscillators: classical $\delta$ Sct and $\gamma$ Dor pulsators, solar-like oscillators which happened to be all red giants, and tidal pulsators. 

We first inspected the output figures produced by the DIT (e.g., Fig. \ref{fig_visualization_tool}) and made our own classification. Then, based on a manual search over the internet for possible earlier studies regarding these systems, we established a list of 303 pulsators associated with an EB, including 163 that were reported for the first time as pulsator and binary candidates. A total of 149 stars are flagged as $\delta$ Sct (100 from this paper), 115 stars as $\gamma$ Dor (69 new), 85 stars as RGs (27 new), 59 as tidally excited oscillators (29 new). There is some overlap among these groups, as some display $\delta$ Sct, $\gamma$ Dor and tidal oscillations altogether. 
Many of these systems are likely to be false positives, i. e., when an EB light curve is blended with a pulsator, and only deeper studies -- often involving complementary RV measurements -- would allow us to make these candidate confirmed pulsators in EBs. False positives are expected to be relatively more common among the shallow-eclipse and short-period systems. 

Among interesting facts are some very short orbit systems ($\leq 0.5$ days) displaying stellar oscillations ($\delta$ Sct and $\gamma$ Dor). Some of them are likely to be false positives, but we know from the literature that $\delta$ Sct pulsators have been identified in almost contact binaries, but not at period under 0.6 days. Some ground based support will be led to determine their nature. The opportunity of detecting oscillations of contact binaries would provide unique views on these types of systems that are very common. As regards red giant oscillators, one of the major current goal is to identify a large sample of oscillators in EBs to test and calibrate asteroseismic inference on stellar masses. We were able to estimate that less than 20 RG/EB/SB2 are present in the whole Kepler dataset, and that hierarchical triple systems may help to bring the sample up to about 30. We also identified new tidally eccentric binaries (HB) with tidal oscillations, as well as detached binaries that display regular coherent pulsations that look like tidal modes, but with periods that are not integer fractions of orbital periods.  

In the context of imagining the future of asteroseismology, the present discovery of stellar pulsators in eclipsing binary candidates constitutes a valuable sample that deserves to be studied in further detail, especially with the help of complementary observations. Priority should be given to the brightest targets for which a fine characterization is obtained more quickly, especially about the rate of false positives. Priority should also be given to systems that are unique, as the three likely RG/EB/SB2s, or some short period classical pulsators. The authors are involved in high-resolution spectroscopic observations with the 3.5-m telescope of the Apache Point Observatory. However, given the large number of systems, such a survey will not be performed only with one telescope. We invite scientists that are interested to contact us in order to coordinate possible future observations.

\section*{Acknowledgments}
P. Gaulme was supported in part by the German space agency (Deutsches Zentrum f\"ur Luft- und Raumfahrt) under PLATO data grant 50OO1501.
P. Gaulme acknowledges NASA grant NNX17AF74G for partial support. The authors thank Jason Jackiewicz at New Mexico State University for comments and text editing. P. Gaulme thanks Cilia Damiani at Max Planck Institut f\"ur Sonnensystemforschung, Jim Fuller at Caltech, Donald W. Kurtz at University of Central Lancashire, Dominic Bowman at KU Leuven, Kosmas Gazeas  at University of Athens, and Giovanni Mirouh at University of Surrey for useful comments on the manuscript. The authors thank the Apache Point Observatory 3.5-meter telescope, which is owned and operated by the Astrophysical Research Consortium, for granting us observing time during quarters 2 and 3, 2019 to support the follow up of the present work.

\longtab{
\begin{tiny}
\begin{longtable}{ p{.05\textwidth}  p{.03\textwidth} p{.03\textwidth} p{.02\textwidth}  p{.01\textwidth}   p{.03\textwidth}  p{.04\textwidth}  p{.02\textwidth}  p{.02\textwidth}  p{.035\textwidth} p{.11\textwidth} p{.1\textwidth} p{.01\textwidth} p{.2\textwidth}} 
\caption{Properties of the systems where pulsations are detected. Systems are sorted by increasing KIC number (first column). Second and third columns are the effective temperatures, and Kepler magnitudes from the KIC. The ``Type'' column indicates the type of binary systems: D for detached, SD for semi-detached, OC for over-contact, and ELL for ellipsoidal. The ``Ecl'' column indicates the number of eclipses that are visible per orbital period (0 for ELL systems, 1 or 2 for the others). Column 6 displays the orbital period $P\ind{orb}$ expressed in days rounded at two digits. Columns 7 to 10 indicate the eclipse depth of the deepest eclipse (primary) expressed in percent, the orbital phase separation ``sep'' in between the secondary and the primary eclipse, the eclipse duration ratio $W_2/W_1$, and the sum of eclipse duration relative to orbital period $(W_2+W_1)/P\ind{orb}$. The pulsator type ``Puls type'' is either $\delta$-Sct for $\delta$ Scuti, $\gamma$ Dor for $\gamma$ Doradus, ``Rossby'', ``RG'' for RG solar-like oscillator, ``tidal'', SPB for slowly pulsating B-type stars, ``WD'' for white dwarf, or sdB for subdwarf B. The type ``rot'' stands for surface stellar ``rotation'' that may mimic stellar pulsations. The literature references are compacted (see below). ``New PB'' indicates whether a target is identified both as a pulsator and a binary system for the first time with a ``Y''. Notes are random relevant specific information regarding each system. ``HB'' stands for heartbeat, FP for False Positive, KOI for Kepler Object of Interest, $P\ind{orb}$ for orbital period, ``ecl.'' for eclipse, ``harm.'' for harmonics, $T_1$ and $T_2$ for star 1 star 2 effective temperatures,  $T\ind{eff}$ refers to the published KIC effective temperature, ``SB1'' and ``SB2'' for single or double-lined spectroscopic binary. When we indicate ``$P\ind{orb} = P\ind{orb}/2$'', it means that the actual orbital period is half of what Villanova database reported.}\\
\hline
KIC &  $T\ind{eff}$& Kmag  & Type& Ecl & $P\ind{orb}$ & Depth & Sep  &  $\frac{W_2}{W_1}$&$\frac{W_1+W_2}{P\ind{orb}}$& Puls type & Literature &New PB  & Notes \\
        & [K]              &             &         &       &     [day]         & [\%] &      &       &  [\%]  & & & \\
\hline
\endfirsthead
\multicolumn{14}{c}%
{\tablename\ \thetable\ -- \textit{Continued from previous page}} \\
\hline
KIC &  $T\ind{eff}$& Kmag  & Type& Ecl & $P\ind{orb}$ & Depth & Sep  &  $\frac{W_2}{W_1}$&$\frac{W_1+W_2}{P\ind{orb}}$& Puls type & Literature &New PB  & Notes \\
        & [K]              &             &         &       &     [day]         & [\%] &      &       &  [\%]  & & & \\
\hline
\endhead
\hline \multicolumn{11}{r}{\textit{Continued on next page}} \\
\endfoot
\hline
\endlastfoot
1295531 & ... & ... & ELL & 0 & 1.69 & 0.97 & ... & ... & ... & $\gamma$ Dor &  & Y &  \\ 
2162283 & 6683 & 9.55 & ELL & 0 & 0.91 & 1.07 & ... & ... & ... & $\gamma$ Dor, $\delta$ Sct & \citetalias{Niemczura_2017} &  &  \\ 
2306740 & 5647 & 13.54 & D & 2 & 10.31 & 31.53 & 0.52 & 0.54 & 6.0 & Rotation(?), Tidal(?) & \citetalias{Yakut_2015} &  &  \\ 
2444348 & 4546 & 10.32 & D & 0 & 103.21 & 0.07 & ... & ... & ... & RG & \citetalias{Beck_2014} &  & HB \\ 
2697935 & 4883 & 10.63 & D & 1 & 21.51 & 0.24 & ... & ... & ... & RG & \citetalias{Beck_2014} &  & HB \\ 
2711114 & 6281 & 12.34 & D & 2 & 2.86 & 0.33 & 0.50 & 0.93 & 6.4 & RG &  & Y & FP or triple \\ 
2720096 & 4832 & 13.05 & D & 0 & 26.67 & 0.00 & ... & ... & ... & RG & \citetalias{Beck_2014} &  & HB \\ 
2720354 & 6513 & 13.12 & D & 2 & 2.82 & 7.35 & 0.50 & 0.57 & 10.8 & $\gamma$ Dor, Rossby(?) &  &  & Villanova Pu \\ 
2970804 & ... & 9.16 & SD & 2 & 0.69 & 0.71 & ... & ... & ... & $\gamma$ Dor, $\delta$ Sct, Rossby(?) &  & Y &  \\ 
2997455 & 4795 & 13.80 & D & 2 & 1.13 & 0.45 & ... & ... & ... & RG & \citetalias{Gaulme_2013} &  & FP or triple \\ 
2998124 & 6374 & 13.12 & D & 2 & 28.60 & 8.94 & 0.47 & 1.21 & 2.4 & $\delta$ Sct &  & Y &  \\ 
3228863 & 6561 & 11.82 & SD & 2 & 0.73 & 48.21 & 0.50 & 0.76 & 27.9 & $\gamma$ Dor & \citetalias{Lee_2014} &  &  \\ 
3327980 & 7321 & 12.12 & D & 2 & 4.23 & 42.54 & 0.50 & 1.04 & 11.5 & $\delta$ Sct, Rossby(?) & \citetalias{Alicavus_2017} & Y &  \\ 
3441784 & ... & 9.73 & D & 1 & 52.57 & 2.44 & ... & ... & ... & $\delta$ Sct &  & Y & KOI 976 \\ 
3547874 & 6371 & 10.95 & D & 0 & 19.69 & ... & ... & ... & ... & Tidal & \citetalias{Thompson_2012,Zimmerman_2017} &  & HB, $P\ind{orb}$ harm., spots \\ 
3556229 & ... & 17.66 & SD & 2 & 0.80 & 15.45 & 0.50 & 0.87 & 27.8 & Tidal? &  & Y &  \\ 
3735597 & 6556 & 12.24 & SD & 2 & 1.97 & 18.77 & 0.50 & 0.90 & 25.7 & $\gamma$ Dor, $\delta$ Sct &  & Y &  \\ 
3749404 & 7144 & 10.57 & D & 0 & 20.31 & ... & ... & ... & ... & Tidal & \citetalias{Hambleton_2016} &  & HB, $P\ind{orb}$ harm., triple \\ 
3761175 & 6890 & 12.83 & ELL & 0 & 5.16 & 0.18 & ... & ... & ... & RG or $\gamma$ Dor &  & Y & $\gamma$ Dor $T\ind{eff}$, RG osc. \\ 
3851949 & 4981 & 13.94 & D & 2 & 54.77 & 0.73 & 0.64 & 0.70 & 5.7 & RG &  & Y & $T_2<T_1$, depleted $\ell=1$, \\ 
3858884 & ... & 9.28 & D & 2 & 25.95 & 37.31 & 0.77 & 1.42 & 5.0 & $\delta$ Sct, Tidal & \citetalias{Maceroni_2014} &  &  \\ 
3863594 & ... & ... & OC & 0 & 0.05 & 0.00 & ... & ... & ... & $\gamma$ Dor/$\delta$ Sct &  & Y & Likely FP (shallow ecl) \\ 
3867593 & 7037 & 13.56 & D & 1 & 73.34 & 9.46 & ... & ... & ... & $\gamma$ Dor & \citetalias{Debosscher_2011} &  &  \\ 
3869326 & 4903 & 11.60 & D & 2 & 8.59 & 0.08 & ... & ... & ... & RG & \citetalias{Bedding_2011} & Y & Blend or triple (shallow ecl) \\ 
3869825 & 6476 & 13.32 & D & 2 & 4.80 & 1.56 & 0.47 & 2.43 & 25.4 & $\gamma$ Dor, $\delta$ Sct &  & Y &  \\ 
4054905 & 4702 & 12.98 & D & 2 & 274.71 & 16.83 & 0.31 & 0.61 & 2.9 & RG & Ben+ &  & SB2 \\ 
4055765 & 6440 & 12.60 & D & 1 & 9.97 & 0.15 & ... & ... & ... & $\gamma$ Dor & \citetalias{Katsova_Nizamov_2018} & Y & KOI 100, flares, $P\ind{orb} = P\ind{orb}/2$ \\ 
4078157 & 5547 & 15.48 & D & 2 & 16.02 & 1.22 & ... & ... & ... & RG & \citetalias{Borkovits_2016} & Y & Triple (ETVs), sec. ecl. \\ 
4142768 & 5401 & 12.12 & D & 1 & 27.99 & 1.75 & ... & ... & ... & $\gamma$ Dor, $\delta$ Sct, Tidal & \citetalias{Manuel_Hambleton_2018} &  & $T\ind{eff}$ cold, HB \\ 
4142768 & 5401 & 12.12 & D & 1 & 27.99 & 1.75 & ... & ... & ... & $\gamma$ Dor, $\delta$ Sct, Tidal & \citetalias{Manuel_Hambleton_2018} &  & HB, $P\ind{orb}$ harm., $T\ind{eff}$ cool \\ 
4150611 & 6623 & 7.90 & D & 3 & 1.52& 0.36 & ... & ... & ... & $\gamma$ Dor, $\delta$ Sct & \citetalias{Helminiak_2017} &  & Quintuple \\ 
 & & & & & 8.65 &  &  &  & ... & &  &  &  \\ 
  & & & & &94.20 &  &  &  & ... & &  &  &  \\ 
4245897 & 6595 & 12.54 & SD & 2 & 11.26 & 75.69 & 0.50 & 0.54 & 11.9 & Tidal?, $\delta$ Sct? &  & Y &  \\ 
4253860 & 6636 & 12.65 & D & 0 & 155.06 & 0.22 & ... & ... & ... & $\gamma$ Dor, $\delta$ Sct &  & Y & HB \\ 
4360072 & 4950 & 11.16 & D & 2 & 1086.50 & 0.61 & 0.44 & 0.90 & 1.0 & RG & Ben+ &  & SB1 \\ 
4544587 & 8255 & 10.80 & D & 2 & 2.19 & 44.79 & 0.65 & 0.75 & 15.1 & $\gamma$ Dor, $\delta$ Sct, Tidal & \citetalias{Hambleton_2013} &  &  \\ 
4570326 & ... & 9.77 & ELL & 0 & 1.12 & 4.73 & ... & ... & ... & $\delta$ Sct & \citetalias{Gaulme_Guzik_2014} &  &  \\ 
4659476 & 6129 & 13.22 & D & 0 & 59.00 & ... & ... & ... & ... & $\gamma$ Dor, Tidal & Guo PhD &  & HB \\ 
4663623 & 4735 & 12.83 & D & 2 & 358.10 & 12.49 & 0.50 & 0.96 & 1.5 & RG & \citetalias{Gaulme_2014,Gaulme_2016a} &  & SB2 \\ 
4677321 & 5979 & 13.48 & SD & 2 & 1.57 & 9.94 & 0.50 & 0.92 & 24.7 & Tidal &  & Y & Two prominent peaks at $(18 \pm 1/4) \nu\ind{orb}$ \\ 
4739791 & 7538 & 14.70 & SD & 2 & 0.90 & 16.14 & 0.50 & 0.95 & 23.9 & $\delta$ Sct, Tidal & \citetalias{Lee_2016b} &  & $P\ind{orb}$ harm., R CMa-type EB \\ 
4758368 & 4594 & 10.80 & D & 2 & 3.75 & 3.62 & 0.50 & 0.96 & 17.7 & RG, $\delta$ Sct, Tidal & \citetalias{Gaulme_2013,Zhang_2017b} &  & Triple (ETVs) \\ 
4769799 & 4911 & 10.95 & D & 2 & 21.93 & 2.22 & ... & ... & ... & RG & \citetalias{Borkovits_2016} & Y & Triple (ETVs), sec. ecl. \\ 
4851217 & 6694 & 11.11 & SD & 2 & 2.47 & 19.80 & 0.48 & 0.97 & 19.4 & $\delta$ Sct & \citetalias{Matson_2017} & Y &  \\ 
4932691 & 7109 & 13.63 & D & 2 & 18.11 & 11.14 & 0.26 & 0.80 & 2.4 & $\gamma$ Dor, Tidal & \citetalias{Kjurkchieva_2016} &  &  \\ 
4947528 & 6828 & 13.91 & OC & 22 & 0.50 & 17.60 & ... & ... & ... & $\gamma$ Dor, Rossby(?) &  & Y &  \\ 
4949187 & 6307 & 13.82 & D & 0 & 11.98 & 0.08 & ... & ... & ... & Tidal & \citetalias{Zimmerman_2017} &  & HB, $P\ind{orb}$ harm. \\ 
4949194 & 8444 & 8.94 & D & 0 & 41.26 & ... & ... & ... & ... & Tidal & \citetalias{Dong_2013} &  & HB, $P\ind{orb}$ harm. \\ 
4949770 & 7907 & 12.57 & ELL & 2 & 1.56 & 13.54 & 0.50 & ... & ... & $\gamma$ Dor, $\delta$ Sct, Rossby(?) &  & Y & $T\ind{eff}$ too hot \\ 
4954113 & 7630 & 11.93 & ELL & 0 & 0.67 & 3.31 & ... & ... & ... & $\delta$ Sct &  & Y &  \\ 
4999260 & 5048 & 9.33 & OC & 22 & 0.38 & 2.35 & ... & ... & ... & RG & \citetalias{Zhou_2010,Gaulme_2013}, SchSt+ &  & Triple \\ 
5006817 & 4935 & 10.87 & D & 0 & 94.81 & 0.18 & ... & ... & ... & RG & \citetalias{Beck_2014} &  & HB \\ 
5024450 & 5945 & 15.06 & D & 2 & 3.05 & 2.23 & 0.50 & 1.06 & 7.6 & $\delta$ Sct & \citetalias{Balona_2013} &  & FP from nearby $\delta$ Sct \\ 
5034333 & 8444 & 11.52 & D & 0 & 6.93 & 0.43 & ... & ... & ... & Tidal & \citetalias{Thompson_2012,Zimmerman_2017} &  & HB \\ 
5039392 & 4152 & 11.65 & D & 0 & 236.73 & 1.44 & ... & ... & ... & RG & \citetalias{Beck_2014} &  & HB \\ 
5179609 & 4777 & 12.78 & D & 2 & 43.93 & 0.97 & ... & ... & ... & RG & \citetalias{Gaulme_2013,Gaulme_2016a} &  & SB1 \\ 
5197256 & 7609 & 11.02 & OC & 0 & 6.96 & 1.04 & ... & ... & ... & $\delta$ Sct & \citetalias{Turner_2015} &  &  \\ 
5198315 & ... & ... & OC & 0 & 3.64 & 3.36 & ... & ... & ... & $\gamma$ Dor? &  & Y & Bump at $[18,28]\ \mu$Hz \\ 
5211385 & 8010 & 12.68 & ELL & 0 & 1.66 & 6.58 & ... & ... & ... & $\gamma$ Dor, $\delta$ Sct &  & Y & $T\ind{eff}$ hot \\ 
5211470 & 4525 & 13.09 & D & 2 & 4.81 & 0.91 & 0.50 & 0.85 & 6.2 & RG &  & Y & Likely FP (no ETVs) \\ 
5217733 & 9116 & 7.39 & D & 2 & 161.25 & 10.09 & 0.17 & 2.03 & 1.9 & SPB or Tidal &  & Y & Hot, flat bottom ecl + high $e$ \\ 
5218014 & 4752 & 12.92 & D & 2 & 10.85 & 0.68 & 0.36 & 1.23 & 3.3 & RG, Tidal(?) & \citetalias{Zhou_2010,Gaulme_2013} &  & Triple? FP? \\ 
5296276 & 5711 & 13.79 & ELL & 0 & 1.89 & 3.47 & ... & ... & ... & Tidal &  & Y &  \\ 
5302006 & 6536 & 15.05 & ELL & 0 & 0.15 & 0.23 & ... & ... & ... & $\delta$ Sct & \citetalias{Balona_2014} &  & Triple \\ 
5308778 & 4812 & 11.78 & D & 2 & 40.57 & 0.21 & 0.50 & 1.05 & 7.8 & RG & \citetalias{Gaulme_2013,Gaulme_2016a} &  & SB1 \\ 
5310435 & 6215 & 13.25 & SD & 2 & 4.93 & 26.57 & 0.50 & 0.64 & 13.4 & $\delta$ Sct &  & Y &  \\ 
5371109 & 7396 & 13.97 & ELL & 0 & 1.19 & 5.15 & ... & ... & ... & $\delta$ Sct &  & Y &  \\ 
5384713 & 3690 & 13.69 & D & 1 & 60.33 & 0.27 & ... & ... & ... & Tidal &  & Y & $P\ind{orb}$ harm., spots, flares \\ 
5467126 & 4683 & 12.37 & D & 2 & 2.85 & 0.00 & 0.50 & 1.00 & 25.1 & RG & \citetalias{Coughlin_2011} & Y & Likely FP \\ 
5475712 & 6370 & 13.28 & SD & 2 & 2.99 & 13.53 & 0.50 & 0.96 & 17.0 & $\delta$ Sct? &  & Y &  \\ 
5479973 & 6169 & 14.89 & SD & 2 & 1.80 & 14.25 & 0.50 & 0.80 & 19.7 & $\delta$ Sct &  & Y & $T\ind{eff}$ cold \\ 
5560831 & ... & 16.09 & SD & 2 & 0.87 & 11.84 & 0.50 & 0.67 & 19.0 & $\gamma$ Dor, $\delta$ Sct, or Tidal &  & Y &  \\ 
5560831 & ... & 16.09 & SD & 2 & 0.87 & 11.84 & 0.50 & 0.67 & 19.0 & $\gamma$ Dor/$\delta$ Sct?, Tidal? &  & Y & Triple (ETVs) \\ 
5565486 & 6471 & 14.96 & D & 2 & 2.83 & 17.41 & 0.50 & 1.09 & 13.8 & $\gamma$ Dor, $\delta$ Sct & \citetalias{Lurie_2017,Zhang_2017} &  &  \\ 
5621294 & 8425 & 13.61 & SD & 2 & 0.94 & 36.47 & 0.50 & 0.93 & 28.8 & $\delta$ Sct & \citetalias{Lee_2015} & Y & $T\ind{eff}$ hot \\ 
5623923 & 8300 & 16.52 & SD & 2 & 1.21 & 8.23 & 0.50 & 0.76 & 19.8 & $\delta$ Sct & \citetalias{Ramsay_2014} &  & $T\ind{eff}$ hot \\ 
5640750 & 4557 & 11.56 & D & 2 & 987.30 & 7.41 & 0.67 & 1.29 & 1.8 & RG & \citetalias{Gaulme_2013,Themessl_2018} &  & SB2 \\ 
5733154 & 6839 & 12.67 & D & 0 & 62.52 & ... & ... & ... & ... & $\delta$ Sct, Tidal &  & Y & HB \\ 
5736537 & 7395 & 13.65 & D & 0 & 1.76 & 1.09 & ... & ... & ... & $\delta$ Sct &  & Y & Peaks are rather evenly spaced \\ 
5786154 & 4743 & 13.53 & D & 2 & 197.92 & 7.83 & 0.28 & 0.77 & 4.7 & RG & \citetalias{Gaulme_2013,Gaulme_2016a} &  & SB2 \\ 
5790807 & 6349 & 9.95 & D & 1 & 80.00 & 2.60 & ... & ... & ... & $\gamma$ Dor & \citetalias{Zimmerman_2017} & Y & HB, rotation \\ 
5809827 & 6076 & 13.54 & SD & 2 & 1.22 & 16.10 & 0.50 & 1.07 & 23.4 & $\gamma$ Dor, Rossby(?) &  & Y &  \\ 
5812701 & 6419 & 11.35 & D & 1 & 17.86 & 1.05 & ... & ... & ... & $\gamma$ Dor & \citetalias{Masuda_2017,Biersteke_2017} & Y & Kepler-448b, $P\ind{orb} = P\ind{orb}/2$ \\ 
5817566 & 7994 & 11.68 & D & 2 & 4.21 & 15.58 & 0.51 & 0.91 & 16.9 & $\delta$ Sct &  & Y & $P\ind{orb} = P\ind{orb}/2$, sec. ecl. \\ 
5866138 & 5078 & 11.07 & D & 2 & 342.30 & 0.43 & ... & ... & ... & RG & Ben+ &  & SB1 \\ 
5872506 & 7571 & 14.73 & OC & 0 & 2.13 & 3.87 & ... & ... & ... & $\delta$ Sct &  & Y &  \\ 
5872696 & ... & 11.85 & ELL & 0 & 0.17 & 0.62 & ... & ... & ... & $\delta$ Sct &  & Y & Very short $P\ind{orb}$ \\ 
5877364 & ... & 8.88 & D & 0 & 89.65 & ... & ... & ... & ... & RG & \citetalias{Shporer_2016} & Y & HB \\ 
5944240 & 7474 & 14.67 & D & 0 & 2.55 & 1.61 & ... & ... & ... & $\delta$ Sct &  & Y & No ecl (ell?) \\ 
5952403 & ... & 6.97 & SD & 2 & 0.91 & 0.32 & ... & ... & ... & Tidal & \citetalias{Southworth_2015} &  & Triply ecl. syst. \\ 
5960989 & 6074 & 12.51 & D & 0 & 50.72 & ... & ... & ... & ... & Tidal & \citetalias{Shporer_2016} &  & HB, $P\ind{orb}$ harm. \\ 
5962514 & 6566 & 14.84 & SD & 2 & 1.58 & 21.44 & 0.50 & 0.95 & 19.4 & $\gamma$ Dor &  & Y &  \\ 
5977736 & 7661 & 13.17 & OC & 0 & 1.58 & 2.13 & ... & ... & ... & $\delta$ Sct &  & Y & No ecl (ell?) \\ 
5983348 & 5680 & 15.05 & D & 2 & 25.15 & 3.61 & 0.50 & 0.78 & 2.0 & Tidal? &  & Y & Sparse spectrum \\ 
5990753 & 4886 & 10.92 & D & 2 & 7.22 & 0.00 & ... & ... & ... & RG & RG/EB \citetalias{Ceillier_2017} &  & Likely FP (shallow ecl) \\ 
5991936 & 8630 & 13.42 & ELL & 0 & 0.81 & 0.10 & ... & ... & ... & $\delta$ Sct? & \citetalias{Balona_2014} &  & Likely FP (shallow ecl), $T\ind{eff}$ hot \\ 
6042191 & 4986 & 13.16 & D & 1 & 43.39 & ... & ... & ... & ... & RG &  & Y & HB, $\ell=1$ depleted \\ 
6063448 & 6416 & 13.76 & D & 2 & 76.02 & 7.78 & 0.78 & 0.81 & 1.1 & $\delta$ Sct, $\gamma$ Dor(?) & \citetalias{Lurie_2017} &  &  \\ 
6109688 & 6845 & 12.27 & D & 2 & 14.09 & 10.73 & 0.55 & 0.52 & 4.1 & $\gamma$ Dor, $\delta$ Sct & \citetalias{Lurie_2017} &  &  \\ 
6117415 & 6282 & 10.54 & D & 1 & 19.74 & 8.47 & ... & ... & ... & Tidal? & Zhao Guo, Valentina Schmid PhDs &  &  \\ 
6145939 & 6090 & 12.37 & D & 1 & 17.75 & 7.98 & ... & ... & ... & $\gamma$ Dor, $\delta$ Sct, Tidal & \citetalias{Lurie_2017} &  & HB \\ 
6220497 & 7254 & 14.75 & SD & 2 & 1.32 & 37.02 & 0.50 & 0.92 & 31.4 & $\delta$ Sct, Tidal & \citetalias{Lee_2016} &  &  \\ 
6262882 & 8317 & 13.98 & OC & 0 & 1.00 & 1.94 & ... & ... & ... & $\delta$ Sct or Tidal &  & Y & $T\ind{eff}$ hot for $\delta$ Sct, not OC but ELL \\ 
6286155 & 5062 & 13.76 & D & 2 & 14.54 & 0.41 & 0.46 & 1.78 & 3.9 & RG &  & Y & Likely FP, $\ell=1$ depleted \\ 
6292398 & ... & ... & D & 0 & 9.24 & 0.04 & ... & ... & ... & $\gamma$ Dor, $\delta$ Sct(?) & \citetalias{Murphy_2015} & Y & HB \\ 
6302592 & 7737 & 13.85 & ELL & 2 & 1.58 & 3.67 & 0.49 & ... & ... & $\delta$ Sct & \citetalias{Zhou_2010} &  &  \\ 
6311681 & 5500 & 15.36 & SD & 2 & 0.84 & 0.60 & ... & ... & ... & RG?, $\gamma$ Dor? &  & Y & Likely FP (shallow ecl) \\ 
6390205 & 8254 & 12.54 & OC & 0 & 0.45 & 5.64 & ... & ... & ... & $\delta$ Sct &  & Y & Very short orbit: triple? \\ 
6525209 & 5207 & 14.66 & D & 2 & 75.13 & 12.40 & 0.22 & 0.51 & 1.6 & RG & \citetalias{Rowe_2015} & Y & $T_2<T_1$, low SNR osc. \\ 
6531496 & 5604 & 16.09 & D & 2 & 14.32 & 5.82 & 0.50 & 1.19 & 2.9 & Tidal(?), Quadruple(?) &  & Y &  \\ 
6549491 & 5597 & 14.91 & OC & 0 & 0.76 & 0.95 & ... & ... & ... & Rotation(?), $\delta$ Sct(?) & \citetalias{Balaji_2015} & Y & $T\ind{eff}$ cool \\ 
6613627 & 7090 & 12.55 & ELL & 0 & 0.15 & 0.67 & ... & ... & ... & $\gamma$ Dor/$\delta$ Sct & \citetalias{Bowman_2016} &  & Very short $P\ind{orb}$: FP, triple? \\ 
6629588 & 6520 & 13.98 & SD & 2 & 2.26 & 20.10 & 0.50 & 0.94 & 18.3 & $\delta$ Sct & \citetalias{Liakos_Niarchos_2017} &  &  \\ 
6669809 & 7239 & 10.76 & SD & 2 & 0.73 & 24.24 & 0.50 & 0.72 & 26.3 & $\delta$ Sct & \citetalias{Borkovits_2016} & Y & Triple (ETVs) \\ 
6757558 & 4742 & 12.87 & D & 1 & 421.16 & 0.35 & ... & ... & ... & RG & Ben+ &  & SB1 \\ 
6762188 & 4801 & 13.67 & D & 1 & 7.16 & 0.27 & ... & ... & ... & RG & \citetalias{Zhou_2010,Gaulme_2013} &  & Triple (ETVs) \\ 
6775034 & 6886 & 13.99 & D & 0 & 10.03 & ... & ... & ... & ... & $\gamma$ Dor(?), Tidal(?) & \citetalias{Shporer_2016,Zimmerman_2017} &  & HB, spots \\ 
6805146 & 6214 & 13.21 & D & 2 & 13.78 & 2.82 & ... & ... & ... & Tidal(?), rotation(?) &  & Y & Sec. ecl., flat bottom ecl. \\ 
6806632 & 9224 & 13.29 & D & 2 & 9.47 & 6.92 & 0.70 & 0.90 & 7.2 & SPB(?),$\delta$ Sct(?) &  & Y & Hot for $\delta$ Sct, cold for SPB, flat bottom ecl, HB \\ 
6847018 & 6211 & 13.34 & D & 2 & 16.66 & 2.75 & ... & ... & ... & Tidal &  & Y & HB, $P\ind{orb}$ harm., sec. ecl. \\ 
6850665 & 4828 & 12.39 & D & 1 & 214.72 & 0.66 & ... & ... & ... & RG &  & Y & HB with ecl. \\ 
6852488 & 7262 & 13.91 & SD & 2 & 2.17 & 17.72 & 0.51 & 0.90 & 18.0 & $\delta$ Sct &  & Y &  \\ 
6889235 & 9288 & 10.96 & D & 2 & 5.19 & 0.16 & 0.50 & 1.14 & 9.9 & WD? SPB? & \citetalias{Rowe_2010} & Y & KOI 74 \\ 
7023917 & ... & 10.12 & OC & 2 & 0.77 & 7.35 & 0.50 & ... & ... & $\delta$ Sct &  & Y & Short orbit \\ 
7037405 & 4605 & 11.88 & D & 2 & 207.15 & 6.86 & 0.40 & 1.16 & 5.1 & RG & \citetalias{Gaulme_2013,Gaulme_2016a} &  & SB2 \\ 
7060333 & 7576 & 9.09 & ELL & 0 & 2.53 & 0.36 & ... & ... & ... & $\gamma$ Dor, $\delta$ Sct & \citetalias{Niemczura_2015} &  &  \\ 
7220322 & 4887 & 11.88 & D & 2 & 0.75 & 0.00 & 0.50 & 1.00 & 20.9 & Tidal &  & Y & $P\ind{orb}$ harm., spots \\ 
7368103 & 7838 & 13.42 & SD & 2 & 2.18 & 12.25 & 0.50 & 0.83 & 16.9 & $\gamma$ Dor, $\delta$ Sct & Zhao Guo PhD &  & $T\ind{eff}$ little hot \\ 
7377422 & 4488 & 13.56 & D & 2 & 107.62 & 1.90 & 0.23 & 0.94 & 4.1 & RG & \citetalias{Gaulme_2013,Gaulme_2016a} &  & SB2, low SNR osc. \\ 
7385478 & 6477 & 11.47 & SD & 2 & 1.66 & 19.01 & 0.50 & 0.83 & 16.0 & $\gamma$ Dor & \citetalias{Ozdarcan_2017} &  &  \\ 
7422883 & 6639 & 11.25 & D & 2 & 11.41 & 4.76 & ... & ... & ... & $\gamma$ Dor & \citetalias{Debosscher_2011} &  & Sec. ecl. \\ 
7431665 & 4729 & 10.97 & D & 1 & 281.40 & 0.58 & ... & ... & ... & RG & \citetalias{Beck_2014} &  & HB with ecl. \\ 
7515679 & 7127 & 12.25 & ELL & 1 & 5.55 & 0.00 & ... & ... & ... & $\gamma$ Dor &  &  & Kepler-1517b, Villanova Pu \\ 
7622486 & ... & 13.09 & SD & 3 & 2.28& 1.92 & ... & ... & ... & $\delta$ Sct(?) & \citetalias{Zhang_2017} & Y & Triple triply eclipsing \\ 
& & & & &40.25 &  &  &  & & & &  &\\ 
7676610 & 8493 & 12.64 & ELL & 0 & 1.23 & 3.73 & ... & ... & ... & $\delta$ Sct & \citetalias{Balaji_2015} & Y & $T\ind{eff}$ hot, short orbit \\ 
7700578 & 6693 & 14.15 & SD & 2 & 1.51 & 47.80 & 0.50 & 0.96 & 22.2 & $\delta$ Sct & \citetalias{Bradley_2015} & Y &  \\ 
7749504 & 11064 & 12.72 & ELL & 0 & 0.57 & 0.27 & ... & ... & ... & Maia var(?), Rot(?) & \citetalias{McNamara_2012,Balona_2015b} & Y & B-type star (ref) \\ 
7765894 & 7233 & 11.96 & ELL & 0 & 3.20 & 4.16 & ... & ... & ... & $\delta$ Sct &  & Y &  \\ 
7767774 & ... & 12.51 & ELL & 0 & 0.17 & 0.33 & ... & ... & ... & $\gamma$ Dor, $\delta$ Sct &  & Y & Very short $P\ind{orb}$: FP, triple? \\ 
7768447 & 4694 & 11.90 & D & 2 & 122.50 & 0.64 & 0.31 & 1.21 & 4.2 & RG &  & Y & Likely FP (shallow ecl) \\ 
7769072 & 4858 & 13.89 & D & 2 & 0.61 & 0.19 & 0.50 & 0.98 & 16.8 & RG & \citetalias{Coughlin_2011} & Y & Likely FP (shallow ecl) \\ 
7799540 & 5177 & 12.39 & D & 0 & 60.00 & 0.00 & ... & ... & ... & RG & \citetalias{Beck_2014} &  & HB \\ 
7833144 & 7724 & 12.56 & D & 0 & 2.25 & 3.32 & ... & ... & ... & $\gamma$ Dor(?), $\delta$ Sct, Tidal &  & Y & Broad excess power at 30-50 $\mu$Hz \\ 
7881722 & 7762 & 13.59 & D & 0 & 0.95 & 1.47 & ... & ... & ... & $\delta$ Sct &  & Y & no ecl. (ell) \\ 
7914906 & 6952 & 11.96 & D & 1 & 8.75 & 4.08 & ... & ... & ... & $\gamma$ Dor, $\delta$ Sct, Tidal & \citetalias{Balona_2014} & Y & HB \\ 
7955301 & 4821 & 12.67 & D & 2 & 15.32 & 1.18 & 0.49 & 1.34 & 6.4 & RG & \citetalias{Coughlin_2011,Gaulme_2013,Rappaport_2013,Borkovits_2016} &  & Triple (ETVs) \\ 
7976783 & 7937 & 11.99 & ELL & 0 & 1.21 & 0.57 & ... & ... & ... & $\gamma$ Dor, $\delta$ Sct &  & Y &  \\ 
8045121 & 6990 & 11.97 & ELL & 0 & 0.26 & 0.53 & ... & ... & ... & $\gamma$ Dor, $\delta$ Sct & Conroy PhD, \citetalias{Borkovits_2016} &  & Triple \\ 
8054233 & 4733 & 11.78 & D & 2 & 1058.00 & 3.81 & 0.41 & 1.29 & 1.2 & RG & \citetalias{Gaulme_2014,Gaulme_2016a} &  & SB1(?) \\ 
8087799 & 7869 & 14.17 & OC & 2 & 0.93 & 5.07 & 0.50 & ... & ... & WD+$\delta$ Sct & \citetalias{Zhang_XB_2017} &  &  \\ 
8095275 & 4683 & 13.61 & D & 0 & 23.01 & 0.34 & ... & ... & ... & RG & \citetalias{Gaulme_2013,Beck_2014} &  & HB \\ 
8112039 & ... & ... & D & 0 & 41.81 & ... & ... & ... & ... & Tidal & \citetalias{Welsh_2011} &  & $P\ind{orb}$ harm., historic HB KOI-54 \\ 
8113154 & 6589 & 12.90 & SD & 2 & 2.59 & 3.08 & ... & ... & ... & $\gamma$ Dor, $\delta$ Sct & \citetalias{Armstrong_2014} & Y & Circumbinary planets \\ 
8129189 & 5080 & 12.48 & D & 2 & 53.65 & 21.57 & 0.63 & 0.97 & 2.7 & RG &  & Y & small RG+RG/SG? \\ 
8143170 & 4957 & 12.85 & D & 2 & 28.79 & 5.09 & 0.46 & 1.13 & 4.2 & RG & \citetalias{Borkovits_2016,Balona_2015} & Y & Triple (ETVs), flares \\ 
8144355 & 4880 & 13.68 & D & 0 & 80.51 & ... & ... & ... & ... & RG & \citetalias{Beck_2014} &  & HB \\ 
8153568 & 6803 & 15.08 & SD & 2 & 3.61 & 82.49 & 0.50 & 0.84 & 21.1 & Tidal & \citetalias{Bradley_2015} & Y &  \\ 
8164262 & 7487 & 13.36 & D & 0 & 87.46 & 0.19 & ... & ... & ... & Tidal g-modes & \citetalias{Hambleton_2018,Saio_2018} &  & HB, $P\ind{orb}$ harm. \\ 
8182360 & 6904 & 15.32 & SD & 2 & 0.70 & 9.22 & 0.50 & 0.96 & 36.7 & $\delta$ Sct & \citetalias{Bradley_2015} & Y &  \\ 
8193315 & 6457 & 13.70 & D & 2 & 2.62 & 22.40 & 0.50 & 1.01 & 12.4 & $\gamma$ Dor, $\delta$ Sct &  & Y &  \\ 
8197761 & 7068 & 10.66 & D & 1 & 19.74 & 0.49 & ... & ... & ... & $\gamma$ Dor, $\delta$ Sct & \citetalias{Sowicka_2017} &  &  \\ 
8210370 & 4793 & 11.17 & D & 0 & 153.70 & 0.64 & ... & ... & ... & RG & \citetalias{Beck_2014} &  & HB \\ 
8219268 & 4712 & 12.49 & SD & 1 & 6.25 & 0.05 & ... & ... & ... & RG & \citetalias{LilloBox_2014} & Y & Kepler-91b,c (no EB) \\ 
8240109 & 7740 & 13.49 & SD & 2 & 2.30 & 11.29 & 0.50 & 1.78 & 18.3 & $\gamma$ Dor(?), $\delta$ Sct &  & Y &  \\ 
8262223 & 7596 & 12.15 & SD & 2 & 1.61 & 15.58 & 0.50 & 1.04 & 16.6 & $\delta$ Sct & \citetalias{Guo_2017} &  & pre-WD+$\delta$ Sct \\ 
8264510 & 7478 & 10.61 & D & 0 & 5.69 & ... & ... & ... & ... & $\delta$ Sct(?), Tidal & \citetalias{Thompson_2012} & Y & HB \\ 
8308347 & 4826 & 15.47 & D & 2 & 164.95 & 6.41 & 0.61 & 0.50 & 3.2 & RG &  & Y & $T_2>T_1$ \\ 
8330092 & 6902 & 13.48 & OC & 22 & 0.32 & 3.71 & ... & ... & ... & $\gamma$ Dor, $\delta$ Sct & \citetalias{Balona_2013} & Y & OC, short orbit \\ 
8410637 & 4682 & 10.77 & D & 2 & 408.32 & 10.47 & 0.23 & 3.48 & 2.8 & RG & \citetalias{Hekker_2010,Frandsen_2013} &  & SB2 \\ 
8429450 & ... & 13.11 & D & 2 & 2.71 & 34.95 & 0.50 & 1.00 & 12.8 & $\gamma$ Dor, $\delta$ Sct & \citetalias{Orosz_2015} & Y & Triple (ETVs) \\ 
8430105 & 4965 & 10.42 & D & 2 & 63.33 & 1.97 & 0.34 & 0.98 & 6.7 & RG & \citetalias{Gaulme_2014,Gaulme_2016a} &  & SB2 \\ 
8430210 & 8636 & 12.45 & ELL & 0 & 1.98 & 0.58 & ... & ... & ... & $\gamma$ Dor, $\delta$ Sct, or Tidal &  & Y & $T\ind{eff}$ hot \\ 
8452840 & 6473 & 12.59 & ELL & 0 & 1.20 & 0.28 & ... & ... & ... & $\gamma$ Dor & \citetalias{Gaulme_2013} &  &  \\ 
8453324 & 4733 & 11.52 & D & 0 & 2.52 & 0.74 & ... & ... & ... & RG & \citetalias{Zhou_2010,Gaulme_2013} &  & Likely FP \\ 
8454250 & ... & 12.78 & D & 1 & 5.08 & 0.23 & ... & ... & ... & $\gamma$ Dor, $\delta$ Sct &  & Y & Exoplanet? \\ 
8455359 & 6645 & 14.21 & OC & 0 & 2.96 & 2.52 & ... & ... & ... & $\gamma$ Dor, $\delta$ Sct &  & Y &  \\ 
8456774 & 6757 & 13.22 & D & 0 & 2.89 & ... & ... & ... & ... & $\delta$ Sct, Tidal & Zhao Guo PhD & Y & HB \\ 
8459354 & 7433 & 11.14 & D & 0 & 53.56 & 0.24 & ... & ... & ... & $\gamma$ Dor, $\delta$ Sct, Tidal & \citetalias{Moya_2017} &  & HB \\ 
8504570 & 6874 & 13.25 & D & 2 & 4.01 & 16.56 & 0.50 & 1.11 & 10.0 & $\gamma$ Dor, $\delta$ Sct &  & Y &  \\ 
8548416 & ... & 13.34 & ELL & 0 & 1.16 & 3.01 & ... & ... & ... & $\gamma$ Dor &  & Y &  \\ 
8553788 & 8045 & 12.69 & SD & 2 & 1.61 & 14.66 & 0.50 & 0.90 & 18.5 & $\delta$ Sct & \citetalias{Liakos_2018} &  & $T\ind{eff}$ hot, Algol type \\ 
8560861 & 7647 & 8.50 & D & 2 & 31.97 & 7.12 & 0.52 & 1.03 & 3.0 & $\gamma$ Dor & \citetalias{Borkovits_2014} &  &  \\ 
8561192 & 6852 & 16.27 & SD & 2 & 2.74 & 65.76 & 0.50 & 0.93 & 21.7 & $\delta$ Sct or Tidal &  & Y & Strong 180-$\mu$Hz signal \\ 
8563964 & ... & 12.94 & ELL & 0 & 0.34 & 1.11 & ... & ... & ... & $\gamma$ Dor & \citetalias{Borkovits_2016} & Y & Triple (ETVs) \\ 
8564976 & 4726 & 13.23 & D & 1 & 152.83 &  0.22 & ... & ... & ... &RG & \citetalias{Kuszlewicz_2019} & & HB, KOI 3890 \\
8569819 & 7137 & 13.04 & D & 2 & 20.85 & 31.31 & 0.50 & 0.42 & 3.3 & $\gamma$ Dor(?), $\delta$ Sct & \citetalias{Kurtz_2015b} &  &  \\ 
8685306 & 6971 & 11.79 & ELL & 0 & 0.81 & 3.02 & ... & ... & ... & $\gamma$ Dor, $\delta$ Sct? & \citetalias{Kouzuma_2018} & Y & OC \\ 
8702921 & 4824 & 11.98 & D & 2 & 19.38 & 0.52 & 0.44 & 4.38 & 26.0 & RG & \citetalias{Gaulme_2014,Gaulme_2016a} &  & SB1 \\ 
8703887 & ... & 11.05 & D & 1 & 14.17 & 1.54 & ... & ... & ... & $\delta$ Sct $\gamma$ Dor, Tidal &  & Y & HB \\ 
8719324 & ... & 11.61 & D & 0 & 10.23 & ... & ... & ... & ... & Tidal & \citetalias{Thompson_2012} &  & HB, $P\ind{orb}$ harm. \\ 
8719419 & 6642 & 12.93 & D & 1 & 12.63 & 3.56 & ... & ... & ... & $\gamma$ Dor &  & Y & Unconfirmed plan cand. \\ 
8800998 & 8616 & 13.72 & ELL & 0 & 0.88 & 1.71 & ... & ... & ... & $\delta$ Sct & \citetalias{George_2018} & Y & $T\ind{eff}$ hot \\ 
8803882 & 5018 & 13.00 & D & 0 & 89.63 & 0.00 & ... & ... & ... & RG & \citetalias{Beck_2014} &  & HB \\ 
8823868 & 9751 & 11.41 & D & 2 & 23.88 & 0.58 & 0.50 & 0.75 & 3.6 & Tidal(?), SPB(?) & \citetalias{Rowe_2010} &  & KOI 81, WD+B \\ 
8848288 & 4624 & 9.84 & SD & 1 & 5.57 & 0.08 & ... & ... & ... & RG & \citetalias{Zhou_2010,Gaulme_2013} &  & Likely FP, exoplanet? \\ 
8894630 & 7117 & 11.54 & ELL & 1 & 1.08 & 3.97 & ... & ... & ... & $\delta$ Sct &  & Y &  \\ 
8912308 & 4872 & 11.43 & D & 0 & 20.17 & ... & ... & ... & ... & RG & \citetalias{Beck_2014} &  & HB \\ 
8912468 & 6194 & 11.75 & ELL & 0 & 0.09 & 0.41 & ... & ... & ... & $\gamma$ Dor, $\delta$ Sct &  & Y & $T\ind{eff}$ cool, Very short $P\ind{orb}$ \\ 
9016693 & 7020 & 11.63 & D & 0 & 26.37 & ... & ... & ... & ... & Tidal & \citetalias{Shporer_2016} &  & HB, $P\ind{orb}$ harm. \\ 
9101279 & 8372 & 13.95 & SD & 2 & 1.81 & 90.09 & 0.50 & 0.82 & 23.9 & $\delta$ Sct &  & Y & $T\ind{eff}$ hot \\ 
9108058 & 6563 & 14.28 & SD & 2 & 2.17 & 61.88 & 0.50 & 0.78 & 21.5 & $\delta$ Sct & \citetalias{Bradley_2015} & Y &  \\ 
9108579 & 6386 & 11.56 & ELL & 0 & 1.17 & 2.02 & ... & ... & ... & $\gamma$ Dor &  & Y &  \\ 
9151763 & 4307 & 11.70 & D & 0 & 438.05 & ... & ... & ... & ... & RG & \citetalias{Beck_2014} &  & HB \\ 
9153621 & 4789 & 12.87 & D & 2 & 305.82 & 3.19 & 0.16 & 3.04 & 3.3 & RG & Ben+ &  & SB2 \\ 
9159301 & 7959 & 12.15 & SD & 2 & 3.04 & 48.88 & 0.50 & 0.89 & 20.7 & $\delta$ Sct & Guo PhD, \citetalias{Matson_2017} &  & SB2 \\ 
9163796 & 5135 & 9.60 & D & 0 & 121.01 & ... & ... & ... & ... & RG & \citetalias{Beck_2014} &  & HB \\ 
9164561 & 8059 & 13.71 & SD & 2 & 1.27 & 4.82 & 0.50 & 0.97 & 20.7 & $\gamma$ Dor, $\delta$ Sct? & \citetalias{Rappaport_2015} & Y & WD, $T\ind{eff}$ hot \\ 
9181877 & 4599 & 11.70 & OC & 22 & 0.32 & 1.11 & ... & ... & ... & RG & \citetalias{Gaulme_2013} &  & Triple (ETVs) \\ 
9207508 & 6718 & 13.71 & SD & 2 & 2.02 & 54.47 & 0.50 & 0.95 & 23.1 & $\delta$ Sct or Tidal & Sumin Tang PhD & Y &  \\ 
9236858 & 6510 & 13.04 & D & 2 & 2.54 & 22.82 & 0.50 & 1.12 & 14.0 & $\gamma$ Dor, $\delta$ Sct(?) & \citetalias{Kjurkchieva_2016b} & Y &  \\ 
9246715 & 4699 & 9.27 & D & 2 & 171.28 & 17.73 & 0.71 & 1.06 & 3.2 & RG & \citetalias{Gaulme_2013,Rawls_2016} &  & RG+RG \\ 
9285587 & 8008 & 12.93 & ELL & 0 & 1.81 & 1.15 & ... & ... & ... & $\gamma$ Dor, $\delta$ Sct & \citetalias{Faigler_2015} &  & dA+WD, $T\ind{eff}$ hot \\ 
9288175 & 6972 & 12.55 & ELL & 0 & 0.26 & 0.58 & ... & ... & ... & $\gamma$ Dor, $\delta$ Sct? &  & Y &  Sparse spectrum\\ 
9291368 & 7889 & 14.01 & SD & 2 & 3.80 & 53.34 & 0.50 & 1.02 & 19.0 & $\delta$ Sct & \citetalias{Bradley_2015} & Y &  \\ 
9343862 & 7709 & 15.01 & SD & 2 & 1.12 & 37.32 & 0.50 & 0.87 & 21.7 & $\delta$ Sct? & \citetalias{Bradley_2015} & Y &  \\ 
9408183 & 4896 & 13.18 & D & 0 & 49.68 & 0.04 & ... & ... & ... & RG & \citetalias{Beck_2014} &  & HB \\ 
9426970 & 6577 & 13.23 & ELL & 0 & 1.77 & 0.99 & ... & ... & ... & $\delta$ Sct &  & Y &  \\ 
9468382 & 4911 & 13.57 & D & 2 & 11.08 & 0.18 & ... & ... & ... & RG &  & Y & Likely FP (shallow ecl) \\ 
9470054 & 7794 & 11.72 & ELL & 0 & 1.47 & 1.78 & ... & ... & ... & $\delta$ Sct &  & Y &  \\ 
9475663 & 7790 & 12.49 & OC & 2 & 0.71 & 4.14 & 0.51 & ... & ... & $\delta$ Sct &  & Y &  \\ 
9480516 & 7051 & 13.39 & ELL & 0 & 1.43 & 0.64 & ... & ... & ... & $\gamma$ Dor(?), $\delta$ Sct &  & Y &  \\ 
9480977 & 7289 & 12.30 & ELL & 0 & 0.87 & 0.31 & ... & ... & ... & $\gamma$ Dor, $\delta$ Sct &  & Y & Sparse spectrum \\ 
9533489 & ... & 12.96 & D & 1 & 197.15 & 0.47 & ... & ... & ... & $\gamma$ Dor, $\delta$ Sct & \citetalias{Bognar_2015} &  &  \\ 
9540226 & 4584 & 11.67 & D & 2 & 175.46 & 2.31 & 0.26 & 0.99 & 5.7 & RG & \citetalias{Gaulme_2013,Beck_2014,Brogaard_2018,Themessl_2018} &  & SB2 \\ 
9592855 & 7290 & 12.22 & SD & 2 & 1.22 & 16.09 & 0.50 & 0.98 & 20.5 & $\gamma$ Dor, $\delta$ Sct & \citetalias{Guo_2017b} &  &  \\ 
9602542 & 8037 & 13.89 & ELL & 2 & 1.46 & 1.71 & 0.50 & ... & ... & $\gamma$ Dor, $\delta$ Sct &  & Y & $T\ind{eff}$ hot \\ 
9612468 & 7202 & 11.53 & ELL & 0 & 0.13 & 0.10 & ... & ... & ... & $\gamma$ Dor, $\delta$ Sct & \citetalias{Demircan_2015} & Y & Very short $P\ind{orb}$: FP, triple? Circumbin. pla. \\ 
9637265 & 7424 & 13.93 & ELL & 0 & 1.86 & 2.75 & ... & ... & ... & $\delta$ Sct &  & Y &  \\ 
9651298 & 7677 & 13.46 & ELL & 0 & 2.16 & 3.56 & ... & ... & ... & $\delta$ Sct &  & Y &  \\ 
9724080 & 7470 & 13.90 & ELL & 0 & 1.17 & 5.91 & ... & ... & ... & $\gamma$ Dor, $\delta$ Sct & \citetalias{Conroy_2014} & Y & Triple (ETVs) \\ 
9777062 & 7466 & 12.24 & D & 2 & 19.23 & 25.89 & 0.60 & 1.38 & 2.8 & $\gamma$ Dor & \citetalias{Sandquist_2016} &  & Am+$\gamma$ Dor \\ 
9786821 & ... & 11.51 & D & 1 & 21.1221.12 & 0.03 & ... & ... & ... & Tidal? &  & Y & HB, spots \\ 
9788457 & 7939 & 13.01 & SD & 2 & 0.96 & 58.46 & 0.50 & 0.77 & 26.6 & $\delta$ Sct & \citetalias{Borkovits_2016} & Y & Triple (ETVs) \\ 
9843435 & 7291 & 14.78 & SD & 2 & 1.68 & 37.13 & 0.50 & 1.09 & 30.0 & $\delta$ Sct & \citetalias{Bradley_2015} & Y &  \\ 
9850387 & 6808 & 13.55 & D & 2 & 2.75 & 17.49 & 0.50 & 1.01 & 13.4 & $\gamma$ Dor, $\delta$ Sct, Rossby(?) &  & Y &  \\ 
9851944 & 6204 & 11.25 & SD & 2 & 2.16 & 20.75 & 0.50 & 1.00 & 23.0 & $\gamma$ Dor, $\delta$ Sct & \citetalias{Guo_2016} &  & $T\ind{eff}$ cool for $\delta$ Sct \\ 
9898401 & 7376 & 12.07 & ELL & 0 & 0.15 & 0.28 & ... & ... & ... & $\gamma$ Dor, $\delta$ Sct, Tidal? &  & Y & Very short $P\ind{orb}$: FP, triple? \\ 
9904059 & 4940 & 13.61 & D & 1 & 102.97 & 0.15 & ... & ... & ... & RG & Ben+ &  & SB1 \\ 
9953894 & 7295 & 11.08 & OC & 2 & 1.38 & 13.55 & 0.50 & ... & ... & $\gamma$ Dor, $\delta$ Sct &  & Y &  \\ 
9955262 & 6478 & 10.14 & D & 1 & 77.48 & 0.14 & ... & ... & ... & $\gamma$ Dor? & \citetalias{Borucki_2011} & Y & Noisy. Exoplanet? \\ 
9970396 & 4716 & 11.45 & D & 2 & 235.30 & 6.21 & 0.41 & 1.15 & 2.9 & RG & \citetalias{Gaulme_2013,Gaulme_2016a} &  & SB2 \\ 
10001167 & 4683 & 10.05 & D & 2 & 120.39 & 2.24 & 0.59 & 1.07 & 7.7 & RG & \citetalias{Gaulme_2013,Gaulme_2016a} &  & SB2 \\ 
10015516 & 5157 & 10.70 & D & 2 & 67.69 & 10.83 & 0.50 & 0.91 & 6.5 & RG+$\gamma$ Dor & Ben+ &  & SB1 \\ 
10031808 & ... & 9.56 & D & 2 & 8.59 & 25.82 & 0.49 & 1.54 & 9.4 & $\gamma$ Dor? & \citetalias{Helminiak_2017b} &  & Triple (SB3) \\ 
10074700 & 5066 & 14.62 & D & 2 & 365.65 & 3.07 & 0.49 & 1.37 & 0.9 & RG & Ben+ &  & SB1 \\ 
10092506 & 6526 & 11.21 & D & 0 & 31.04 & 0.05 & ... & ... & ... & $\delta$ Sct, Tidal & \citetalias{Dimitrov_2017} &  & HB, $P\ind{orb}$ harm., SB1 \\ 
10149845 & ... & 12.10 & ELL & 0 & 4.06 & 4.51 & ... & ... & ... & $\gamma$ Dor, $\delta$ Sct &  & Y &  \\ 
10275747 & 7362 & 12.80 & SD & 2 & 0.66 & 53.88 & 0.50 & 0.78 & 27.8 & $\gamma$ Dor, $\delta$ Sct &  & Y &  \\ 
10275887 & ... & 13.04 & D & 2 & 9.73 & 43.23 & 0.50 & 0.94 & 12.9 & $\delta$ Sct &  & Y &  \\ 
10383620 & 7470 & 12.83 & SD & 2 & 0.73 & 23.27 & 0.50 & 1.01 & 33.2 & $\delta$ Sct & \citetalias{Borkovits_2016} & Y & Triple (ETVs) \\ 
10417135 & 8421 & 13.25 & ELL & 0 & 1.20 & 2.49 & ... & ... & ... & $\gamma$ Dor, $\delta$ Sct & \citetalias{Zhou_2010} &  & $T\ind{eff}$ hot \\ 
10417986 & ... & 9.13 & ELL & 0 & 0.07 & 0.04 & ... & ... & ... & $\gamma$ Dor, $\delta$ Sct &  & Y & Very short $P\ind{orb}$: FP, triple? \\ 
10454725 & 6911 & 14.24 & SD & 2 & 0.83 & 32.61 & 0.50 & 1.04 & 41.4 & $\gamma$ Dor, $\delta$ Sct &  & Y & Low-amplitude spectrum \\ 
10485250 & 4957 & 15.79 & D & 2 & 16.47 & 0.72 & ... & ... & ... & RG &  & Y & $T_2<T_1$, spots, low SNR osc., sec. ecl. \\ 
10486425 & 7018 & 12.46 & D & 2 & 5.27 & 10.39 & 0.50 & 1.02 & 6.5 & $\gamma$ Dor & \citetalias{Alicavus_Soydugan_2014} &  &  \\ 
10491544 & 4835 & 13.44 & D & 2 & 22.77 & 3.08 & 0.71 & 1.09 & 3.1 & RG & \citetalias{Coughlin_2011} & Y & $T_2\sim T_1$ \\ 
10556068 & 7901 & 11.56 & ELL & 0 & 2.12 & 1.71 & ... & ... & ... & $\gamma$ Dor, $\delta$ Sct &  & Y & Sparse, low SNR \\ 
10581918 & 7252 & 12.80 & SD & 2 & 1.80 & 72.62 & 0.50 & 0.92 & 18.9 & $\delta$ Sct & \citetalias{Liakos_2017} &  &  \\ 
10614012 & 4859 & 9.71 & D & 1 & 132.17 & 1.00 & ... & ... & ... & RG & \citetalias{Beck_2014} &  & HB \\ 
10619109 & 7028 & 11.70 & SD & 2 & 2.05 & 26.88 & 0.50 & 1.05 & 21.2 & $\delta$ Sct? & \citetalias{Liakos_2017} &  &  \\ 
10661783 & 7887 & 9.59 & SD & 2 & 1.23 & 21.78 & 0.50 & 1.08 & 33.0 & $\gamma$ Dor, $\delta$ Sct & \citetalias{Southworth_2011,Lehmann_2013} &  &  \\ 
10684673 & 7106 & 11.12 & ELL & 0 & 0.19 & 1.06 & ... & ... & ... & $\gamma$ Dor, $\delta$ Sct & \citetalias{Turner_Maynard_2016,Turner_2019} &  & Sparse spectrum, Very short $P\ind{orb}$: FP, triple? \\ 
10735331 & 6694 & 13.38 & D & 1 & 213.40 & 1.29 & ... & ... & ... & $\gamma$ Dor, Rossby(?) & \citetalias{Santerne_2016} & Y &  \\ 
10735519 & 4881 & 11.78 & SD & 1 & 0.91 & 0.19 & ... & ... & ... & RG & \citetalias{Zhou_2010,Gaulme_2013,Bell_2019} &  & Likely FP (shallow ecl) \\ 
10736223 & 7797 & 13.65 & SD & 2 & 1.11 & 69.23 & 0.50 & 0.93 & 21.6 & $\delta$ Sct & \citetalias{Conroy_2014,Gies_2015} & Y & Triple (ETVs) \\ 
10789421 & ... & 11.85 & ELL & 0 & 0.78 & 0.54 & ... & ... & ... & $\gamma$ Dor &  & Y &  \\ 
10809677 & 4995 & 13.94 & D & 2 & 7.04 & 0.53 & 0.50 & 0.94 & 3.6 & RG & \citetalias{Zhou_2010,Coughlin_2011,Gaulme_2013} &  & Likely FP (shallow ecl) \\ 
10858117 & 5354 & 14.32 & D & 2 & 606.11 & 16.51 & 0.08 & 1.36 & 0.3 & RG? &  & Y & Low SNR, high ecc. \\ 
10905804 & 8000 & 14.42 & SD & 1 & 0.75 & 17.00 & ... & ... & ... & $\delta$ Sct & \citetalias{Conroy_2014} & Y & Triple (ETVs) \\ 
10920813 & ... & 13.76 & D & 2 & 53.74 & 31.75 & 0.46 & 1.25 & 5.6 & RG &  & Y & $T_2>T_1$, $\ell=1$ depleted \\ 
10972830 & ... & 10.17 & ELL & 0 & 0.68 & 0.02 & ... & ... & ... & Tidal? &  & Y & True binary? (very shallow ELL) \\ 
10979669 & 6697 & 12.33 & ELL & 0 & 0.93 & 6.47 & ... & ... & ... & $\gamma$ Dor, $\delta$ Sct? &  & Y &  \\ 
10989032 & 8620 & 13.87 & SD & 2 & 2.31 & 0.95 & 0.50 & 1.36 & 5.9 & $\delta$ Sct & \citetalias{Zhang_XB_2017} &  & WD+$\delta$ Sct \\ 
10991989 & 5021 & 10.28 & SD & 2 & 0.97 & 0.88 & 0.50 & 0.93 & 18.3 & RG & \citetalias{Gaulme_2013,Helminiak_2016} &  & Triple (ETVs) \\ 
11044668 & 4959 & 12.35 & D & 0 & 139.45 & 0.31 & ... & ... & ... & RG & \citetalias{Beck_2014} &  & HB \\ 
11135978 & 5004 & 12.33 & OC & 22 & 0.29 & 0.76 & ... & ... & ... & RG & \citetalias{Zhou_2010,Gaulme_2013} &  & Triple (ETVs)? \\ 
11147460 & 4855 & 13.91 & D & 2 & 4.11 & 0.47 & 0.50 & 1.02 & 8.6 & RG & \citetalias{Zhou_2010,Gaulme_2013,Lurie_2017} &  & Likely FP, asynchronous EB \\ 
11179657 & ... & 17.07 & ELL & 0 & 0.79 & 1.80 & ... & ... & ... & sdB g-modes & \citetalias{Pablo_2012} &  &  \\ 
11180361 & 8330 & 7.75 & SD & 1 & 0.53 & 0.16 & ... & ... & ... & $\gamma$ Dor, $\delta$ Sct & \citetalias{Murphy_2018} &  & $T\ind{eff}$ hot \\ 
11197853 & 4981 & 13.59 & D & 1 & 0.70 & 0.01 & ... & ... & ... & RG & \citetalias{Gaulme_2013} &  & FP (very shallow ecl.) \\ 
11285625 & 6882 & 10.14 & D & 2 & 10.79 & 11.57 & 0.50 & 0.98 & 5.0 & $\gamma$ Dor & \citetalias{Debosscher_2013} &  &  \\ 
11295347 & 7620 & 11.73 & ELL & 0 & 0.89 & 1.87 & ... & ... & ... & $\gamma$ Dor, $\delta$ Sct &  & Y &  \\ 
11401845 & 7590 & 14.36 & SD & 2 & 2.16 & 41.04 & 0.50 & 1.00 & 20.3 & $\delta$ Sct, Tidal & \citetalias{Lee_2017} &  &  \\ 
11408810 & 7555 & 12.61 & OC & 0 & 0.75 & 1.08 & ... & ... & ... & $\gamma$ Dor, Rossby(?) &  & Y &  \\ 
11494130 & 6330 & 10.99 & D & 0 & 18.96 & 0.06 & ... & ... & ... & Tidal & \citetalias{Thompson_2012,Smullen_2015} &  & HB, $P\ind{orb}$ harm., SB1 \\ 
11566064 & 6679 & 13.37 & D & 1 & 152.11 & 0.38 & ... & ... & ... & $\gamma$ Dor? & \citetalias{Dodson_Robinson_2012} & Y & Low SNR, planet candidate \\ 
11572363 & 6069 & 12.43 & D & 0 & 19.03 & ... & ... & ... & ... & Tidal &  & Y & HB, $P\ind{orb}$ harm. \\ 
11671429 & 7363 & 10.97 & D & 2 & 112.46 & 23.26 & 0.73 & 0.62 & 1.7 & $\gamma$ Dor, $\delta$ Sct & \citetalias{Uytterhoeven_2011} &  &  \\ 
11768970 & 5038 & 12.66 & D & 2 & 15.54 & 0.95 & 0.65 & 2.23 & 1.9 & RG & \citetalias{Coughlin_2011} & Y & Grazing eclipses \\ 
11817750 & 6930 & 12.25 & D & 2 & 9.76 & 3.85 & 0.50 & 1.08 & 3.9 & $\gamma$ Dor & \citetalias{Lurie_2017} &  &  \\ 
11820830 & 7007 & 12.09 & D & 1 & 12.73 & 1.14 & ... & ... & ... & $\gamma$ Dor &  & Y & HB \\ 
11825198 & ... & ... & OC & 0 & 2.16 & 1.56 & ... & ... & ... & $\gamma$ Dor, $\delta$ Sct, Rossby(?) &  & Y &  \\ 
11874338 & 5041 & 13.93 & D & 2 & 15.98 & 0.06 & ... & ... & ... & RG & \citetalias{Gaulme_2013} &  & FP (very shallow ecl.) \\ 
11923819 & 7724 & 11.56 & D & 2 & 33.16 & 21.09 & 0.34 & 1.31 & 2.2 & $\gamma$ Dor & \citetalias{Lurie_2017} &  &  \\ 
11968514 & 4940 & 11.45 & D & 2 & 1.04 & 0.00 & 0.50 & 1.00 & 17.9 & RG & \citetalias{Zhou_2010,Gaulme_2013} &  & Likely FP (shallow ecl) \\ 
11973705 & 7404 & 9.12 & ELL & 0 & 6.77 & 0.75 & ... & ... & ... & $\gamma$ Dor, $\delta$ Sct, Rossby(?) & \citetalias{Balona_2011} &  & SPB+$\delta$ Sct \\ 
12071006 & 7338 & 13.53 & SD & 2 & 6.10 & 88.68 & 0.50 & 1.00 & 19.1 & $\gamma$ Dor, $\delta$ Sct & \citetalias{Conroy_2014,Gies_2015} & Y & Triple (ETVs) \\ 
12167361 & 8017 & 10.38 & D & 1 & 47.93 & 0.21 & ... & ... & ... & $\gamma$ Dor & \citetalias{Lurie_2017} &  & $T\ind{eff}$ hot, KOI 980 \\ 
12216706 & 8530 & 15.01 & SD & 2 & 1.47 & 16.48 & 0.50 & 0.75 & 15.1 & $\delta$ Sct &  & Y & $T\ind{eff}$ hot \\ 
12216817 & 6681 & 10.66 & ELL & 0 & 0.25 & 4.10 & ... & ... & ... & $\gamma$ Dor, $\delta$ Sct & \citetalias{Turner_Maynard_2016} &  &  \\ 
12268220 & 7826 & 11.43 & D & 2 & 4.42 & 4.47 & ... & ... & ... & $\gamma$ Dor, $\delta$ Sct &  & Y &  \\ 
12367310 & 4965 & 13.84 & D & 2 & 8.63 & 3.44 & 0.51 & 1.12 & 4.1 & RG &  & Y & Triple (ETVs), spots \\ 
12470041 & 7290 & 13.41 & D & 1 & 14.67 & 2.90 & ... & ... & ... & $\gamma$ Dor, Tidal(?) &  & Y & HB \\ 
12645761 & 4844 & 13.37 & D & 2 & 5.425.42 & 1.25 & 0.50 & 1.10 & 8.2 & RG & \citetalias{Zhou_2010,Gaulme_2013} &  & FP or triple \\ 
12785282 & 6924 & 13.52 & SD & 2 & 0.79 & 15.40 & 0.50 & 0.96 & 26.0 & $\gamma$ Dor &  & Y &  \\ 
\label{tab:maintable}
\end{longtable}
\end{tiny}
}


\bibliographystyle{aa} 
\bibliography{bibi}


\appendix
\section{Table of pulsators in eclipsing binaries}

Literature abbreviation used in Table \ref{tab:maintable}: 

\citetalias{Alicavus_Soydugan_2014} \citet{Alicavus_Soydugan_2014}, \citetalias{Alicavus_2017} \citet{Alicavus_2017}, \citetalias{Armstrong_2014}, \citet{Armstrong_2014}, \citetalias{Balaji_2015} \citet{Balaji_2015}, \citetalias{Balona_2011} \citet{Balona_2011}, \citetalias{Balona_2013} \citet{Balona_2013}, \citetalias{Balona_2014} \citet{Balona_2014}, \citetalias{Balona_2015} \citet{Balona_2015},
\citetalias{Balona_2015b} \citet{Balona_2015b},
 \citetalias{Beck_2014} \citet{Beck_2014},
 \citetalias{Bedding_2011} \citet{Bedding_2011},
 \citetalias{Bell_2019} \citet{Bell_2019},
 \citetalias{Bognar_2015} \citet{Bognar_2015},
 \citetalias{Bowman_2016} \citet{Bowman_2016},
 \citetalias{Biersteke_2017} \citet{Biersteke_2017},
 \citetalias{Borucki_2011} \citet{Borucki_2011},
 \citetalias{Borkovits_2014} \citet{Borkovits_2014},
 \citetalias{Borkovits_2016} \citet{Borkovits_2016},
 \citetalias{Bradley_2015} \citet{Bradley_2015},
 \citetalias{Brogaard_2018} \citet{Brogaard_2018},
 \citetalias{Ceillier_2017} \citet{Ceillier_2017},
 \citetalias{Conroy_2014} \citet{Conroy_2014},
 \citetalias{Coughlin_2011} \citet{Coughlin_2011},
 \citetalias{Debosscher_2011} \citet{Debosscher_2011},
 \citetalias{Debosscher_2013} \citet{Debosscher_2013},
 \citetalias{Demircan_2015} \citet{Demircan_2015},
 \citetalias{Dimitrov_2017} \citet{Dimitrov_2017},
 \citetalias{Dodson_Robinson_2012} \citet{Dodson_Robinson_2012},
 \citetalias{Dong_2013} \citet{Dong_2013},
 \citetalias{Frandsen_2013} \citet{Frandsen_2013},
 \citetalias{Gaulme_2013} \citet{Gaulme_2013},
 \citetalias{Gaulme_2014} \citet{Gaulme_2014},
 \citetalias{Gaulme_2016a} \citet{Gaulme_2016a},
 \citetalias{Gaulme_Guzik_2014} \citet{Gaulme_Guzik_2014},
 \citetalias{Gies_2015} \citet{Gies_2015},
 \citetalias{Guo_2016} \citet{Guo_2016},
  \citetalias{Guo_2017} \citet{Guo_2017},
 \citetalias{Guo_2017b} \citet{Guo_2017b},
 \citetalias{Faigler_2015} \citet{Faigler_2015},
 \citetalias{George_2018} \citet{George_2018},
 \citetalias{Hekker_2010} \citet{Hekker_2010},
 \citetalias{Hambleton_2013} \citet{Hambleton_2013},
 \citetalias{Hambleton_2016} \citet{Hambleton_2016},
 \citetalias{Hambleton_2018} \citet{Hambleton_2018},
 \citetalias{Helminiak_2016} \citet{Helminiak_2016},
 \citetalias{Helminiak_2017} \citet{Helminiak_2017},
 \citetalias{Helminiak_2017b} \citet{Helminiak_2017b},
 \citetalias{Katsova_Nizamov_2018} \citet{Katsova_Nizamov_2018},
  \citetalias{Kjurkchieva_2016} \citet{Kjurkchieva_2016},
 \citetalias{Kjurkchieva_2016b} \citet{Kjurkchieva_2016b},
 \citetalias{Kouzuma_2018} \citet{Kouzuma_2018},
 \citetalias{Kurtz_2015b} \citet{Kurtz_2015b},
 \citetalias{Kuszlewicz_2019} \citet{Kuszlewicz_2019} 
 \citetalias{Lee_2014} \citet{Lee_2014},
 \citetalias{Lee_2015} \citet{Lee_2015},
 \citetalias{Lee_2016} \citet{Lee_2016},
 \citetalias{Lee_2016b} \citet{Lee_2016b},
 \citetalias{Lee_2017} \citet{Lee_2017},
 \citetalias{Lehmann_2013} \citet{Lehmann_2013},
 \citetalias{LilloBox_2014} \citet{LilloBox_2014},
 \citetalias{Lurie_2017} \citet{Lurie_2017},
 \citetalias{Liakos_2017} \citet{Liakos_2017},
 \citetalias{Liakos_2018} \citet{Liakos_2018},
 \citetalias{Liakos_Niarchos_2017} \citet{Liakos_Niarchos_2017},
 \citetalias{Maceroni_2014} \citet{Maceroni_2014},
 \citetalias{Manuel_Hambleton_2018} \citet{Manuel_Hambleton_2018},
 \citetalias{Masuda_2017} \citet{Masuda_2017},
 \citetalias{Matson_2017} \citet{Matson_2017},
 \citetalias{McNamara_2012} \citet{McNamara_2012},
 \citetalias{Moya_2017} \citet{Moya_2017},
 \citetalias{Murphy_2015} \citet{Murphy_2015},
 \citetalias{Murphy_2018} \citet{Murphy_2018},
 \citetalias{Niemczura_2015} \citet{Niemczura_2015},
 \citetalias{Niemczura_2017} \citet{Niemczura_2017},
 \citetalias{Ozdarcan_2017} \citet{Ozdarcan_2017},
 \citetalias{Orosz_2015} \citet{Orosz_2015},
 \citetalias{Pablo_2012} \citet{Pablo_2012},
  \citetalias{Ramsay_2014} \citet{Ramsay_2014},
 \citetalias{Rappaport_2013} \citet{Rappaport_2013},
 \citetalias{Rappaport_2015} \citet{Rappaport_2015},
  \citetalias{Rawls_2016} \citet{Rawls_2016},
 \citetalias{Rowe_2010} \citet{Rowe_2010},
 \citetalias{Rowe_2015} \citet{Rowe_2015},
 \citetalias{Sowicka_2017} \citet{Sowicka_2017},
 \citetalias{Saio_2018} \citet{Saio_2018},
  \citetalias{Sandquist_2016} \citet{Sandquist_2016},
 \citetalias{Santerne_2016} \citet{Santerne_2016},
 \citetalias{Shporer_2016} \citet{Shporer_2016},
 \citetalias{Smullen_2015} \citet{Smullen_2015},
 \citetalias{Southworth_2011} \citet{Southworth_2011},
 \citetalias{Southworth_2015} \citet{Southworth_2015},
 \citetalias{Themessl_2018} \citet{Themessl_2018},
 \citetalias{Thompson_2012} \citet{Thompson_2012},
 \citetalias{Turner_Maynard_2016} \citet{Turner_Maynard_2016},
 \citetalias{Turner_2015} \citet{Turner_2015},
 \citetalias{Turner_2019} \citet{Turner_2019},
 \citetalias{Uytterhoeven_2011} \citet{Uytterhoeven_2011},
 \citetalias{Welsh_2011} \citet{Welsh_2011},
 \citetalias{Yakut_2015} \citet{Yakut_2015},
 \citetalias{Zhang_2017} \citet{Zhang_2017},
 \citetalias{Zhang_2017b} \citet{Zhang_2017b},
 \citetalias{Zhang_XB_2017} \citet{Zhang_XB_2017},
 \citetalias{Zhou_2010} \citet{Zhou_2010},
 \citetalias{Zimmerman_2017} \citet{Zimmerman_2017}

\end{document}